\documentstyle[fps]{book}

\newcommand{\be}{\begin{equation}}
\newcommand{\ee}{\end{equation}}
\newcommand{\bd}{\begin{displaymath}}
\newcommand{\ed}{\end{displaymath}}
\newcommand{\ba}{\begin{eqnarray}}
\newcommand{\ea}{\end{eqnarray}}
\newcommand{\bi}{\begin{itemize}}
\newcommand{\ei}{\end{itemize}}
\newcommand{\noi}{\noindent}

\renewcommand{\Re}{\mathop{\rm Re}\nolimits}

\newcommand{\tr}{\mathop{\rm{tr}}}

\newcommand{\LP}{\lambda{\bf \Phi}^4}
\newcommand{\HT}{\mathcal {HT}}
\newcommand{\HS}{\mathcal {HS}}

\newlength{\baselineskipsave}
\newlength{\blss}
\def\dsl{\raise.15ex\hbox{/} \kern-.57em\partial}
\def\psl{\raise.15ex\hbox{/} \kern-.57em p}
\def\ksl{\raise.15ex\hbox{/} \kern-.57em k}
\def\qsl{\raise.15ex\hbox{/} \kern-.57em q}

\begin{document}

\title{\begin{center} {\bf Estudio en la red de transiciones
d\'ebiles de primer orden en dimensi\'on 4.} 
\end{center} }

\author{Isabel Campos Plasencia}
\date{Junio 1998}

\pagenumbering{arabic}
\pagestyle{plain}


\thispagestyle{empty}

\begin{center}

\begin{figure}[!t]
\hspace*{4.4cm}
\leavevmode
\fpsxsize=3cm
\def\fpsangle{0}
\fpsbox[70 90 579 760]{escudo.ps} 
\end{figure}


{\LARGE Universidad de Zaragoza}

\vspace{0.25cm}

{\Large Facultad de Ciencias}

\vspace{0.25cm}

{\large Departamento de F\'{\i}sica Te\'orica}

\vspace{1.5cm}

{\LARGE\bf ESTUDIO EN EL RET\'ICULO DE

\vspace{0.25cm}

TRANSICIONES D\'EBILES

\vspace{0.25cm}

DE PRIMER ORDEN

\vspace{0.4cm}

EN DIMENSI\'ON 4.}

\end{center}

\vspace{2.5cm}

\begin{center}
{\sl Memoria de tesis doctoral}  \\[2mm]
{\sl presentada por}  \\[2mm]
{\large\bf Isabel M. Campos Plasencia}  \\[2mm]
{\sl Zaragoza, Junio 1998.} 
\end{center}

\cleardoublepage

\thispagestyle{empty}


\noi ALFONSO TARANCON LAFITA, Profesor titular de F\'{\i}sica Te\'orica de
la Universidad de Zaragoza,

\vspace{2mm}

\noi CERTIFICA

\vspace{2mm}

\noi que la presente memoria, {\em ``Estudio en el ret\'{\i}culo de 
transiciones d\'ebiles de primer orden en dimensi\'on 4''}, 
ha sido realizada en el 
Departamento de F\'{\i}sica Te\'orica de la Universidad de Zaragoza bajo su 
direcci\'on, y autoriza su presentaci\'on para que sea calificada como Tesis
Doctoral.

\vspace{2mm}

\begin{flushright} Zaragoza, JUNIO 1998. \end{flushright}

\vspace{3cm}

\begin{flushright} Fdo: Alfonso Taranc\'on Lafita. \end{flushright}

\cleardoublepage


\chapter*{Agradecimientos \label{capthanks}}
\thispagestyle{empty}

Lleg\'o la hora de los agradecimientos, y realmente no sabe una
por donde empezar. A lo largo de estos a\~nos muchas personas
me han prestado su ayuda a nivel cient\'{\i}fico y su apoyo a
nivel personal, a todos ellos quiero dejar patente mi m\'as
profundo reconocimiento.

En primer lugar, gracias a Alfonso Taranc\'on, por querer asumir la 
res\-ponsabilidad de dirigir mi trabajo durante estos a\~nos, 
por todo lo que me ha ense\~nado, por las largas horas de explicaciones,
y en especial por su esfuerzo en inculcarnos a todos una forma de
trabajo en concordancia con el esp\'{\i}ritu cient\'{\i}fico.

Gracias en general al resto de miembros de la colaboraci\'on RTN:
Jose, Carlos, David, H\'ector, V\'{\i}ctor, Juan Jes\'us, Antonio,
Luis Antonio, Andr\'es y Jose Luis. Se
espera del doctorando que destaque a las personas y no a las entidades
pero realmente es muy dif\'{\i}cil. 

Sin embargo no ser\'{\i}a justa si no menciono especialmente a 
Luis Antonio Fern\'andez para darle las gracias por su
generosidad a la hora de transmitir lo mucho que sabe, por su paciencia
a la hora de contestar a mis muchas preguntas, y por sus valiosos
consejos.

Gracias tambi\'en a Jose Luis Alonso por tratar de contagiarnos a
todos su inter\'es y dinamismo a la hora de dedicarnos a la
investigaci\'on cient\'{\i}fica, y por los \'animos que me ha dado siempre.

A Andr\'es Cruz, por la colaboraci\'on mantenida durante este tiempo,
durante la cual me ha hecho llegar grandes dosis de sensatez.

Por supuesto el ambiente de trabajo creado por los miembros del
departamento de F\'{\i}sica Te\'orica ha ayudado a la feliz consecuci\'on
de esta tesis. Quiero destacar especialmente a Isabel y a Pedro, 
sin los cuales, esto se nos vendr\'{\i}a encima sin duda, y a mis
sucesivos compa\~neros de despacho Peppe Bimonte, Arjan Van der Sijs 
y Jes\'us Salas.

Vamos ahora de paseo por Europa:

No tengo palabras para agradecer a Richard Kenway su hospitalidad, su
encanto, su ayuda durante y despu\'es de mi estancia en Edimburgo.
 
Como ya no tengo m\'as palabras, no se como decirle a Christian Lang
que le estoy muy agradecida por su amabilidad
extrema, su simpat\'{\i}a, sus clases de Ingl\'es, (y de Austr\'{\i}aco)
y en general por la acogida en su departamento durante mi estancia en Graz, 
tan importante para mi formaci\'on tanto en el plano cient\'{\i}fico
como en el humano, en resumen, gracias Christian.

Los comentarios (de todo tipo, remarco) de Jiri Jers\'ak significaron
mucho para mi en un momento dif\'{\i}cil, y vinieron a despejar
muy negros nuba\-rrones. Gracias por esos d\'{\i}as en Aachen y por las
\'utiles discusiones que hemos mantenido desde entonces. Es una pena
que no nos pongamos de acuerdo, pero todo llegar\'a.

Por supuesto hay siempre que mirar hacia el futuro, y as\'{\i} he 
agradecer a Istv\'an Montvay
la magn\'{\i}fica oportunidad que me brinda para continuar 
mi formaci\'on a nivel post-doctoral.

Ya en el plano personal, no puedo olvidarme de los amigos que han estado
siempre al otro lado de la pantalla, todos ellos por 
esos mundos de Dios: Mar\'{\i}a
Pilar y Juan Pablo, vaya mails hijos m\'{\i}os;
Bel\'en, Nuria y s\'olo falta Silvia, siempre en el otro lado del
mundo, pero afortunadamente siempre tan cercana.

Y a Paula, y a Ana, por los buenos ratos que hemos pasado, y los que nos quedan
por pasar, of course.

A esa panda que montamos en Edimburgo aquel glorioso verano
del 95, y que siguen en mi vida desde entonces: Juan,
Mar\'{\i}a, Dora, gracias por ser tan majos; Maria Grazia, gracias
por ser tan sensata siempre; por supuesto al inefable Tim,
que no tiene arreglo (al menos que yo sepa) pero que tambi\'en
merece estar aqu\'{\i}.

Mis compa\~neros de piso han sido los que me han aguantado d\'{\i}a 
a d\'{\i}a, lo cual algunos d\'{\i}as no era trivial. 
Estos pobres han soportado pacientemente todas
mis veleidades, todas: Jano, Javier, Juan, Oscar, Alfonso y last but not least,
{\sl moquito}, el c\'ocker m\'as guapo del mundo.

Ya para el final el agradecimiento a quien m\'as se lo merece, 
a mi familia, por proporcionarme la mejor educaci\'on posible, 
y por todo su apoyo. A mi madre, a mi hermano, y
especialmente a mi padre, a cuya memoria est\'a
dedicado el trabajo de todos estos a\~nos.

\begin{flushright}
{\sl Isabel, Junio 1998}
\end{flushright}

\newpage

\chapter*{Presentaci\'on \label{cappresen}}
\thispagestyle{empty}

Desde que Feynman en 1948 introdujera la
integral de camino en Teor\'{\i}a Cu\'antica de Campos (TCC),
una gran parte de los f\'{\i}sicos te\'oricos de part\'{\i}culas
han trabajado usando este formalismo.

Las predicciones obtenidas estuvieron practicamente limitadas
al regimen perturbativo (peque\~nas fluctuaciones en torno al
campo libre) en los primeros a\~nos, puesto que en este l\'{\i}mite
las integrales son Gaussianas y por tanto se pueden resolver por
m\'etodos anal\'{\i}ticos. 

No perturbativamente, una posible regularizaci\'on se obtiene
reemplazando el espacio-tiempo continuo por una red discreta (Wilson, 1974).
Por supuesto la pregunta a responder es si las observaciones
f\'{\i}sicas que tenemos en el continuo se corresponden de alguna
manera con la formulaci\'on de la teor\'{\i}a en la red. Si se puede
probar que el l\'{\i}mite continuo de una teor\'{\i}a de campos en la
red existe, uno podr\'{\i}a usar la definici\'on de la teor\'{\i}a
en la red como definici\'on de la teor\'{\i}a en el continuo.

El trabajo pionero de Wilson desencaden\'o inter\'es
por atacar la soluci\'on de las integrales funcionales numericamente.
El impresionante auge de la inform\'atica en la \'ultima d\'ecada, 
con el consiguiente aumento de los recursos accesibles de potencia 
de calculo, ha favorecido sin duda el avance de este tipo de estudios,
gracias a los cuales se ha obtenido conocimiento sobre los aspectos
no pertubativos de TCC.

A lo largo de esta memoria se van a estudiar las propiedades
de distintos modelos relevantes para la F\'{\i}sica de Part\'{\i}culas 
mediante su formulaci\'on en una red espacio-temporal. 
Los an\'alisis estar\'an basados en resultados que se obtienen 
con la ayuda de ordenadores.
Para ello se van a usar ideas provenientes de la
Mec\'anica Estad\'{\i}stica Cl\'asica.
As\'{\i} utilizaremos el
concepto de transici\'on de fase y de par\'ametro de orden,
aplic\'andolos en diferentes contextos.

Cualquiera que est\'e interesado en el estudio num\'erico de
fen\'omenos cr\'{\i}ticos debe considerar como fundamental el hecho 
de que las transiciones de fase ocurren s\'olo en el l\'{\i}mite 
termodin\'amico. Cuando se hacen si\-mulaciones num\'ericas se est\'a
forzado a trabajar con un n\'umero finito de grados de libertad,
por tanto no hay transiciones de fase. 
Sin embargo los efectos que se observan en redes de tama\~no finito
son precursores del verdadero comportamiento l\'{\i}mite. La teor\'{\i}a
de {\sl Finite Size Scaling} dicta la forma concreta en que esto ocurre
y por lo tanto, 
mediante el estudio de la evoluci\'on de los observables
de inter\'es con el tama\~no de la red se pueden obtener predicciones
sobre el comportamiento en el l\'{\i}mite termodin\'amico.
Las t\'ecnicas de tama\~no finito son por lo tanto una herramienta 
fundamental en los an\'alisis que se realizan a lo largo de esta
memoria.

La estructura de la memoria es la siguiente. Se ha querido,
por motivos de completitud, hacer una breve introducci\'on
al estudio de fen\'omenos cr\'{\i}ticos en la red en un cap\'{\i}tulo
preliminar. Tr\'as este preliminar,
en primer lugar se describe el modelo O(4) Anti-Ferromagn\'etico.
La motivaci\'on inicial de este trabajo fue la busqueda de puntos
fijos no triviales en cuatro dimensiones en el sector Anti-Ferromagn\'etico
del Modelo Est\'andar.

El segundo cap\'{\i}tulo corresponde al estudio de una transici\'on
que podr\'{\i}amos decir cl\'asica en el marco de las simulaciones
en la red de TCC, la transici\'on de fase
en el modelo SU(2)-Higgs a temperatura cero. 

En el tercer cap\'{\i}tulo se ha abordado un problema que durante
los \'ultimos a\~nos est\'a sometido a una vigorosa discusi\'on:
la influencia de las condiciones de contorno
en el orden de la transici\'on de fase de la teor\'{\i}a U(1) compacta
pura gauge.

Debido a la conexi\'on existente entre Mec\'anica Estad\'{\i}stica
y TCC es \-natural realizar incursiones
en problemas puramente de Mec\'anica Estad\'{\i}stica, dentro
de la interacci\'on con otros grupos de investigaci\'on.
En mi caso esta incursi\'on se hizo estudiando 
un problema podr\'{\i}amos decir que de {\sl moda} como es la formulaci\'on
de modelos adaptados a la descripci\'on de procesos que envuelven
un flujo de informaci\'on. Este trabajo se describe en el cap\'{\i}tulo
final de esta memoria.

La estructura de cada cap\'{\i}tulo consta de 
una introducci\'on para motivar y presentar el trabajo
realizado, seguida de una exposici\'on del m\'etodo utilizado y 
de los resultados obtenidos.

Finalmente se comentan las conclusiones obtenidas y se detalla
la lista de publicaciones.

\newpage

\chapter{Preliminares \label{capPRE}} 
\thispagestyle{empty}
\markboth{\protect\small CAP\'ITULO \protect\ref{capPRE}}
{\protect\small {\sl Preliminares}}

\cleardoublepage

\section{Regularizaci\'on no perturbativa en Teor\'{\i}a Cu\'antica
de Campos}

La Teor\'{\i}a Cu\'antica de Campos es el marco m\'as adecuado
para describir las interacciones fuerte y electrod\'ebil. En el
Modelo Est\'andar (SM) \cite{SM0} la teor\'{\i}a unificada que describe estas
interacciones est\'a basada en el grupo gauge $SU(3) \otimes SU(2) 
\otimes U(1)_{\rm Y}$. 

Uno de los posibles formalismos que permiten estudiar 
las propiedades de una TCC es el de la integral de camino. A esta
formulaci\'on se llega a trav\'es de la generalizaci\'on del concepto
de integral de camino introducido por Feynmann para la Mec\'anica
Cu\'antica, y su extensi\'on a tiempos imagi\-narios (integral de
camino Eucl\'{\i}dea). En este contexto las funciones de Green Eucl\'{\i}deas
o funciones de Schwinger a partir de las cuales se construye la TCC
\cite{OSTER0}, se pueden escribir como los momentos de una determinada
medida de probabilidad:
\be
{\mathcal S}_{\rm n} (x_1, \dots ,x_n) = \int [d \Phi] 
\Phi(x_1) \cdots \Phi(x_n) e^{-S[\Phi]} \ ,
\label{SWING}
\ee

donde $S[\Phi]$ es la acci\'on Eucl\'{\i}dea \cite{SYM0}:
\be
S[\Phi] = \int dt {\mathcal L}(\Phi, \partial_{\mu} \Phi)
\ee

La formulaci\'on de la integral de camino no s\'olo proporciona una
visi\'on f\'{\i}sica llamativa de la evoluci\'on de un sistema
cu\'antico como una suma sobre caminos cl\'asicos, sino que 
abre el camino a c\'alculos tanto anal\'{\i}ticos como num\'ericos.
Adem\'as ayuda a comprender la relaci\'on entre TCC y fen\'omenos 
cr\'{\i}ticos en Mec\'anica Estad\'{\i}stica (ME) cl\'asica. Para ver
esta relaci\'on consideremos la funci\'on de correlaci\'on a $n$ puntos
de una TCC escalar en el espacio Eucl\'{\i}deo:
\be
\langle 0 |{\mathcal T} ( \Phi(x_1) \cdots \Phi(x_n) ) | 0 \rangle =
\frac{1}{Z} \int [d \Phi] \Phi(x_1) \cdots \Phi(x_n) e^{-S[\Phi]}  \ ,
\label{CORREL}
\ee 
donde hemos incluido la normalizaci\'on $Z$ de manera que el valor
medio de la identidad sea uno:
\be
Z = \int [d \Phi] e^{-S[\Phi]}   \ .
\label{PARTI}
\ee
Estas ecuaciones se pueden interpretar como la funci\'on 
de correlaci\'on y la funci\'on de
partici\'on respectivamente de un sistema 
de ME cl\'asica. La TCC en $d_s$ dimensiones
espaciales se puede hacer equivalente a un sistema de ME cl\'asica
en equilibrio en $d= d_s + 1$ dimensiones Eucl\'{\i}deas. 
La acci\'on Eucl\'{\i}dea juega el papel de $E/k_b T$, donde con $E$ se
denota la energ\'{\i}a cl\'asica de la configuraci\'on \footnote{
Se ha usado el convenio $k_b=1$ a lo largo de esta memoria}.

Una primera aproximaci\'on a la
soluci\'on de este tipo de integrales funcionales es considerar peque\~nas
fluctuaciones en torno al campo libre (integral gausiana), 
este es el contexto de la Teor\'{\i}a de Perturbaciones. La
resoluci\'on perturbativa pasa por hacer un desarrollo en serie
de la parte no gausiana de (\ref{CORREL}). Sin embargo al calcular
la contribuci\'on de los distintos t\'erminos de la serie perturbativa
aparecen divergencias asociadas a los grandes momentos
como consecuencia de trabajar con un numero infinito de grados
de libertad (la variable $x$ es continua). En los c\'alculos
perturbativos estas divergencias ultravioletas (UV) han de ser compensadas
mediante la introducci\'on de contrat\'erminos en el lagrangiano
de partida, lagrangiano desnudo, expresado en funci\'on de par\'ametros
desnudos, acoplamientos y masas.

Para hacer c\'alculos hay que dar sentido a estas integrales divergentes,
mediante un {\sl {cut-off}} en momentos, trabajando en dimensi\'on menor que 4,
etc. Este procedimiento se denomina regularizaci\'on. Una vez que se han
hecho los c\'alculos se suprime la regularizaci\'on, para ello se
fijan un cierto n\'umero de par\'ametros (acoplamientos y masas) a sus
valores renorma\-lizados y se hace tender el {\sl {cut-off}} a infinito o
la dimensi\'on a 4. En este proceso, los infinitos que aparecen al calcular
los t\'erminos de la serie perturbativa son absorbidos en los par\'ametros
desnudos y las constantes de renormali\-zaci\'on que aparecen en los
contrat\'erminos. Una teor\'{\i}a se dice renormali\-zable cuando bastan
un n\'umero finito de contrat\'erminos para hacer la teor\'{\i}a finita
a todos los ordenes de teor\'{\i}a de perturbaciones. En dimensi\'on 4
las teor\'{\i}as gauge, como el SM, son renormalizables \cite{THOOF0},
y por tanto no hay ning\'un problema en dar sentido perturbativo a
la definici\'on (\ref{SWING}).

Resumiendo, las integrales funcionales del tipo (\ref{SWING}) no est\'an
definidas apropiadamente, 
su expresi\'on es puramente formal y para darle sentido
hay que regularizar la teor\'{\i}a de manera que no aparezcan infinitos
en el UV. En teor\'{\i}a de perturbaciones son posibles varias 
regularizaciones que nos permiten dar sentido a las integrales. 

Sin embargo, si se quiere dar cuenta de fen\'omenos no accesibles al
c\'alculo perturbativo, como son por ejemplo la rotura espont\'anea
de la simetr\'{\i}a quiral, o la generaci\'on de masas para los hadrones,
necesitamos una regularizaci\'on que permita dar sentido al formalismo
de la integral de camino en el regimen no perturbativo.
Se obtiene una regularizaci\'on no perturbativa
introduciendo una longitud elemental
``$a$'' en el espacio-tiempo \cite{WILSON0}. En este esquema las coordenadas
del espacio-tiempo $x_{\mu}$ est\'an discretizadas:
\be
x_{\mu} = a n_{\mu}, \quad n_{\mu} \in {\mathcal Z}  \ .
\ee
Debido a la discretizaci\'on,
la transformada de Fourier de una funci\'on G(x)
\be
{\tilde G}(k) = a^4 \sum_{n_{\mu}} e^{i a k_{\mu} n_{\mu}} G(x) \ ,
\ee
es peri\'odica, es decir invariante bajo
\be
k_{\mu} \rightarrow k_{\mu} + 2\pi/a \ .
\ee
Por lo tanto se puede restringir $k_{\mu}$ a todo intervalo de longitud
$2\pi/a$, por ejemplo $- \pi/a < k_{\mu} \leq \pi/a$,
que recibe el nombre de primera zona de Brillouin.
Es decir, la consecuencia de haber introducido una longitud elemental
``$a$'', es que hemos regularizado la teor\'{\i}a en el UV puesto que ahora
los momentos est\'an restringidos a los valores del interior de una
caja de arista $2\pi /a$.
Esta regularizaci\'on es m\'as potente que las
regularizaciones perturbativas, puesto que se espera que sirva
como definici\'on de la teor\'{\i}a completa, no s\'olo de su
aspecto perturbativo.

\section{Formulaci\'on en la red de la TCC}

Para una teor\'{\i}a escalar hemos visto que si colocamos las
variables din\'amicas en los nodos de la red, el sistema es equivalente
a un modelo de ME donde las variables son los valores de los campos.

En general en los nodos de la red se colocan
los campos escalares y los fermi\'onicos, las derivadas se sustituyen 
por diferencias finitas y los campos vectoriales se colocan en los 
segmentos que unen nodos vecinos (links).
En cuanto a estos \'ultimos, parecer\'{\i}a que
la elecci\'on natural es colocar los campos vectoriales (potenciales
gauge) en las links. Sin embargo
hacerlo as\'{\i} rompe la invariancia gauge poniendo en peligro
la renormalizabilidad. La forma de poner los campos gauge en la red
es poner en las links no el mismo campo gauge, sino los elementos
del grupo gauge. La link que conecta el site ${\rm n}$ con el site 
de coordenadas ${\rm n} + \hat{\mu}$ se denota por $U_{\mu}({\rm n})$,
\be
U_{\mu} ({\rm n}) = e^{iag A_{\rm n, \mu}^b {\mathcal T}_b} \ ,
\ee
donde $\hat{\mu}$ es el vector unitario en la direcion espacio-temporal
$\mu$, y ${\mathcal T}_b$ son los generadores del grupo gauge
que estemos considerando.

En la red es conveniente trabajar con magnitudes adimensionales
de manera que los operadores definidos en el continuo se redefinen
en la red de la siguiente forma:
\be
Q_{\rm {cont}}  \rightarrow {\hat Q}_{\rm {lat}} = a^s Q_{\rm {cont}} \ ,
\ee
donde $s$ es la dimensi\'on de $Q_{\rm {cont}}$ en unidades de energ\'{\i}a.

La formulaci\'on de la integral de camino en TCC y su regularizaci\'on
en la red, permite la evaluaci\'on num\'erica de las integrales
funcionales. En particular estaremos interesados en calcular el promedio
de operadores invariantes gauge:
\be
\langle {\mathcal O}(\Phi, A) \rangle =
\frac{1}{Z} \int [d \Phi][d A] {\mathcal O}(\Phi,A) e^{-S[\Phi,A]}  \ ,
\label{OPE}
\ee
en una red finita cuadridimensional de lado $L$ mediante integraci\'on
num\'erica. Para obtener informaci\'on sobre (\ref{OPE}) numericamente
definimos
\be
\bar{\mathcal O} \equiv \lim_{\Theta \rightarrow \infty}
\int_0^{\Theta} dt {\mathcal O}
(\Phi(t),A(t))/{\Theta} \ ,
\label{MEDIA}
\ee
donde $t$ es un tiempo de ordenador, definido mediante la
din\'amica que nosotros hayamos elegido.

En estas condiciones, se evalua el operador ${\mathcal O}$ y se suma sobre
las dife\-rentes configuraciones de los campos $[\Phi(\tau),A(\tau)]$
generadas en el curso de la din\'amica, que en general es discreta,
es decir,  se generan configuraciones numeradas 1,2,3 etc..., con lo
cual la integral (\ref{MEDIA}) ser\'a en realidad una suma del operador
${\mathcal O}$ evaluado sobre las diferentes configuraciones.

La din\'amica se suele elegir definiendo una matriz de probabilidad
de transici\'on de una configuraci\'on a otra, que sea erg\'odica
(para que el sistema no quede atrapado en un subconjunto del espacio
fase), y que deje invariante la distribuci\'on de probabilidad de Boltzmann. 
Una tal din\'amica se llama proceso de Markov y la secuencia de
configuraciones generadas se llama cadena de Markov. Es posible demostrar
que una din\'amica que cumple estas condiciones converge a la distribuci\'on
de equilibrio de Boltzmann definida en (\ref{OPE}), y
por tanto se tiene que:
\be
\bar{\mathcal O} = \langle {\mathcal O} \rangle
\label{IGUAL}
\ee
En la practica hay dos problemas a tener en cuenta. El primero
es que no se puede generar un n\'umero infinito de configuraciones.
Cuando se considera un n\'umero finito
de configuraciones $N$, la expresi\'on (\ref{IGUAL}) es cierta
con un error que es del orden de $1/\sqrt{N}$
(teorema del l\'{\i}mite central). El segundo problema es que la
din\'amica que usamos genera configuraciones que est\'an correlacionadas.
El tiempo que tarda el sistema en perder la memoria de la configuraci\'on
de la que viene se denomina tiempo de autocorrelaci\'on y se denota
por $\tau$. Hablando cualitativamente, si se generan $N$ configuraciones
el n\'umero de configuraciones independientes es $N/\tau$. Para una
exposici\'on detallada sobre los errores intr\'{\i}nsecos a 
una simulaci\'on de Monte Carlo vease \cite{SOKAL0}.

Del mismo modo que se hace en teor\'{\i}a de perturbaciones, una vez
que hemos dado sentido a las integrales hay que eliminar la regularizaci\'on.
El proceso de eliminar la regularizaci\'on impuesta por la formulaci\'on
en la red se denomina tomar el l\'{\i}mite continuo.
La forma precisa en que hay que tomar este l\'{\i}mite la dicta el
Grupo de Renormalizaci\'on (RG). En particular cuando se est\'a suficientemente
proximo al l\'{\i}mite continuo las magnitudes f\'{\i}sicas no deben depender
de los acoplamientos desnudos ni de ''$a$''. Es decir, la renormalizabilidad
de la teor\'{\i}a implica la existencia de trayectorias de f\'{\i}sica
constante en el espacio de acoplamientos desnudos. 

En este contexto el
proceso a seguir ser\'{\i}a el siguiente. Se eval\'ua en la red el
operardor ${\hat Q}(g)$ correspondiente al observable $Q$ para distintos
valores de los acoplamientos $g$. Despu\'es se restauran las dimensiones
multiplicando ${\hat Q}(g)$ por la potencia adecuada de ``$a$'', obtenemos
as\'{\i} $Q(g,a)$. La forma en que depende $g$ de ``$a$'' la sabemos
perturbativamente, de manera que al menos en la regi\'on perturbativa
sabemos por donde pasan las trayectorias de f\'{\i}sica constante. Usando
pues la predicci\'on perturbativa podemos conocer $Q(g(a),a)$. Si el
observable $Q$ no depende de ``$a$'' podremos decir que estamos estudiando
correctamente la f\'{\i}sica del continuo mediante nuestra formulaci\'on
en la red. Esto es lo que se conoce como {\sl scaling asint\'otico}. 
Una condici\'on m\'as suave que la anterior es lo que se conoce 
como {\sl scaling}. Aqu\'{\i} se exige que cocientes adimensionales de
cantidades con dimensiones (por ejemplo $m_{0^{++}}/m_{2^{++}}$)
permanezcan constantes al variar los par\'ametros y acercarnos al l\'{\i}mite
cont\'{\i}nuo. Por supuesto scaling asint\'otico implica scaling pero
no a la inversa.

Supongamos que queremos calcular el espectro de masas de la teor\'{\i}a,
calculando para ello el decaimiento en la red de las funciones de correlaci\'on
apropiadas para largos tiempos Eucl\'{\i}deos. La masa m\'as peque\~na de
la teor\'{\i}a est\'a determinada por la longitud de correlaci\'on $\xi$ m\'as
larga, que se define como la escala de longitud sobre la cual la part\'{\i}cula
se puede propagar con una amplitud significativa. La renormalizabilidad
de la teor\'{\i}a exige que la masa f\'{\i}sica sea finita, por lo tanto
la masa en unidades de la red, ${\hat m} = a m$, tiene que ser cero. Esto
implica que la longitud de correlaci\'on medida en unidades de la red
${\hat \xi}$ debe diverger. 

Resumiendo, el l\'{\i}mite continuo de
una teor\'{\i}a de campos en la red puede tener lugar s\'olo en un
punto cr\'{\i}tico del sistema de ME al cual es equivalente la TCC,
y que est\'a definido mediante la funci\'on de partici\'on (\ref {PARTI}).
Esta condici\'on es por otra parte l\'ogica, puesto que s\'olo cuando
$\xi \rightarrow \infty$ el sistema pierde memoria de la red subyacente.

\section{Transiciones de fase en el ret\'{\i}culo}

Los mecanismos de transici\'on de fase aparecen frecuentemente en
f\'{\i}sica de part\'{\i}culas para dar cuenta de las propiedades
observadas en la naturaleza. 
As\'{\i} por ejemplo, la generaci\'on
de masa para los bosones vectoriales se entiende a trav\'es del
mecanismo de Higgs, que se origina en una transici\'on de fase producida
por la ruptura espont\'anea de la simetr\'{\i}a del campo de Higgs.

La ausencia de una transici\'on de fase en SU(2) y SU(3) entre la regi\'on de
acoplamiento fuerte y acoplamiento d\'ebil, sugerida 
por los trabajos pioneros de Creutz \cite{CREUTZ0}, se toma como criterio
para pensar que la propiedad de confinamiento
que puede ser demostrada en el r\'egimen de acoplamiento
fuerte \cite{SEILER0}, se mantiene en el r\'egimen de 
acoplamiento d\'ebil, y por tanto
dan fundamento a la creencia de que la QCD es una teor\'{\i}a
razonable para estudiar la interacci\'on fuerte.

Como se ha apuntado anteriormente la construcci\'on de una teor\'{\i}a
en el cont\'{\i}nuo, implica la existencia
de un punto cr\'{\i}tico en el sentido de la teor\'{\i}a de transiciones
de fase en ME.

Por lo tanto es fundamental el estudio del diagrama de fases de las
TCC en el ret\'{\i}culo, tanto desde el punto de vista de la caracterizaci\'on
de las fases en que se manifiesta la teor\'{\i}a, como para encontrar
puntos cr\'{\i}ticos en los cuales la teor\'{\i}a es susceptible de
definir una TCC en el continuo.

Las propiedades de un sistema f\'{\i}sico dependen
de par\'ametros como la temperatura, la presi\'on, el campo magn\'etico
aplicado, etc...
Sin embargo un cambio en las condiciones
a que est\'a sometido puede hacer que
las propiedades del sistema var\'{\i}en bruscamente. Este cambio brusco
se denomi\-na transici\'on de fase. 
Un cambio brusco, en el lenguaje
de las matem\'aticas es una discontinuidad, la cual se manifestar\'a en
las medidas de las funciones termodin\'amicas.

Vamos a centrarnos en las transiciones inducidas por la competici\'on
orden-desorden. Para tener un soporte concreto la discusi\'on se
har\'a para el modelo de Ising. 
A pesar de su simplicidad las conclusiones que se obtengan ser\'an
aplicables con gran generalidad.
La acci\'on del modelo de Ising en presencia de un campo magn\'etico 
externo es:
\be
S = - \beta \sum_{\langle i,j \rangle} \sigma_i \sigma_j - h \sum_{i} \sigma_i \ ,
\label{SI}
\ee

donde $\beta = J/T$ y $h=H/T$, siendo $J$ la intensidad del acoplamiento
entre espines y $H$ el campo magn\'etico externo aplicado. 
El esp\'{\i}n $\sigma$ toma los valores +1 \'o -1. La notaci\'on
$\langle i,j \rangle$ indica que la suma se extiende a las parejas de
primeros vecinos. 

La funci\'on de partici\'on del sistema esta dada por:
\be
Z_N = \sum_{\rm {conf}} e^{- S({\rm conf})}  \ ,
\ee

donde el sumatorio en configuraciones significa:
\be
\sum_{\rm {conf}} = \sum_{\sigma_1 = \pm 1}  \sum_{\sigma_2 = \pm 1} \cdots
 \sum_{\sigma_N = \pm 1} \ ,
\ee

siendo $N$ el n\'umero total de espines.
Por tanto el n\'umero de t\'erminos en la funci\'on de partici\'on es $2^N$.

La densidad de energ\'{\i}a libre est\'a dada por:
\be
F(\beta,h) = - \frac{1}{N}\log Z_N \ ,
\ee
cuyas derivadas con respecto a $\beta$ y/\'o a $h$ nos dar\'an las
magnitudes termodin\'amicas.

Definimos el valor medio del esp\'{\i}n $\sigma_i$ como:
\be
\langle \sigma_i \rangle = \frac{1}{Z} \sum_{\rm {conf}} \sigma_i 
e^{-S({\rm conf})}  \ .
\ee 
Debido a la invarianza traslacional de la acci\'on, y trabajando
en una red con condiciones de contorno peri\'odicas, este
promedio no depende del esp\'{\i}n $i$ particular para el que se calcula.

La magnetizaci\'on del sistema estar\'a dada por el valor medio
de la suma de los espines individuales:
\be
M(T,h) = N^{-1}  \langle \sum_i \sigma_i \rangle  \ ,
\ee

Denotando por $w(E)$ el n\'umero de configuraciones con energ\'{\i}a 
interna $E$, podemos reagrupar la funci\'on de partici\'on, y
expresarla como una suma sobre energ\'{\i}as en lugar de como una 
suma sobre configuraciones:
\be
Z = \sum_E w(E) e^{-E} = 
\sum_E e^{-[E - T \Omega(E)]} \ ,
\ee
donde hemos usado la definici\'on de entrop\'{\i}a, $\Omega = \log w(E)$.
 La probabilidad de
aparici\'on de una configuraci\'on con energ\'{\i}a interna $E$ es
proporcional a su factor de Boltzman multiplicado por la degeneraci\'on
$w(E)$. Si el estado fundamental tiene una degeneraci\'on finita,
la configuraci\'on correspondiente a $T=0$
ser\'a aquella que corresponda al estado de m\'{\i}nima
energ\'{\i}a (todos los espines alineados) pues es infinitamente
m\'as probable que cualquier otra. 

El modelo de Ising presenta una l\'{\i}nea de transiciones de fase
a campo magn\'etico cero: 
$\{h=0, 0 \leq T \leq T_c \}$. Que las transiciones de fase tienen
que ocurrir a campo magn\'etico nulo es una consecuencia de la simetr\'{\i}a
de la acci\'on \cite{LEE0}:
\be
\sigma_i \rightarrow -\sigma_i, \qquad h \rightarrow - h  \ .
\ee

Por encima de $T_c$ las fluctuaciones t\'ermicas desordenan por completo
al sistema y ya no hay transiciones de fase.

El par\'ametro conveniente para estudiar la aparici\'on de la transici\'on
de fase es la magnetizaci\'on, en particular la {\sl magnetizaci\'on
espont\'anea} definida como:
\be
M_0 (T) = \lim_{h \rightarrow 0} M(T,h) \ .
\ee

Por encima de $T_c$ la magnetizaci\'on espontanea se anula,
mientras que es distinta de cero para $T<T_c$.

Las transiciones de fase se clasifican de acuerdo con el
comportamiento cualitativo de $M_0$ en el punto de la transici\'on.
En el modelo de Ising
$M_0$ es continua en $T=T_c$, y acorde a ello la transici\'on se
dice que es continua en el punto \{$h=0,T=T_c$\}.

Sin embargo, en el resto de puntos de transici\'on de fase
de la l\'{\i}nea,
es decir $\{h=0, 0 \leq T < T_c \}$, $M_0$ se comporta de forma discontinua:
\ba
\lim_{h \rightarrow 0^-} M(T,h) &=& - M_0(T) \nonumber \\
\lim_{h \rightarrow 0^+} M(T,h) &=& + M_0(T) \ .
\ea

Correspondientemente se dice que las transiciones de fase son 
de {\sl primer orden} o discontinuas.

Por ser la magnetizaci\'on espont\'anea la magnitud que indica en que
fase se encuentra el sistema, recibe el nombre de {\sl par\'ametro
de orden de la transici\'on}.

Hablando con generalidad, se dice que una transici\'on de fase es
de primer orden cuando hay una discontinuidad en la primera derivada
del potencial termodin\'amico, en el caso que estamos considerando,
de la energ\'{\i}a libre:
\be
\begin{array}{llll}
\hbox{{\rm Energ\'{\i}a interna:}} & U & = & \partial F / \partial \beta\\[2mm]
\hbox{{\rm Magnetizaci\'on:}} & M & = & -(\partial F/\partial h)
\end{array}
\ee

Si las primeras derivadas son continuas, pero aparecen discontinuidades
en las derivadas segundas (al menos una de ellas es discontinua)
la transici\'on se dice de {\sl segundo orden}. En general se habla
de transiciones de fase continuas cuando la discontinuidad aparece
en una derivada de orden igual o superior a dos, siendo todas las
primeras derivadas cont\'{\i}nuas.

En el caso Ising en el punto ($h=0,T=T_c$) la transici\'on es de segundo orden
por lo que aparecen divergencias en el calor espec\'{\i}fico
y en la susceptibilidad magn\'etica:
\be
\begin{array}{llllll}
\hbox{{\rm Calor Espec\'{\i}fico:}} & C_v & = & \partial U /\partial \beta & = & (\partial^2 F /\partial \beta^2)  \\[2mm]
\hbox{{\rm Susceptibilidad:}} & \chi & = & \partial M / \partial h & = & - (\partial^2 F /\partial h^2)
\end{array}
\ee
La diferencia fundamental entre ambos tipos de transiciones es que las
transiciones continuas presentan ordenamientos de largo alcance
en el punto de la transici\'on. Las divergencias que aparecen est\'an
ligadas a la divergencia de la longitud de correlaci\'on. 
Sin embargo en las transiciones de primer orden el sistema no anticipa
la transici\'on y la longitud de correlaci\'on permanece
finita cuando nos aproximamos al punto de la discontinuidad.

Desde el punto de vista
de la TCC las transiciones interesantes son por tanto las de segundo
orden pues s\'olo en estos puntos se puede obtener una teor\'{\i}a
estrictamente renormalizable.
Puesto que en esta memoria estaremos especialmente interesados
en la diferenciaci\'on de transiciones de fase de primer
y segundo orden, veamos m\'as detalladamente las caracter\'{\i}sticas
de ambas.

\subsection{Transiciones de fase continuas}

Para una descripci\'on cualitativa de los fen\'omenos que ocurren
en estas transiciones de fase tomemos como ejemplo el modelo de Ising
(\ref{SI}) en ausencia de campo magn\'etico externo ($h=0$). 

A temperatura elevada se observan espines hacia arriba y espines hacia
abajo formando islotes cuyo tama\~no promedio es la longitud de correlaci\'on
entre espines. La correlaci\'on entre dos espines situados a una
distancia $r_{ij}$ tiende a cero exponencialmente con la separaci\'on:
\be
\langle \sigma_i \sigma_j \rangle \sim e^{- r_{ij}/\xi(T)} \ ,
\ee
con una longitud de correlaci\'on caracter\'{\i}stica de la 
temperatura, $\xi(T)$. En estas condiciones la magnetizaci\'on $M$
es nula.

A medida que disminuye $T$, el tama\~no de los islotes aumenta pues la
correlaci\'on entre espines aumenta al disminuir la agitaci\'on
t\'ermica. Cuando se alcanza la temperatura cr\'{\i}tica, $T=T_c$
los islotes tienen todos los tama\~nos posibles. Es decir, a $T_c$ las
fluctuaciones del sistema son de todas las longitudes, ya no hay una
escala t\'{\i}pica para el sistema, por eso se dice que la f\'{\i}sica es
invariante de escala en el punto de la transici\'on, que recibe el
nombre de punto cr\'{\i}tico.

As\'{\i} mismo por debajo de $T_c$ la funci\'on de correlaci\'on entre espines
situados a distancia $r_{ij}$ es el parametro de orden al cuadrado:
\be
\lim_{r_{ij} \rightarrow \infty} \langle \sigma_i \sigma_j \rangle =
\langle \sigma_i \rangle \langle \sigma_j \rangle = M^2  \ .
\ee

Se define la funci\'on de correlaci\'on entre
dos espines de la siguiente forma:
\be
\Gamma(\epsilon,r_{ij}) = \langle \sigma_i \sigma_j \rangle -
\langle \sigma_i \rangle \langle \sigma_j \rangle \ ,
\ee

donde $\epsilon = (T-T_c)/T_c$ recibe el nombre de temperatura reducida.

Con esta definici\'on para $T>T_c$ y $T<T_c$ , $\Gamma(\epsilon,r_{ij})$
decrece exponencialmente con la distancia: $\Gamma(\epsilon,r_{ij}) \sim
\exp(-r_{ij}/\xi)$, siendo $\xi$ la longitud de correlaci\'on asociada
a la funci\'on de correlaci\'on entre dos espines. La definici\'on
precisa viene dada por:
\be
\xi \equiv \lim_{|r_{ij}| \rightarrow \infty} \frac{-|r_{ij}|}{\log 
\Gamma(\epsilon,r_{0i})} \ .
\ee 

Que exista una magnetizaci\'on espont\'anea a campo exterior
nulo, es de por si un hecho remarcable puesto que uno no esperar\'{\i}a
en estas condiciones que el sistema elija una direcci\'on privilegiada
hacia la que apuntar. Sin embargo la m\'as peque\~na inhomegeneidad en
los islotes, o un campo residual no nulo pueden hacer que una direcci\'on
resulte privilegiada frente a las dem\'as, o dicho de otra forma el 
vac\'{\i}o con $M=0$ no es estable. Este fen\'omeno se llama
{\sl ruptura espont\'anea de la simetr\'{\i}a}.

El hecho de que la transici\'on se produzca como
un fen\'omeno coopera\-tivo a gran
escala induce a pensar que ciertas de sus caracter\'{\i}sticas s\'olo
dependen de las propiedades del sistema a gran escala, esto es, de sus
propiedades generales, tales como la dimensi\'on $d$ del espacio, 
la dimensi\'on del par\'ametro de orden o las simetr\'{\i}as de los
acoplamientos, y no de los detalles de la interacci\'on. Esta
es la idea sobre la que descansa el concepto de {\sl Universalidad}, que
nos llevar\'a a definir magnitudes para describir el comportamiento
cr\'{\i}tico de los sistemas dependiendo s\'olo de estas caracter\'{\i}sticas
generales, por ejemplo los exponentes cr\'{\i}ticos.

Los exponentes cr\'{\i}ticos caracterizan el comportamiento 
de los par\'ame\-tros
de orden, susceptibilidades, etc, en el entorno del punto cr\'{\i}tico.
La hip\'otesis que se hace es que el comportamiento de los observables en el
entorno del punto cr\'{\i}tico se puede describir como una potencia
de la temperatura reducida, $\epsilon$. Esta potencia
se llama exponente cr\'{\i}tico y cada observable f\'{\i}sico dependiente
de la temperatura tiene asociado uno. De esta manera
dado un observable $f(\epsilon)$, se hace la hip\'otesis de que $f(\epsilon)$
es cont\'{\i}\-nuo y positivo para valores peque\~nos y positivos del parametro
$\epsilon$. Se supone adem\'as que el l\'{\i}mite
\be
\lambda = \lim_{\epsilon \rightarrow 0} \frac{\log f(\epsilon)}{\log \epsilon} \ ,
\ee
existe y est\'a bien definido. A $\lambda$ se le llama exponente cr\'{\i}tico
asociado a $f(\epsilon)$. 

Para sistemas magn\'eticos como el modelo de Ising, los exponentes
se definen de la siguiente forma:
\be
\begin{array}{lll}
C_v(T) & \sim & \epsilon^{- \alpha} \ ; \\[2mm]
M(T) & \sim & (- \epsilon)^{\beta}  \ ; \\[2mm]
\chi (T) & \sim & \epsilon^{-\gamma}  \ ; \\[2mm]
\xi (T) & \sim & \epsilon^{-\nu} \ .
\end{array} 
\label{EXPO}
\ee

Por \'ultimo se definen tambi\'en exponentes para el comportamiento
en el punto cr\'{\i}tico ($\epsilon$=0) de la funci\'on de correlaci\'on
de dos espines situados a distancia $r$:
\be
\Gamma(0,r) \sim 
r^{-(d - 2 + \eta)}  \ ,
\ee

En general la funci\'on de correlaci\'on se puede expresar como:
\be
\Gamma(\epsilon,r) = \frac{g(r,\xi)}{r^{d-2+\eta}} \ .
\ee

Para $T \neq T_c$,  $g(r,\xi) \sim \exp(-r/\xi)$ mientras que en 
el entorno del punto cr\'{\i}tico ($\xi = \infty$) la funci\'on de 
correlaci\'on se comporta como una potencia de $r$.

Se introduce tambi\'en un exponente cr\'{\i}tico asociado  
al comportamiento del par\'ametro de orden en el punto cr\'{\i}tico
a campo externo distinto de cero:
\be
M(T_c,h) \sim |h|^{1/\delta} \ .
\ee

En la dimensi\'on cr\'{\i}tica superior estas predicciones est\'an
modificadas por la existencia de correcciones logar\'{\i}tmicas \cite{LOG0}.

Hay dos motivos esencialmente por los cuales estamos interesados
en estas cantidades a pesar de que posean menos informaci\'on
que la funci\'on completa $f(\epsilon)$:

\begin{enumerate}
\item Cerca del punto cr\'{\i}tico ($T \approx T_c$) domina el t\'ermino
con $\epsilon$ elevado a la potencia m\'as baja. Esto se confirma 
experimentalmente en gr\'aficas log-log puesto que se obtienen rectas en torno
al punto cr\'{\i}tico cuya pendiente es el exponente cr\'{\i}tico. Por otra
parte los fen\'omenos f\'{\i}sicos en los que estamos interesados, tales
como las existencia de largas correlaciones, ocurren en el entorno del
punto cr\'{\i}tico.

\item Existen relaciones entre los exponentes que trascienden a cualquier
sistema particular pues dependen de caracter\'{\i}sticas muy generales,
en la l\'{\i}nea del concepto de Universalidad mencionado anteriormente.
Estas relaciones no han sido probada con la mayor generalidad sino s\'olo
en el marco de ciertas hip\'otesis que conciernen a la forma de los
potenciales termodin\'amicos.

\end{enumerate}

 En particular se supone que el potencial de
Gibbs es una funci\'on homogenea generalizada ({\sl hip\'otesis de escala})
es decir:
\be
G(\lambda^{a_{\epsilon}} \epsilon, \lambda^{a_H} H) = \lambda G(\epsilon,H) \ .
\label{ESCALA}
\ee

Esta hip\'otesis permite establecer relaciones entre los
exponentes cr\'{\i}ticos:

\be
\begin{array}{l}
\gamma = \nu(2 - \eta) \ ; \\[2mm]
\alpha + 2\beta +\gamma = 2 \ ; \\[2mm]
\gamma = \beta(\delta - 1)  \ ; \\[2mm]
\gamma = 2 \beta \delta + \alpha - 2 \ .
\end{array}
\ee

Si adem\'as se asume la {\sl hip\'otesis de hiperescala} que consiste
en suponer que
las fluctuaciones que contribuyen a la parte
singular de la densidad de energ\'{\i}a libre van como:
\be
F_{\rm {sing}} (\epsilon) \sim \frac{1}{\xi(\epsilon)^{d}} \sim 
\epsilon^{\nu d} \ ,
\ee
teniendo en cuenta que el calor
espec\'{\i}fico es la segunda derivada de la energ\'{\i}a libre,
podemos escribir su ley de escalado como:
\be
C_v \sim \epsilon^{\nu d - 2 } \ .
\ee

Como consecuencia se obtiene la {\sl relaci\'on de Josephson}:
\be
\alpha = 2 - \nu d  \ .
\ee

Si bien la hip\'otesis de escala no ha sido demostrada con rigurosidad,
\'esta surge de un modo natural en el marco de las hip\'otesis de trabajo
del grupo de renormalizaci\'on. En este contexto vamos a 
describir la l\'{\i}nea argumental de Kadanoff que lleva a la plausibilidad 
de la hip\'otesis de escala. 

Supongamos un Hamiltoniano tipo Ising como el descrito por la acci\'on
(\ref{SI}), es decir un sistema de $N$ espines en una red de dimensi\'on
$d$. Con\-sideremos la red dividida en celdas de lado $aL$,
tenemos por tanto ${\rm n} = N/L^d$ celdas cada una con $L^d$
espines. Los argumentos de Kadanoff se aplican s\'olo a la situaci\'on
en que $\xi \gg aL$, es decir, en un entorno suficientemente peque\~no
del punto cr\'{\i}tico. En estas condiciones dentro de un mismo islote
de espines correlacionados hay un gran n\'umero de celdas.

Vamos a asociar a cada celda $\alpha$ ($\alpha=1,2, \dots ,{\rm n}$) un
esp\'{\i}n ${\tilde \sigma}_{\alpha}$ de acuerdo con la regla de la
mayor\'{\i}a. Puesto que cada celda est\'a enteramente
dentro de un islote de espines correlacionados, es razonable asumir que
los ${\tilde \sigma}_{\alpha}$ se comportar\'an como los espines individuales
$\sigma_i$ en el sentido de que actuar\'an como si tomasen los valores
$\pm 1$. 

Podemos pensar que la acci\'on escrita en t\'erminos de estos 
momentos ${\tilde \sigma}_{\alpha}$, tenga la misma forma que la
acci\'on original, excepto que los par\'ametros $\beta$ y $h$ ser\'an en
general distintos. Asociamos pues unos nuevos $\tilde{\beta}$ y $\tilde{h}$
a la acci\'on construida con las ${\tilde \sigma}_{\alpha}$.
Puesto que la temperatura cr\'{\i}tica es una medida de la intensidad
de la interacci\'on $J$, no haremos la discusi\'on en funci\'on 
de $\tilde{\beta}$ 
sino de la temperatura reducida $\tilde{\epsilon}$.

Ahora bien, si la acci\'on tiene la misma forma funcional, es razonable 
pensar que lo mismo ocurrir\'a con los potenciales termodin\'amicos. As\'{\i}
por ejemplo el potencial de Gibbs de cada celda, 
$G(\tilde{\epsilon},\tilde{h})$, ser\'a la misma funci\'on de $\tilde{h}$
y $\tilde{\epsilon}$ que el potencial de Gibbs correspondiente a cada site.
Por ser \'este una magnitud extensiva se tiene:
\be
G(\tilde{\epsilon},\tilde{h}) = L^d G(\epsilon,h) \ .
\label{GIB}
\ee
La dependencia de $\tilde{h}$ con $h$ y $\tilde{\epsilon}$ con $\epsilon$
se puede pensar que ser\'a en general una funci\'on de $L$:
\be
\begin{array}{l}
\tilde{h} = H(L) h \\
\tilde{\epsilon} = T(L) \epsilon  \ .
\end{array}
\ee

En estas condiciones ya se puede argumentar el cumplimiento de la hip\'otesis
de escala, sin embargo siguiendo con el argumento de Kadanoff, 
se asume que $H(L)$ y $T(L)$ se pueden expresar como una potencia de
$L$: $H(L)=L^x$ y $T(L)=L^y$, con $x$ e $y$ arbitrarios. Esto nos permite
reescribir (\ref{GIB}) como:
\be
\begin{array}{rrr}
G(L^y \epsilon, L^x h)& =& L^d G(\epsilon,h) \\[2mm]
\Rightarrow G(L^{y/d} \epsilon, L^{x/d} h)&=&L G(\epsilon,h)  \ ,
\end{array}
\label{GIB2}
\ee

que tiene la misma forma funcional que (\ref{ESCALA}) excepto que $L$ no
es un par\'ametro libre como lo es $\lambda$ en (\ref{ESCALA}).
En efecto $L$ est\'a sujeto a la condici\'on $1 \ll L \xi/a$, sin embargo,
puesto que $\xi$ diverge en el punto de la transici\'on siempre podemos
ponernos tan proximos al punto cr\'{\i}tico como queramos para lograr que
$L$ est\'e en ese intervalo. Es decir que (\ref{GIB2}) significa que
el potencial de Gibbs es una funci\'on homog\'enea generalizada.

Una l\'{\i}nea de argumentaci\'on an\'aloga aplicada a la funci\'on
de correlaci\'on a dos puntos, $\Gamma (\epsilon,r)$,
muestra que \'esta es tambi\'en una funci\'on homog\'enea generalizada. 
Teniendo en cuenta que la longitud de correlac\'on es funci\'on s\'olo
de la temperatura reducida, podemos escribir
$\Gamma (\epsilon,r)$ sustituyendo la dependencia
en $\epsilon$ por la dependencia en $\xi$: $\Gamma (\xi,r)$. Puesto que
es una funci\'on homog\'enea tenemos:
\be
\Gamma (\lambda \xi,\lambda r) = \lambda^u \Gamma(\xi,r) \ .
\ee
 
Tomando $\lambda = 1/\xi$:
\be
\Gamma(1,r/\xi) = \xi^{-u} \Gamma(\xi,r)  \ .
\ee

Puesto que uno de los dos par\'ametros est\'a fijo podemos escribir la
funci\'on de correlaci\'on como dependiente de un solo par\'ametro,
es decir $\Gamma(1,r/\xi) \equiv F(\xi,r)$. Es decir que podemos
escribir:
\be
\Gamma(\xi,r) = \xi^p \Gamma(1,r/\xi) \ .
\ee

Esta igualdad significa que la correlaci\'on entre dos espines situados
a distancia $r$, depende de $r$ s\'olo a trav\'es del cociente $r/\xi$. 
Dicho de otra forma, que s\'olo existe una longitud caracter\'{\i}stica
en el problema y que esta es la longitud de correlaci\'on.

Sin embargo a la hora de extraer informaci\'on mediante m\'etodos
num\'ericos hay que tener en cuenta que las divergencias que marcan
la existencia de una transici\'on de fase ocurren s\'olo cuando el
n\'umero de grados de li\-bertad tiende a infinito. En un volumen finito,
es decir dentro de los ordenadores, todas las cantidades se obtienen 
mediante sumas finitas siendo por tanto funciones anal\'{\i}ticas en
todo el espacio de acoplamientos. Como consecuencia aparecen dos fen\'omenos
fundamentalmente: las divergencias son sustituidas
por picos de altura finita, creciente con el tama\~no de la red; la 
localizaci\'on de estos picos se mueve en el espacio de acoplamientos
conforme el tama\~no de la red cambia.

La teor\'{\i}a de {\sl Finite Size Scaling} (FSS) \cite{BARBER0} 
est\'a basada en que estos
efectos que aparecen al aumentar el tama\~no de la red son precursores
del comportamiento del sistema en el l\'{\i}mite termodin\'amico, y
que pueden ser explotados para extraer las propiedades de la transici\'on
de fase. Considere\-mos un ret\'{\i}culo de lado $L$.
La teor\'{\i}a de FSS descansa sobre la hip\'otesis de que en
el entorno del punto cr\'{\i}tico el comportamiento del sistema se puede
describir enteramente en funcion de la variable reescalada:
\be
y = L/\xi(T) \ .
\ee

En una transici\'on de fase continua $\xi(T)$ crece conforme nos acercamos
a $T_c$. Es razonable suponer que los efectos de tama\~no finito se
har\'an palpables cuando \'esta alcance el tama\~no de la red $L$.
Definimos la temperatura cr\'{\i}tica aparente $T_c^{\ast}(L)$ como
aquella en la que
$\xi(T_c^{\ast}(L)) \sim L$. En estas condiciones podemos estimar
como depende de $L$ la temperatura aparente puesto que:
\ba
\xi(T_c^{\ast}(L)) \propto L & \sim & |T_c^{\ast}(L) - T_c|^{\nu} \nonumber \\[2mm]
\Rightarrow |T_c^{\ast}(L) - T_c| & \sim & L^{-1/\nu}  \ .
\ea

En general si $Q_L(T)$ es un observable f\'{\i}sico en el sistema finito,
cuyo comportamiento en el l\'{\i}mite termodin\'amico es:
\be
Q_{\infty} (T) \sim C_{\infty} \epsilon^{- \rho} \quad \epsilon 
\rightarrow 0 \ .
\label{INFTY}
\ee

En un volumen finito el comportamiento ser\'a del tipo:
\be
Q_L(T) \sim {\hat C}_L {\hat \epsilon}^{-\rho} \ ,
\ee

con ${\hat \epsilon}  = |T - T_c^{\ast}(L)| \sim L^{-1/\nu}$.

Si introducimos ahora la hip\'otesis de FSS se tiene:
\be
Q_L(T) \sim L^{\omega} f(y) \ .
\label{FINITO}
\ee

Para determinar $\omega$ basta imponer que (\ref{FINITO}) reproduzca
(\ref{INFTY}) en el l\'{\i}mite de $L \rightarrow \infty$:
\be
f(y) \sim C_{\infty} \epsilon^{- \rho} \quad L \rightarrow \infty \ ,
\ee
para ello $\omega = \rho/\nu$.

Por otra parte puesto que en el sistema finito no hay transici\'on
de fase:
\be
f(y) \rightarrow f_0 \quad \epsilon \rightarrow 0 \ ,
\ee

es decir $f(y)$ debe aproximar una constante finita en el punto
cr\'{\i}tico. 

Para resumir, el resultado significativo de este an\'alisis,
desde el punto de vista de extraer informaci\'on num\'erica, es que
la forma en que var\'{\i}an las magnitudes termodin\'amicas con el
tama\~no de la red est\'a determinada por los exponentes cr\'{\i}ticos
del sistema en el l\'{\i}mite termodin\'amico: dado un observable $Q(\epsilon)$
que en $L=\infty$ diverge en el punto de la transici\'on 
como $Q(\epsilon) \sim \epsilon^{-\rho}$, en $L$ finito su comportamiento
ser\'a de la forma:
\be
Q_L(T_c^{\ast}(L)) \sim f_0 L^{\rho/\nu} \ .
\ee

\subsection{Transiciones de fase de primer orden.}

Como ya se ha se\~nalado anteriormente, seg\'un el criterio b\'asico
de clasificaci\'on termodin\'amica las transiciones de fase de primer
orden son aquellas en las que la primera derivada de la energ\'{\i}a
libre presenta una discontinuidad.

Cuando un sistema realiza una transici\'on de fase de primer orden
de una fase de alta temperatura a otra de baja temperatura, \'este
desprende una cantidad no nula de calor (el {\sl calor latente} $C_l$)
a medida que se enfr\'{\i}a
en un intervalo infinitesimal de temperaturas alrededor de la
temperatura de la transici\'on.
El valor de $C_l$ se relaciona con el salto en
la entrop\'{\i}a que se produce en las transiciones
de primer orden de la siguiente forma:
\be
\Omega = -(\partial F / \partial T) \quad \Rightarrow \quad  C_l = T_c 
\Delta \Omega \ .
\ee

Se tiene mucho menor conocimiento te\'orico sobre el comportamiento de
las transiciones de fase de primer orden que sobre los fen\'omenos
asociados a las transiciones continuas. Esto es debido a que al no haber
una longitud de correlaci\'on divergente, no podemos restringir el estudio
a los fen\'omenos asociados a largas longitudes de onda, por lo tanto 
el concepto de \-Universalidad no es aplicable. Las singularidades
que aparecen en las transiciones de primer orden se deben exclusivamente
a la coexistencia de fases, no hay regi\'on cr\'{\i}tica, ni exponentes
cr\'{\i}ticos.
No podemos usar por tanto la teor\'{\i}a de FSS para estudiar el comportamiento
de estas transiciones en un volumen finito.

Debido a la existencia de calor latente,
el calor espec\'{\i}fico es una funci\'on $\delta(\epsilon)$.
Analogamente a lo que ocurre con las transiciones continuas,
en una red finita de lado $L$, las singularidades
de las funciones en el punto de la transici\'on, en este caso las
$\delta^{\prime}$s, ser\'an suavizadas
y sustituidas por picos de altura finita. De la misma forma esperamos
un desplazamiento de la temperatura a la cual aparecen estos picos con
el tama\~no de la red, $T_c^{\ast}(L)$.

Para discutir cuantitativamente los efectos de tama\~no finito asociados
a una transici\'on de primer orden vamos a utilizar la aproximaci\'on 
fenomenol\'ogica introducida por Binder \cite{BINDER0}.
Se asumir\'a que el tama\~no de la red es mayor
que la longitud de correlaci\'on de cada fase, de tal manera que los
observables toman los valores asint\'oticos de la transici\'on.

En una red de lado $L$ y a una cierta temperatura $T$, la 
distribuci\'on de probabilidad de la energ\'{\i}a, $P_L(E)$, es
gausiana:
\be
P_L(E) = \frac{A}{\sqrt{C}} \exp \{ - \frac{(E - E_0)^2 L^d}{2T^2 C}  \} \ ,
\label{UNA}
\ee

donde $E_0$ es la energ\'{\i}a del sistema en el l\'{\i}mite de volumen 
infinito, y es caracter\'{\i}stica de la temperatura $T$. La anchura de
la distribuci\'on es proporcional al calor espec\'{\i}fico en este 
l\'{\i}mite, denotado por $C$.

En el punto de la transici\'on de fase la caracter\'{\i}stica fundamental
de las transiciones de primer orden es la coexistencia entre fases. 
Supongamos que ocurre una transici\'on en $T=T_c$ de un estado desordenado
a un estado ordenado con degeneraci\'on $q$. 
Denotemos por $E_+$ ($E_-$) la energ\'{\i}a interna de la fase desordenada
(ordenada) a $T=T_c$, de manera que el calor latente es $E_+ - E_-$.

La generalizaci\'on de (\ref{UNA}) al caso en que hay coexistencia de fases
es considerar que la distribuci\'on de probabilidad a $T_c$ 
es una doble gaussiana. En el intervalo $\Delta T = T - T_c$ las gaussianas
estar\'an centradas respectivamente en $E_+ + C_+ \Delta T$ y 
$E_- + C_- \Delta T$:

\ba
P_L(E) & \propto & \frac{a_+}{\sqrt{C_+}} 
\exp \{ -\frac{(E - E_+ - C_+ \Delta T)^2 L^d}{2T^2 C_+} \}  \\
& + & \frac{a_-}{\sqrt{C_-}}   
\exp \{ -\frac{(E - E_- - C_- \Delta T)^2 L^d}{2T^2 C_-} \}  \ , 
\label{DOS}
\ea

siendo $C_+$ y $C_-$ el calor espec\'{\i}fico de las fases desordenada
y ordenada respectivamente:
\ba
\lim_{T \rightarrow T_c^- } C_v(T) &=& C_- \\
\lim_{T \rightarrow T_c^+ } C_v(T) &=& C_+ \ , 
\ea

Puesto que se est\'a considerando el comportamiento del sistema cerca
de $T_c$ se puede considerar que $C_+$ y $C_-$ son constantes
a lo largo de esta regi\'on.
Los dos picos est\'an pesados de acuerdo con la diferencia de energ\'{\i}a
libre entre las dos fases $\Delta F(T) = F_+ - F_-$, de manera que lejos
de $T_c$ s\'olo uno de los picos sobrevive, mientras
que en el punto de la transici\'on $\Delta F(T_c)=0$ y ambos t\'erminos
contribuyen con el mismo peso:
\ba
a_+ & =& \sqrt{C_+} e^x \\[2mm]
a_- &= & q \sqrt{C_-} e^{-x}  \ ,
\ea

donde 
\be
x= \frac{- \Delta F L^d}{2T} \ .
\ee

Se puede aproximar $x$ expandiendo en serie $F_+$ y $F_-$ entorno
a $T_c$ hasta orden $\Delta T$ (orden mayor significar\'{\i}a considerar
la variaci\'on de $C_+$ y $C_-$). Teniendo en cuenta que 
$F_{\pm}=U_{\pm} - TS_{\pm}$ y que $dF_{\pm}= -S_{\pm} dT$ se obtiene:
\be
\Delta F = \frac{-(E_+ - E_-)\Delta T}{T_c}
\ee

El valor medio de la energ\'{\i}a vendr\'a
dado por el primer momento de la distribuci\'on de probabilidad (\ref{DOS}):

\be
\langle E \rangle = \frac{a_+ E_+ + a_- E_-}{a_+ + a_-}
+ \Delta T \frac{a_+ C_+ + a_- C_-}{a_+ + a_-} \ ,
\ee

y por lo tanto el calor espec\'{\i}fico:
\ba
C_v(T,L) & = &  \frac{\partial \langle E \rangle_L}{ \partial T} =
\frac{a_+ C_+ + a_- C_-}{a_+ + a_-} \\
& + &  \frac{a_+ a_- L^d}{T^2} \frac{[(E_+ - E_-) + (C_+ - C_-)\Delta T]^2}
{(a_+ + a_-)^2}  \ ,
\ea

cuyo m\'aximo ocurre para:
\be
\frac{T_c^{\ast}(L) - T_c}{T_c} = \frac{T_c \log q}{E_+ - E_-}
\frac{1}{L^d} \ ,
\ee
con una altura de:
\be
C_v(L)^{\rm {max}} = \frac{(E_+ - E_-)^2}{4T_c^2}L^d + \frac{C_+ + C_-}{2} \ .
\ee

Cuando se est\'a interesado en simulaciones de Monte Carlo hay errores
asociados a la simulaci\'on que pueden hacer m\'as aceptable una descripci\'on 
de los efectos de tama\~no finito en funci\'on de exponentes efectivos:
\ba
T_c^{\ast}(L) - T_c(\infty) & \propto & L^{-\lambda} \\
C_v^{\rm {max}}(L) & \propto & L^{\alpha_m} \ , 
\ea

donde hemos introducido los \'{\i}ndices $\lambda$ y $\alpha_m$
para estudiar esta dependencia con $L$. 
Es decir, el resultado que se encuentra es que
los exponentes $\lambda$ y $\alpha_m$ son igual a la dimensi\'on
del espacio, $d$, asintoticamente, es decir para $L$
suficientemente grande ($L \gg \xi$).

Sin embargo la definici\'on de estos exponentes es puramente formal,
y no tienen nada que ver con las definiciones que hemos visto 
En una transici\'on de primer orden $L$ aparece s\'olo porque el
volumen es $L^d$ en $d$ dimensiones: los m\'aximos de las susceptibilidades
y del calor espec\'{\i}fico crecen como $L^d$ y la funci\'on $\delta$
se obtiene en el l\'{\i}mite de volumen infinito porque la anchura de
estas funciones decrece como $L^{-d}$.

Desde el punto de vista del estudio num\'erico, en
las transiciones de primer orden, en general, cualquier 
observable termodin\'amico
muestra en la evoluci\'on de MC saltos entre las dos fases 
({\sl flip-flops}) generando una distribuci\'on de doble pico.

Las transiciones fuertes (peque\~na $\xi$) son f\'aciles de detectar. 
Las discontinuidades
se observan ya a nivel de los ciclos de histeresis que muestran ramas
metaestables. Las discontinuidades en la magnitudes termodin\'amicas
son relativamente f\'aciles de detectar, y tanto el calor espec\'{\i}fico
como las susceptibilidades en transiciones magn\'eticas divergen como
$L^d$ para redes de tama\~no no muy grande. 

La situaci\'on num\'erica es menos clara en las llamadas transiciones 
de primer orden d\'ebiles. La longitud de correlaci\'on es muy grande,
y por tanto los efectos transitorios se prolongan hasta redes de
tama\~no respetable para nuestro actuales ordenadores. En estos casos
uno debe simular redes de $L$ creciente y tratar de ver si el
comportamiento de primer orden anteriormente descrito se alcanza en el
l\'{\i}mite de volumenes muy grandes. En particular se pueden usar
las t\'ecnicas desarrolladas de FSS 
para estudiar los exponentes cr\'{\i}ticos efectivos
y ver como evolucionan con $L$ las magnitudes termodin\'amicas.
Dentro de este contexto se dice que en la regi\'on asint\'otica
las transiciones de primer orden tienen asociado un 
exponente cr\'{\i}tico ``$\nu$''$=1/d$ o 
que ``$\alpha/\nu$'' $=d$, pero entendiendo bien que no son
exponentes cr\'{\i}ticos definibles en sentido extricto.

\newpage

\chapter{El Modelo O(4) Anti-Ferromagn\'etico \label{capO4}}
\thispagestyle{empty}
\markboth{\protect\small CAP\'ITULO \protect\ref{capO4}}
{\protect\small {\sl El Modelo O(4) Anti-Ferromagn\'etico} }

\cleardoublepage        

\section{Introducci\'on}

En el Lagrangiano del Modelo Est\'andar el campo de Higgs est\'a
descrito por un campo escalar de cuatro componentes, el cual, aparte
de interaccionar con los campos gauge y con los fermiones, interacciona
consigo mismo a trav\'es de un autoacoplo cu\'artico. Si consideramos
el l\'{\i}mite en el que los acoplos gauge y los de Yukawa tienden
a cero, los campos gauge y los fermi\'onicos se desacoplan. La
acci\'on correspondiente a este l\'{\i}mite es el modelo $\LP$:
\be
S_{\rm cont} = \int_{\Omega}  \frac{1}{2} \sum_{\mu} (\partial_{\mu} 
{\bf \Phi}_{\rm B} ({\bf{r}}))^2 + \frac{1}{2} m_{\rm B}^2 
{\bf \Phi}_{\rm B} ({\bf{r}})^2 +
\frac{g_{\rm B}}{4!} {\bf \Phi}_{\rm B} ({\bf{r}})^4   \ .
\ee

Una forma conveniente 
de escribir la acci\'on cuando estamos interesados en simulaciones
de MC o expansiones de alta T es:
\be
S_{\rm lat} = \sum_{\bf{r}}(-\kappa 
       \sum_{\mu} {\bf \Phi}_{\bf{r}} {\bf \Phi}_{\bf{r}+\hat{\bf{\mu}}} 
    + {\bf \Phi}_{\bf{r}}^2 + \lambda [{\bf \Phi}_{\bf{r}}^2 - 1]^2 )  \ ,
\label{O4ACC}
\ee

donde $\kappa$ recibe el nombre de {\sl hopping parameter}.

La correspondencia entre los par\'ametros en la red y en el cont\'{\i}nuo
es la siguiente:
\be
\begin{array}{lll}
{\bf \Phi}_{\rm B}({\bf r}) a & = & \sqrt{2 \kappa} 
{\bf \Phi}_{\bf{r}} \ ; \\[2mm]
m_{\rm B}^2 a^2 & = & (1 - 2 \lambda) / \kappa - 8 \ ; \\[2mm]
g_{\rm B} & = & 6 \lambda / \kappa^2  \ .
\end{array}
\ee

El diagrama de fases del modelo (\ref{O4ACC}) 
presenta una linea de transiciones
de segundo orden $\kappa_{\rm c}(\lambda)$ separando la fase sim\'etrica
de la fase en que la simetr\'{\i}a O(4) est\'a rota a O(3)$\otimes$O(3).
El l\'{\i}mite cont\'{\i}nuo hay que tomarlo aproxim\'andose a esta l\'{\i}nea
cr\'{\i}tica.

En $\lambda=0$ se tiene el punto fijo Gaussiano infrarojo puesto
que la teor\'{\i}a en este l\'{\i}mite se reduce al campo escalar libre.
El l\'{\i}mite cont\'{\i}nuo en el entorno de este punto fijo es trivial
en el sentido de que la constante de acoplamiento renormalizada
$g_{\rm R}$ se hace cero.

Para poder definir un l\'{\i}mite cont\'{\i}nuo no trivial ($g_{\rm R} \neq 0$) 
es necesario encontrar un punto fijo ultravioleta.

En dimensi\'on $d > 4$ est\'a probado rigurosamente \cite{AIZ1,FRO1}
que $\LP$ no tiene ning\'un punto fijo ultravioleta, la teor\'{\i}a
es por tanto trivial para todo valor de $\lambda$.

En dimensi\'on $d=4$ las evidencias anal\'{\i}ticas \cite{KOGUT1,BAKER1}
y num\'ericas \cite{LUS11,LUS21,BLOCK1} acumuladas soportan la
conjetura de Wilson \cite{KOGUT1} apuntando a la trivialidad de las
teor\'{\i}as $\LP$.
Sin embargo una prueba rigurosa como la obtenida en $d > 4$ no ha
sido obtenida, a pesar de los serios intentos llevados a cabo (ver
por ejemplo \cite{AIZ21}).

Los teoremas existentes cubren s\'olo ciertos
casos especiales, que por argumentos de plausibilidad, se supone que las
QFT deben cumplir \cite{FERNANDEZ1}. En particular se asume generalmente
que la autointeracci\'on del campo escalar se puede tratar perturbativamente.
Sin embargo las teor\'{\i}as con acoplos negativos generan fuertes oscilaciones
en los campos, que podr\'{\i}an dar lugar a que
se generen no perturbativamente nuevos puntos fijos, 
donde eventualmente se podr\'{\i}an definir l\'{\i}mites cont\'{\i}nuos 
no triviales \cite{GALLA1}. 
Desde este punto de vista el estudio de teor\'{\i}as Anti-ferromagn\'eticas
(AF) es interesante puesto que nada impide formular una
una QFT, ``honesta'' en el sentido de que cumple los axiomas
de Osterwalder-Schrader, usando modelos antiferromagn\'eticos (AF)
\cite{GALLA21}.

El antiferromagnetismo ha sido considerado en una amplia variedad
de modelos con el objetivo de encontrar propiedades no presentes
en teor\'{\i}as puramente ferromagn\'eticos. En el contexto de la
superconductividad de alta temperatura el AF parece jugar un papel
esencial \cite{SUPERT1}. En dimensi\'on $d=4$, interacciones competitivas
como posibles agentes de nuevas clases de universalidad han sido
investigadas para estudiar el punto multicr\'{\i}tico de modelos tipo
Yukawa. Este punto multicr\'{\i}tico es el punto de encuentro de cuatro fases
distintas (FM, AF, Ferrimagn\'etica y Paramagn\'etica (PM)). La cuesti\'on
de si es posible o no definir un l\'{\i}mite cont\'{\i}nuo no trivial
en este punto permanece como un problema abierto 
\cite{BOCK1,BOCK21,EBI1,JL1}.

As\'{\i} pues el estudio de modelos AF se presenta como una posibilidad
interesante de estudiar en detalle. Estos modelos suelen presentar
diagramas de fases muy ricos, y presumiblemente nuevas clases de universalidad
podr\'{\i}an ser encontradas \cite{POL1}. La existencia de acoplamientos
de signo o\-puesto influencia el vac\'{\i}o de la teor\'{\i}a, en concreto
en el estado fundamental se pueden observar fenomenos de frustraci\'on
(la energ\'{\i}a no puede ser mi\-nimizada simultaneamente para todos
los acoplamientos) o de desorden (la entrop\'{\i}a del vac\'{\i}o es distinta
de cero).

En el trabajo que se presenta a continuaci\'on se considera
la teor\'{\i}a $\LP$ en el l\'{\i}mite en que $\lambda \rightarrow \infty$,
lo cual equivale a fijar el m\'odulo del campo de Higgs. Para
introducir Antiferromagnetismo se ha a\~nadido
un acoplo negativo a segundos vecinos.
El objetivo es la b\'usqueda y caracterizaci\'on de puntos de transici\'on
de fase de segundo orden en el diagrama de fases.

\newpage

\begin{center}
{\bf Abstract}
\end{center}

{\sl
We study the phase diagram of the four dimensional O(4) model with
first ($\beta_1$) and second ($\beta_2$) neighbor couplings,
specially in the $\beta_2 <0$ region, where we find a line of
transitions which is compatible with second order.
We also compute the critical
exponents on this line at the point $\beta_1 =0$ ($\rm{F}_4$ lattice) by
Finite Size Scaling techniques up to a lattice size $L=24$, being these
exponents different from the Mean Field ones.}

\section{Description of the model}

Our starting point is the non-linear $\sigma$ model, with action:

\begin{equation}
S_{\sigma} = - \beta \sum_{\bf{r},\mu} 
        {\bf \Phi}_{\bf{r}} {\bf \Phi}_{\bf{r}+\hat{\bf{\mu}}}\ . 
\end{equation} 

Where ${\bf \Phi}$ is a 4-component vector with fixed modulus 
${\bf \Phi}_{\bf{r}}\cdot{\bf \Phi}_{\bf{r}}=1$.

The naive way to introduce AF in the non-linear $\sigma$ model is to consider
a negative coupling. In this case the state with minimal energy for large
$\beta$ is a staggered vacuum. On a hypercubic lattice, if we denote the
coordinates of site $\bf{r}$ as $(r_x,r_y,r_z,r_t)$, making the 
transformation
\begin{equation}
{\bf \Phi}_{\bf{r}}\to
        (-1)^{r_x+r_y+r_z+r_t}{\bf \Phi}_{\bf{r}}\ ,
\label{MAPPING}
\end{equation}
the system with negative $\beta$ is mapped onto the positive $\beta$
one, both regions being exactly equivalent.

Therefore to consider true AF we must take into account either different
geometries or more couplings, in order to break the symmetry under the
transformation
(\ref{MAPPING}). In four dimensions the simplest
option is to add more couplings, we have chosen to add a coupling between
points at a distance of $\sqrt{2}$ lattice units.

Following this we will consider a system of spins
\{$\bf{\Phi}_{\bf{r}}$\} taking values in the hyper-sphere 
${\rm S}^3 \subset \mathbf{R}^4$ and placed in the nodes of a cubic lattice.  
The interaction is defined by the action
\begin{equation}
S = - \beta_1 \sum_{\bf{r},\mu} 
        {\bf \Phi}_{\bf{r}} {\bf \Phi}_{\bf{r}+\hat{\bf{\mu}}} 
    - \beta_2 \sum_{\bf{r},\mu<\nu} 
        {\bf \Phi}_{\bf{r}} {\bf \Phi}_{\bf{r}+\hat{\bf{\mu}} + 
 \hat{\bf{\nu}}}\ ,
\label{ACCION}
\end{equation}

The transformation (\ref{MAPPING}) maps the semi-plane 
$\beta_1 >0$ onto the $\beta_1 <0$, and therefore
only the region with $\beta_1 \geq 0$ will be considered. On the line
$\beta_1 =0$ the system decouples in two $\rm{F}_4$ independent sublattices.

When $\beta_2 =0$ the model is known to present a continuous transition
between a disordered phase, where O(4) symmetry is exact, to an ordered
phase where the O(4) symmetry is spontaneously broken to
O(3).
This transition is second order, being the critical
exponents those of MFT: $\alpha=0$, $\nu=0.5$, $\beta=0.5$,
$\eta=0$ and $\gamma =1$ up to logarithmic corrections.
The critical coupling for this case can be studied analytically by an expansion
in powers of the coordination number ($q=2d$), being 
$\beta^{\rm c} = 0.6055 + O(q^{-2d})$ \cite{FISH1}.

From a Mean Field analysis, we observe that
for $\beta_2 > 0$ the behavior of the system will not change 
qualitatively from the $\beta_2 =0$ case but with higher coordination number. 
In fact, taking into account that the energy (for non-frustrated systems)
is approximately proportional to the coordination 
number, there will be a transition phase line whose approximate equation
is
\begin{equation}
\beta_1^{\rm c} + Q\beta_2^{\rm c} = \beta^{\rm c}\ ,
\label{RECTA}
\end{equation}
where $Q$ is the quotient between the number of second and first
neighbors, $2d(d-1)$ and $2d$, respectively.
This line can be thought as a prolongation of the critical point
at $\beta_2 =0$ so the transitions on this line are expected
to be second order with MFT exponents.
This is also the behavior of the two couplings Ising model in this
region \cite{ISING1}.

When $\beta_2 <0$, the presence of two couplings with opposite sign makes 
frustration to appear, and very different vacua are possible.

\section{Observables and order parameters}

We define the energy associated to each coupling:
\begin{equation}
E_1 \equiv \frac{\partial \log Z}{\partial \beta_1} =
 \sum_{\bf{r},\mu} \bf{\Phi}_{\bf{r}}\cdot \bf{\Phi}_{\bf{r} +
 \hat{\bf{\mu}}}\ ,
\end{equation}
\begin{equation}
E_2 \equiv \frac{\partial \log Z}{\partial \beta_2} =
 \sum_{\bf{r},\mu<\nu} \bf{\Phi}_{\bf{r}}\cdot
  \bf{\Phi}_{\bf{r} + \hat{\bf{\mu}} + \hat{\bf{\nu}}}\ .
\label{ENERGIAS}
\end{equation}

In terms of these energies, the action
reads
\begin{equation}
S = -\beta_1 E_1 -\beta_2 E_2\ .
\end{equation}

It is useful to define the energies per bound as
\begin{equation}
e_1 = \frac{1}{4V} E_1,\ e_2 = \frac{1}{12V} E_2\ ,
\end{equation}
where $V=L^4$ is the lattice volume. With this normalization $e_1$ , $e_2$
belong to the interval $[-1,1]$.

We have computed the configurations which minimize
the energy for several asymptotic values of the parameters. We have only
considered configurations with periodicity two. More complex structures have
not been observed in our simulations.

Considering only the $\beta_1\ge 0$ case, 
we have found the following regions:
\begin{enumerate}

\item{} Paramagnetic (PM) phase or disordered phase, for small absolute
values of $\beta_1,\beta_2$.    

\item{} Ferromagnetic (FM) phase. It appears when $\beta_1+6\beta_2$
is large and positive.

When the fluctuations go to zero, the vacuum takes the form
$\bf{\Phi}_{\bf{r}}= \bf{v}$, where $\bf v$ is an
arbitrary element of the hyper-sphere. 

Concerning the definition of the order parameter let us remark that
because of tunneling phenomena in finite lattice we are forced to use
pseudo-order parameters for practical purposes. Such quantities behave
as true order parameters only in the thermodynamical limit. In the FM
phase, we define the standard (normalized) magnetization as

\begin{equation}
{\bf M}_{\rm F} =\frac{1}{V} \sum_{\bf{r}} {\bf \Phi}_{\bf{r}}\ ,
\end{equation}
and we use as pseudo-order parameter the square root of the norm of
the magnetization vector
\begin{equation}
M_{\rm F} = \langle \sqrt{{\bf M}_{\rm F}^2} \, \rangle\ .
\end{equation}
This quantity has the drawback of being non-zero in the symmetric phase
but it presents corrections to the bulk behavior order $1/\sqrt{V}$.

\item{} Hyper-Plane Antiferromagnetic phase (HPAF). It corresponds
to large $\beta_1$, with $\beta_2$ in a narrow interval 
($[-\beta_1/2,-\beta_1/6]$ in the Mean Field approximation).
In this region the vacuum correspond to spins aligned in three
directions but anti-aligned in the fourth ($\mu$).

In absence of fluctuations the associated vacuum would be
${\bf \Phi}_{\bf{r}} = (-1)^{r_{\mu}} \bf{v}$, 
where $\mu$ can be any direction, and $\bf{v}$ any vector on S$^4$.

We  define an {\em ad hoc} order parameter for this phase as 
\begin{equation}
{\bf M}_{\rm{HPAF},\mu} = \frac{1}{V}
\sum_{\bf{r}}(-1)^{r_{\mu}} {\bf \Phi}_{\bf{r}}\ .
\end{equation}

${\bf M}_{\rm{HPAF},\mu}$ will be different from
zero only in the HPAF phase, where the system becomes antiferromagnetic on the
$\mu$ direction. From the four order parameters (one for every possible value
of $\mu$) only one of them will be different from zero in the HPAF phase.
So, we define as the pseudo order parameter:
\begin{equation}
M_{\rm HPAF} = \sqrt{ \sum_{\mu} {\bf M}_{\rm{HPAF},\mu}^2}\ .
\end{equation} 

\item{} Plane Anti-Ferromagnetic (PAF) phase for $\beta_2$ large and
negative.
In this region the ground state 
is a configuration with spins aligned in two directions and
anti-aligned in the remaining two. It is characterized with by
one of the six combinations of two different directions ($\mu,\nu$),
and an arbitrary spin $\bf v$:
${\bf \Phi}_{\bf{r}} = (-1)^{r_{\mu} + r_{\nu}} \bf{v}$.
For the PAF region we first define 
\begin{equation}
{\bf M}_{\rm{PAF},\mu,\nu} 
= \frac{1}{V} \sum_r (-1)^{r_{\mu} + r_{\nu}} {\bf \Phi}_{\bf{r}}\ ,
\end{equation}
 
and the quantity we measure is
\begin{equation}
M_{\rm PAF} = \sqrt{ \sum_{\mu < \nu} {\bf M}_{\rm{PAF},
(\mu,\nu)}^2 }
\label{MPAF}
\end{equation}

\end{enumerate}

In order to avoid undesirable (frustrating) boundary
effects for ordered phases, we work with even lattice side $L$ as
periodic boundary conditions are imposed.

From this data we can compute the derivatives of any observable with
respect to the couplings as the connected correlation function with
the energies
\begin{equation}
\frac{\partial O}{\partial \beta_j} =
 \langle O E_j \rangle - \langle O \rangle \langle E_j \rangle
\end{equation}

An efficient method to determine $\beta_{\rm c}$ for a second order transition 
is to measure the Binder cumulant \cite{BINDER1} for various lattice
size and to locate the cross point in the space of $\beta$.

For O($N$) models $U_L(\beta)$ takes the form \cite{BREZIN1}:
\begin{equation}
U_L(\beta) = 1 + 2/N - \frac{\langle ({\bf m}^2)^2 \rangle}
        {{\langle {\bf m}^2 \rangle}^2}
\label{BINDERPAR}
\end{equation} 
where $\bf{m}$ is an order parameter for the transition.

It can be shown \cite{BINDER1,BREZIN1} that $U_L(0) \rightarrow O(1/V)$ and
$U_L(\infty) \rightarrow 2/N$. 
The slope of $U_L(\beta)$ at $\beta_{\rm c}$ increases with  $L$.

The value of the Binder cumulant is closely related
with the triviality of the theory since the renormalized coupling
(in the massless thermodynamical limit)
at zero momentum can be written as:
\begin{equation}
g_{\rm R} = \lim_{L \rightarrow \infty} g_{\rm R}(L) = 
\lim_{L \rightarrow \infty} (L/\xi_L)^d U_L(\beta_{\rm c})
\label{GR}
\end{equation}
where $\xi_L$ is the correlation length in the size $L$ lattice. 

From this point of view triviality is equivalent to have a vanishing
$g_{\rm R}$ in the thermodynamical limit. In this context it is clear that
we can use the value of $g_{\rm R}$ to classify the universality class.
Out of the upper critical dimension, $L/\xi_L$ is a constant at
$\beta_{\rm c}$ since $\xi \sim L$, and we could use the Binder cumulant
for the same purpose \cite{PAR1}. At the upper critical dimension,
$\xi_L$ presents logarithmic corrections and $L/\xi_L$ is no longer a 
constant at $\beta_{\rm c}$. For the FM O(4) model in $d = 4$ (upper critical
dimension) we have perturbatively $L/\xi_L\sim (\ln L)^{-1/4}$ \cite{BRE1}. 
In order to have a non trivial theory, the Binder cumulant should
behave as a positive power of $\ln L$, 
but from its definition \cite{BINDER1} we see
that $U_L(\beta) \leq 1$. This is just another way of stating the
perturbative triviality of the FM O(4) model. 

\subsection{Symmetries on the F$_{\mathrm 4}$ lattice}

In the $\beta_1=0$ case the system decouples in two independent lattices, each
one constituted by the first neighbors of the other. So we consider
two lattices with $\rm{F}_4$ geometry.
There are several reasons to choose the point $\beta_1 = 0$ for a careful
study of the PM-PAF transition. The region with $\beta_1 > 1.5$ evolve
painfully with our local algorithms; For small $\beta_1$ 
we expect very large correlation in MC time because the interaction between 
both sublattices is very small, and the response of one lattice to changes
in the other is very slow. We also remark that the presence of two almost
decoupled lattices is rather unphysical.

We also have the experience from a previous
work for the Ising model \cite{ISING1} that the correlation length at
its first order transition is smaller in the $\rm{F}_4$ 
lattice, that means, we can find asymptotic critical behavior in smaller
lattices.

However we should point out that the results in the $\rm{F}_4$
lattice cannot be easily extrapolated to a neighborhood of the
$\beta_1$ axis. Certainly, the geometry of the model is very modified
when $\beta_1 \neq 0$, and perhaps continuity arguments present
problems. Nevertheless, we have run also the case $\beta_1 \sim 0$, and
as occurs in the Ising model we have not found qualitative differences.

In the following when we refer to the size of the lattice $L$ on the
$\rm{F}_4$ lattice we mean a lattice with $L^4/2$ sites. 

We have to find the configurations that maximize $E_2$ in order to define 
appropriate order parameters for the phase transition.

The system has a very complex structure. As starting point we have studied
numerically the vacuum with $\beta_2 \ll 0$. For this 
values we have found in the simulation:

\begin{enumerate}
\item{} The vacuum has periodicity two. 
To check this, we have defined: 
\begin{equation}
{\bf V}_i = \frac{1}{L^d/2^d} \sum_I {\bf \Phi}_{I_i}\ ,
\label{VMAG}
\end{equation}
where $i=0, \dots ,7$ stands for the $i^{th}$ vertex of each $2^4$
hypercube belonging to the $\rm{F}_4$ lattice, and with $I$ we denote the
$2^4$ hypercubes themselves.

From these vectors we can define the 8 magnetizations associated to the
elementary cell,
\begin{equation}
V_i = \langle \sqrt{{\bf V}_i^2} \rangle\ ,
\end{equation}

We have checked that all $V_i$ tends to 1 for the ordered phase in the
thermodynamical limit, so we conclude that the ordered vacua have
periodicity two.

Let us remark for the sake of completeness that all order parameters we have
defined can be written as an appropriate linear combination of the 
${\bf V}_i$.
\item{} In the elementary cell,
$\bf{\Phi}_{\bf{r}+\hat{\bf{\mu}}+ \hat{\bf{\nu}}} =
\bf{\Phi}_{\bf{r}}$ $\forall \mu, \nu$ with $\mu<\nu$. So, in
this section we will restrict the study of the vacuum structure to the
four sites ($i=0,1,2,3$) belonging to the cube in the hyper-plane
$r_t=0$.
\item{} We have measured the energy per bound associated to the
second neighbors coupling. We check that in the thermodynamical
limit $e_2 = -1/3$.
\item{} If we choose the symmetry breaking direction by keeping fix 
one vector, (eg. ${\bf \Phi}_0$) we find:
\begin{equation}
\sum_{i=1}^{3} \left((\bf{\Phi}_0\cdot\bf{\Phi}_i)\bf{\Phi}_0
        -\bf{\Phi}_i\right) = 0 \ ,
\end{equation}

\end{enumerate}

The vacuum structure is not completely fixed by these three conditions
since different symmetry breaking patterns are possible.
For instance, a configuration
${\bf \Phi}_0 = (1,0,0,0)$,  
${\bf \Phi}_1 = (-1/3,\frac{2\sqrt{2}}{3}{\bf v_1})$,
 ${\bf \Phi}_0 =(-1/3,\frac{2\sqrt{2}}{3}{\bf v_2}) $,  
${\bf \Phi}_0 =(-1/3,\frac{2\sqrt{2}}{3}{\bf v_3}) $, 
with {\bf $v_i$} a 3-component
unitary vector with the constraint $\sum_{i \neq j}{\bf v_i v_j} = 0$, 
breaks O(4), but an O(2) symmetry remains (for the different {\bf $v_i$}).

To determine which is the vacuum in presence of fluctuations,
we consider four independent fields in a $2^4$ cell with
periodic boundary conditions.
Let us first consider an O($2$) group.
 We can study the four vectors as a mechanical system of mass-less
links of length unity, rotating in a plane around the same point, whose
extremes are attached with a spring of natural length zero. The
energy for the system is:

\begin{equation}
E = - \sum_{i,j=0, i>j}^{3} \cos(\theta_i - \theta_j)\ .
\label{ENERO2}	
\end{equation}

We consider the fluctuation matrix, $H = \partial E^2 / \partial \theta_i
\partial \theta_j$ in order to find the normal modes. 
The matrix elements of $H$ take the form:

\begin{equation}
H_{i,j} = \delta_{ij} \sum_{k \neq i} \cos(\theta_i - \theta_k) -
                \cos(\theta_i - \theta_j)(1 - \delta_{ij})\ ,
\end{equation}

In the FM case the minimum correspond to $\theta_i=\phi$, 
for all $i$. The fluctuation matrix is

\[ H_{FM} = \left( \begin{array}{cccc}
-3   &1   &1   &1 \\
1    &-3  &1   &1 \\
1    &1   &-3  &1 \\
1    &1   &1   &-3
\end{array} \right)  \]

\noindent which has a single zero mode, and a three times degenerated
non-zero mode with eigenvalue $\lambda = -4$. 

For the AF (maximum energy) case, the maximum energy is found, up to
permutations, at $\theta_0=\phi$, $\theta_1 = \phi + \pi$, $\theta_2 =
\phi + \alpha$ and $\theta_3 = \phi + \pi + \alpha$, $\forall \alpha$. 
In addition to
the $\phi$ freedom that corresponds to the global O(2) symmetry, 
there is a degeneration of the vacuum in the $\alpha$ angle
and this zero mode is double $\forall \alpha$.

The fluctuation matrix in the AF case is

\[ H_{AF} = \left( \begin{array}{cccc}
1    &-1   &\cos\alpha   &-\cos\alpha \\
-1   &1    &-\cos\alpha   &\cos\alpha \\
\cos\alpha    &-\cos\alpha   &1  &-1 \\
-\cos\alpha    &\cos\alpha   &-1   &1
\end{array} \right)  \]

The  other two eigenvalues are $\lambda_{1,2} = 2(1 \pm \cos\alpha)$,
so, an additional zero mode appears when $\alpha=0$, obtaining
in this case a three fold degenerated zero mode corresponding to:
$\theta_0 = \theta_1 = \theta_2 + \pi = \theta_3 + \pi$. 

The O(4) case is qualitatively similar. We have 12 degrees of
freedom. Of all configurations that minimize the energy, that with a
largest degeneration (9-times) consist of 2 spins aligned and 2
anti-aligned that correspond to a PAF vacuum.
We consider this degeneration as the main difference with the FM sector,
and could be relevant to obtain different critical exponents.

In presence of fluctuations the configurations with largest 
degeneration are favored by phase space considerations, so we expect
that the real vacuum is a PAF one. This statement will be checked
below with Monte Carlo data in the critical region.

\section{Finite Size Scaling analysis}

Our measures of critical exponents 
are based on the FSS ansatz \cite{ITZ1,CARDY1}. Let be
$\langle O(L,\beta) \rangle$ the mean value of an observable measured
on a size $L$ lattice at a coupling $\beta$. If 
$O(\infty,\beta) \sim \vert \beta - \beta_{\rm c} \vert^{x_O}$, from the FSS
ansatz one readily obtains \cite{CARDY1}
\begin{equation}
\langle O(L,\beta) \rangle = L^{x_O/\nu} F_O(L/\xi(\infty,\beta)) +
\ldots \ ,
\label{FSS}
\end{equation}
where $F_O$ is a smooth function and the dots stand for corrections to
scaling terms. 

To obtain $\nu$ we apply equation (\ref{FSS}) to the
operator $\rm{d} \log M_{\rm{PAF}}/\rm{d} \beta$ 
whose related $x$ exponent is
1. As this operator is almost constant in the critical region, 
we just measure at the extrapolated critical point or any 
definition of the apparent critical point in a finite lattice, the
difference being small corrections-to-scaling terms.

For the magnetic critical exponents the situation is more involved
as the slope of the magnetization or the unconnected susceptibility
is very large at the critical point.

We proceed as follows (see refs. \cite{RP21} for other applications
of this method). Let be $\Theta$ any operator
with scaling law $x_{\Theta} = 1$ (for instance
the Binder parameter or a correlation length defined in a finite
lattice divided by $L$).
Applying eq. (\ref{FSS}) to an arbitrary operator, $O$,  and to $\Theta$
we can write
\begin{equation}
\langle O(L,\beta) \rangle = 
        L^{x_O/\nu} f_{O,\Theta}(\langle\Theta(L,\beta)\rangle)+
        \ldots \ .
\end{equation}
Measuring the operator $O$ in a pair of lattices of sizes $L$ and $sL$
at a coupling where the mean value of $\Theta$ is the same, one
readily obtains
\begin{equation}
\left.\frac{\langle O(sL,\beta) \rangle}{\langle O(L,\beta) \rangle}
\right|_{\Theta(L,\beta)=\Theta(sL,\beta)} = s^{x_O / \nu} + \ldots \ .
\label{QUOTIENT}
\end{equation}
The use of the spectral density method (SDM) \cite{FS1} avoids an 
exact a priori knowledge of the 
coupling where the mean values of $\Theta$ cross. 
We remark that usually the main source
of statistical error in the measures of magnetic exponents is the
error in the determination of the coupling where to measure. However, 
using eq. (\ref{QUOTIENT}) we can take into account the correlation
between the measures of the observable and the measure of the
coupling where the cross occurs. This allows to reduce
the statistical error in an order of magnitude.

\subsection{FSS at the upper critical dimension: logarithmic corrections}

Being $d=4$ the upper critical dimension of the FM O(4) model,
logarithmic corrections to the Mean Field predictions are expected
to set in.
In particular, FSS in its standard formulation breaks at $d = 4$ because
the essential assumption, namely $\xi_L(\beta_{\rm c}) \sim L$ 
is no longer true. In fact, in four dimensions \cite{BELLAC1}:
\begin{equation}
\xi (\infty, t)  \sim  \vert t \vert^{-1/2} 
\vert \ln \vert t \vert \vert^{\frac{1}{4}}   \ .
\end{equation}

The FSS formula for the correlation length was calculated by
Brezin \cite{BRE1}. At the critical point one gets:
\begin{equation}
\xi (L, \beta_{\rm c}) \sim L(\ln L)^{1/4} \ .
\end{equation}

It has been suggested \cite{LANG1} that the usual FSS statement should
be replaced by the more general one:
\begin{equation}
\frac{O(L, \beta_{\rm c})}{O(\infty, \beta)} =
F_O\left(\frac{\xi(L, \beta_{\rm c})}{\xi(\infty, \beta)}\right)\ .
\end{equation}

When applying the quotient method described above to systems in four
dimensions one has to take into account the logarithmic corrections,
so that the modified formula reads:
\begin{equation}
\left.\frac{\langle O(sL,\beta) \rangle}{\langle O(L,\beta) \rangle}
\right|_{\Theta(L,\beta)=\Theta(sL,\beta)} = 
s^{x_O / \nu}\left(1 + \frac{\ln s}{\ln L}\right)^{1/4} \ .
\label{QUOLOG}
\end{equation}

This point is particularly important when measuring the magnetic critical
exponents because as we have already mentioned, the slope of the 
magnetization and susceptibility are very large, and special
care has to be taken when locating the coupling where to measure.

\section{Numerical Method}

We have simulated the model in a $L^4$ lattice with periodic
boundary conditions. The biggest lattice size has been $L=24$.
For the update we have employed a combination of Heat-Bath and Over-relaxation
algorithms (10 Overrelax sweeps followed by a Heat-Bath sweep).

The dynamic exponent $z$ we obtain is near 1.
Cluster type algorithms are not
expected to improve this $z$ value. In systems with competing
interactions the cluster size average is a great fraction of the whole
system, loosing the efficacy they show for ferromagnetic spin systems.

We have used for the simulations ALPHA processor based machines. The total
computer time employed has been the equivalent of two years of
ALPHA AXP3000. We measure every 10 sweeps and store the individual measures
to extrapolate in a neighborhood of the simulation coupling by using
the SDM.

In the $\rm{F}_4$ case, we have run 
about $2\times10^5\tau$ for each lattice size, 
being $\tau$ the
largest integrated autocorrelation time measured, that corresponds to
$M_{\rm PAF}$, and ranges from 2.3 measures for $L=6$ to 8.9 for $L=24$.
 We have discarded more than $10^2\tau$ for
thermalization.  The errors have been estimated with the jack-knife
method.

\section{Results and Measures}

\subsection{Phase Diagram}

We have studied the phase diagram of the model using a $L=8$ lattice. We
have done a sweep along the parameter space of several thousands
of iterations, finding the transition lines shown in Figure 1.
The symbols represent the coupling values where a peak in the order parameter
derivative appears.

\begin{figure}[t]
\centerline{
\fpsxsize=8cm
\def\fpsangle{90}
\fpsbox[70 90 579 760]{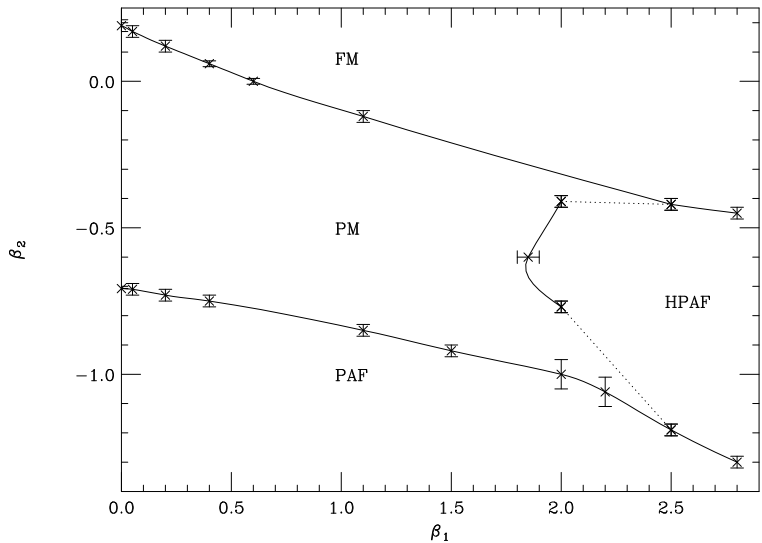} }
\caption{\small Phase diagram obtained from the MC simulation on 
a $L=8$ lattice } 
\label{PHASE}
\end{figure}

\begin{figure}[!t]
\centerline{
\fpsxsize=8cm
\def\fpsangle{90}
\fpsbox[70 90 579 760]{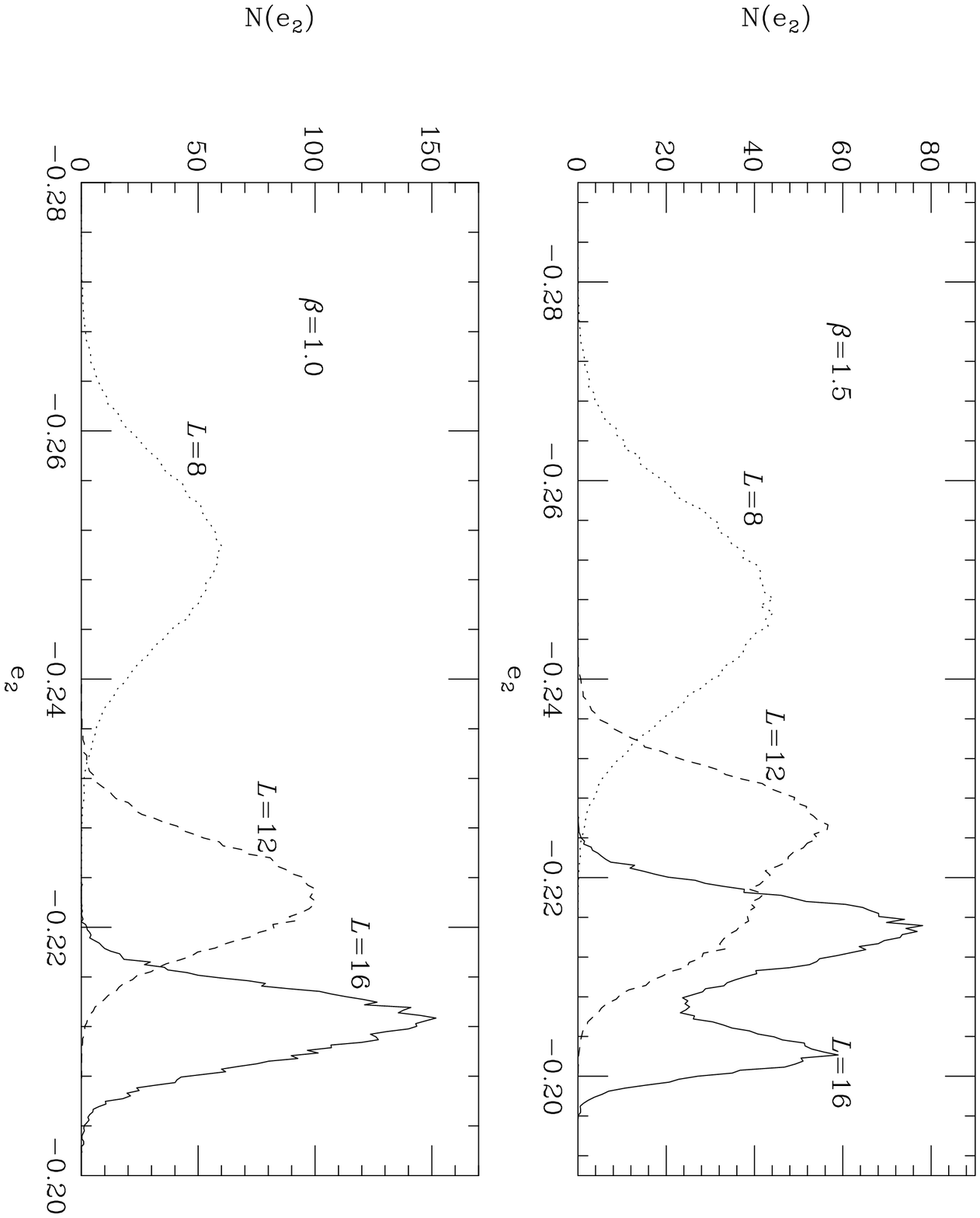} }
\caption{\small Energy distribution for $L$=8,12,16 
measured at the peak of the specific heat at $\beta_1=1.5$ (upper window)
and at $\beta_1=1.0$ (lower window).} 
\label{fignew}
\end{figure}      

The line FM-PM has a clear second order behavior.
It contains the critical point for the O(4) model with first neighbor
couplings ($\beta_1\approx 0.6$, $\beta_2=0$) with classical exponents
($\nu=0.5$, $\eta=0$).
In the $\beta_1 = 0$ axis, we have computed the critical coupling
($\beta_2^{\rm c} \approx 0.18$) and the critical exponents as a test 
for the method in
the $\rm{F}_4$ lattice. We have also considered the
influence of the logarithmic corrections when computing the exponents.

The lines FM-HPAF, HPAF-PAF and PM-HPAF show clear metastability,
indicating a first order transition.

The regions between the lower dotted line and the PAF transition line,
and between the upper dotted line and the FM transition line, are
disordered up to our numerical precision. We could expect always a PM
region separating the different ordered phases, however, from a MC
simulation it is not possible to give a conclusive answer since the
width of the hypothetical PM region decreases when increasing
$\beta_1$, and for a fixed lattice size there is a practical limit in
the precision of the measures of critical values.

The line PM-PAF has a very interesting behavior. To get ride of the influence
of the HPAF region we have started the study at $\beta_1 = 1.5$. The
energy distributions encountered at  $\beta_1 = 1.5$ and  $\beta_1 = 1.0$
are displayed in figure \ref{fignew}.
We do not obeserve latent heat anymore in lattices up to L=16
in $\beta_1=0$. Two scenarios are suggested by this fact:

\begin{enumerate}
\item{} There exist a value of $\beta_1$ in which the order of
the phase transition changes from being first order, to be continuous.
\item{} The phase transition line is first order everywhere, though increasily
weak as the limit $\beta_1 = 0$ is approached.
\end{enumerate}

\begin{figure}[!b]
\centerline{
\fpsxsize=8cm
\def\fpsangle{90}
\fpsbox[70 90 579 760]{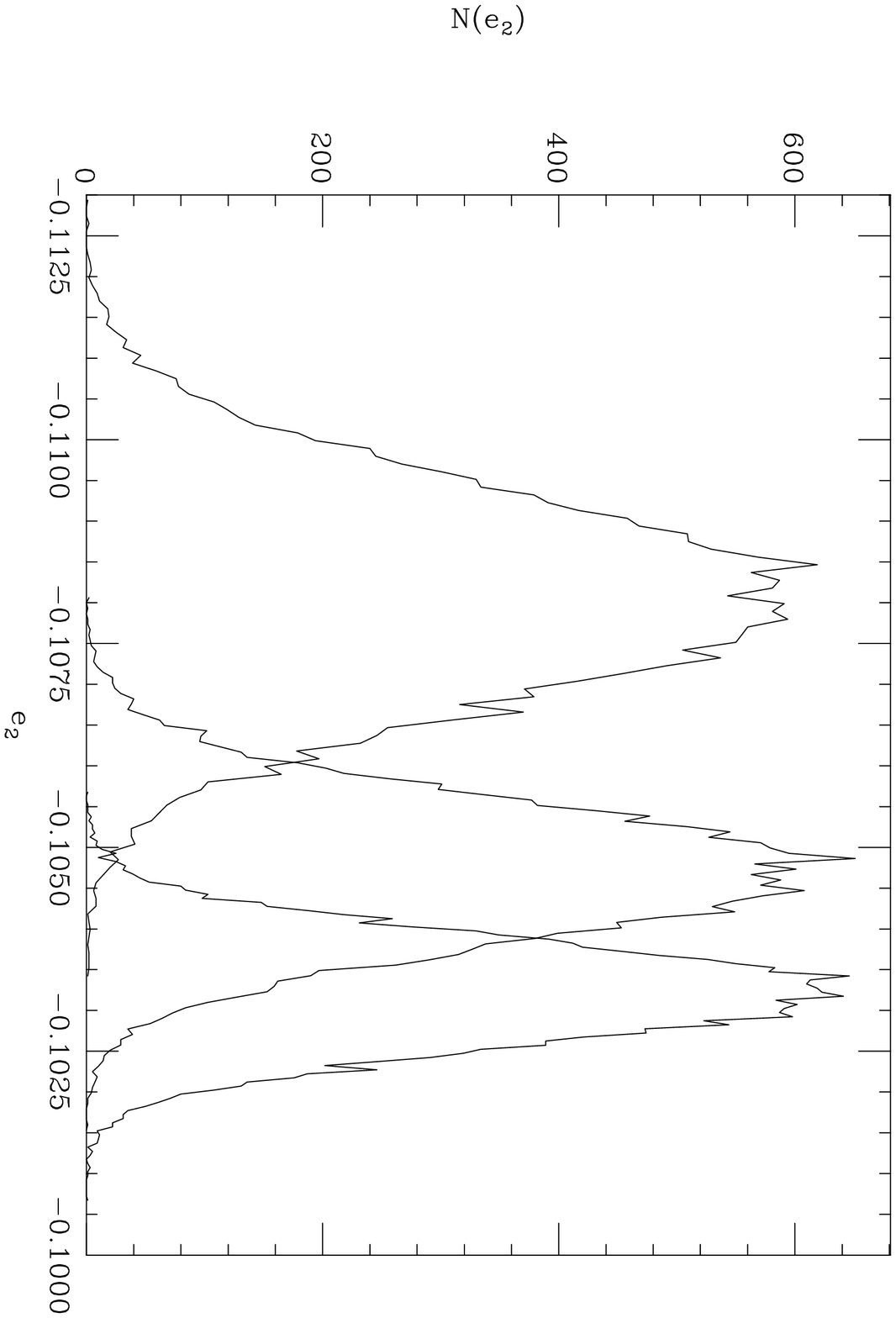} }
\caption{\small Energy distribution for $L$=16,20 and 24 at the peak
of the specific heat on the F$_4$ lattice. } 
\label{TODOSH}
\end{figure}

The possibility of a second order behavior of the PM-PAF transition 
line contrast with the first order one found in the Ising model 
with two couplings in the analogous region \cite{ISING1}. 
This would not be that surprising because we
are dealing now with a global continuous symmetry. The spontaneous 
symmetry breaking of such symmetries manifest in the appearance of soft
modes or low energy excitations (long wavelength), the Goldstone bosons
in QFT terminology \cite{BELLAC1}. In general, these low energy modes 
will perturb the mechanism of long distance ordering, softening in this way the
phase transitions.

Regarding the differences with the FM case, the most remarkable
feature is the different vacuum structures appearing, specially the very large
degeneration in the PAF transition, in contrast with the single
degeneration of the FM O(4) mode.

For the reasons mentioned in section 1.3.1,
the simpler point for studying the properties of the transition,
namely the critical exponents is the $\rm{F}_4$ limit. Most of the MC
work has been done for this case.

\subsection{Results on the F$_{\mathrm 4}$ lattice}

\subsubsection{Results on the FM region}
Firstly, we have checked our method on the FM region of the $\rm{F}_4$
lattice. In Figure 3 the crossing points of the Binder cumulant for
various lattice sizes are displayed. The prediction for the critical
coupling $\beta_{\rm c} \sim 0.1831(1)$ agrees with an earlier study by
Bhanot \cite{BHA21}.

\begin{figure}[t]
\centerline{
\fpsxsize=8cm
\def\fpsangle{90}
\fpsbox[70 90 579 760]{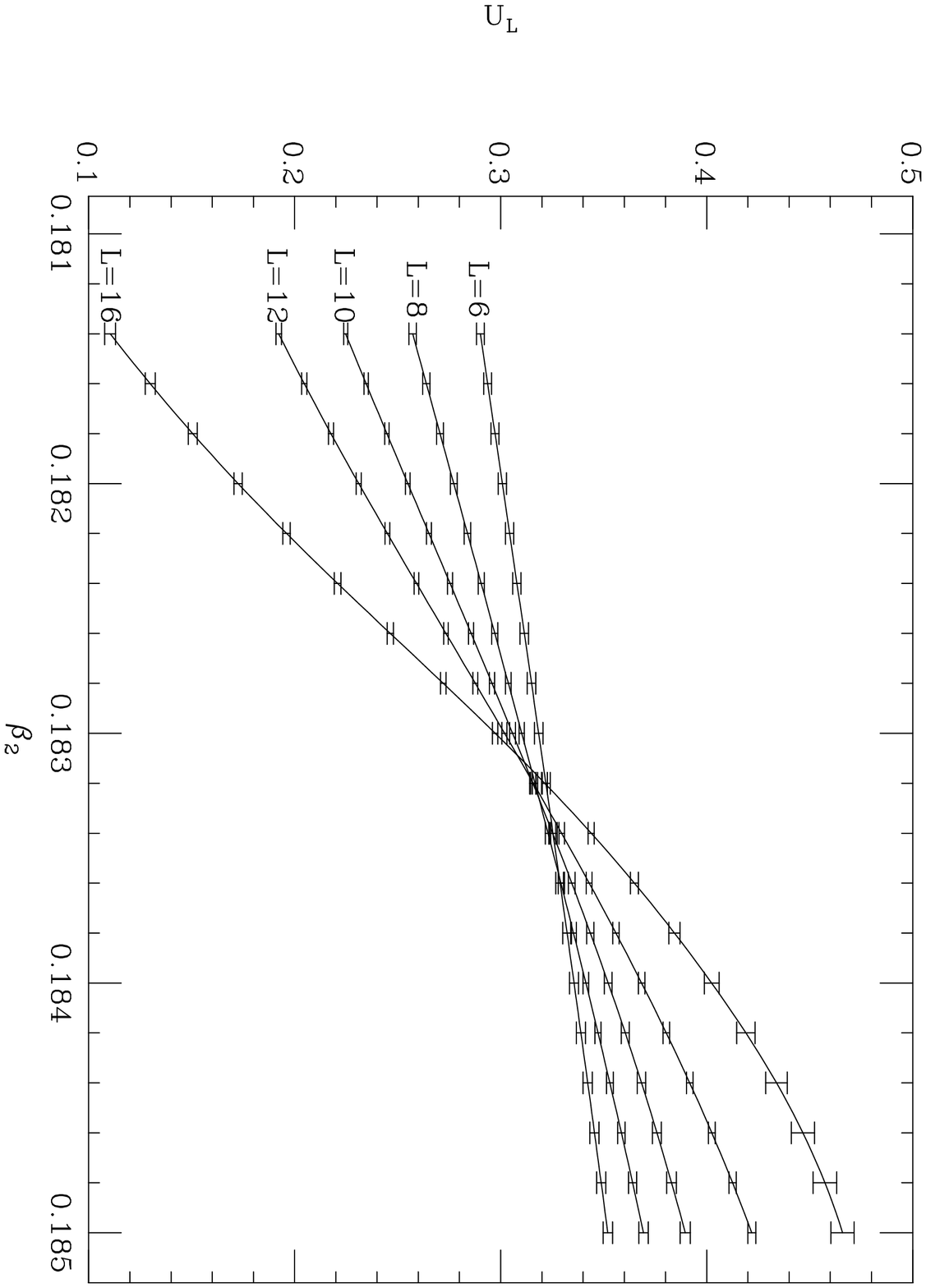} }
\caption{\small Crossing points of the Binder Cumulant for various lattice
sizes on the FM-PM phase transition } 
\label{BINDER_F}
\end{figure}

Concerning the measures of critical exponents, we have applied the
quotient method, described in section 1.4 . In table \ref{TABLE_LOG}
we quote the results when logarithmic corrections are included
(formula (\ref{QUOLOG})), and also for sake of comparison, when they
are neglected (formula (\ref{QUOTIENT})). We see how in fact the agreement
of the critical exponents with the MFT predictions is better when the
logarithmic corrections are taken into account.

\begin{table}[h]
{

\begin{center}
{
\begin{tabular}{|c|c|c|c|}\hline
L values  &$8/16$  &$12/16$  &$10/12$  \\ \hline
\multicolumn{4}{|c|}{(without logarithmic corrections)} \\ \hline
$\alpha$/$\nu$                    &0.08(5)    &0.02(2)  &0.13(12) \\
\hline
$\beta$/$\nu$
  &0.92(3)   &0.94(3) &0.87(4)  \\ \hline
$\gamma$/$\nu$
  &2.16(2)  &2.12(2) &2.24(4)  \\ \hline
\multicolumn{4}{|c|}{(with logarithmic corrections)} \\ \hline
$\alpha$/$\nu$                      &0.0   &0.0  &0.03(8) \\
\hline
$\beta$/$\nu$
  &1.04(3)  &1.06(3) &1.04(2)  \\ \hline
$\gamma$/$\nu$
  &1.94(3)  &1.90(4) &1.93(3)  \\ \hline
\end{tabular}
}
\end{center}
}
\caption[a]{Critical exponents for the FM-PM phase transition in the
$\rm{F}_4$ lattice.}
\protect\label{TABLE_LOG}

\end{table}

From now on we will focus on the transition between the PM phase
and the PAF phase on the $\rm{F}_4$ lattice.

\subsubsection{Vacuum symmetries on the PAF region}

We will check using MC data that the ordered vacuum in the critical
region is of type PAF.

Let us define
\begin{equation}
A_{ij}={\bf V}_i\cdot{\bf V}_j\ .
\end{equation}
The leading ordering corresponds to the eigenvector associated to the
maximum eigenvalue of the matrix $A$, that should scale as
$L^{-2\beta/\nu}$ at the critical point.
The scaling law of the biggest eigenvalue agrees with the $\beta/\nu$
value reported in Table \ref{TABLE_EXPO}, and the associated eigenvector
is, within errors, (1,1,-1,-1).

We also have found that the other eigenvalues scale as
$L^{-4}$.  This is the expected behavior if just the O(4) symmetry is
broken, and it remains an O(3) symmetry in the subspace orthogonal to
the O(4) breaking direction.

\subsubsection{Critical Coupling}

To obtain a precise determination of the critical point, $\beta_{\rm c}$,
we have used the data for the Binder parameter (\ref{BINDERPAR}).

In Figure 4 we plot the crossing points of the Binder
cumulants for the simulated lattices sizes. Extrapolations have been done
using SDM from simulations at $\beta_2 = -0.7090$ for
$L=6,8,10,12$ and 16; $\beta_2 =-0.7078$ for $L=20$, and
$\beta_2=-0.7070$ for $L=24$.

\begin{figure}[!t]
\centerline{
\fpsxsize=8cm
\def\fpsangle{90}
\fpsbox[70 90 579 760]{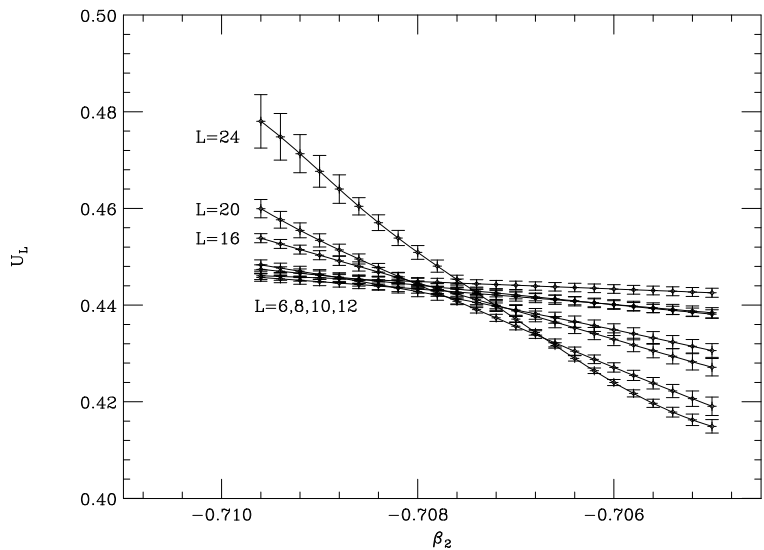} }
\caption{\small Crossing points of the Binder cumulant for various  lattice
sizes.  } 
\label{BINDER}
\end{figure}

The shift of the crossing point of the curves  can be explained through the
finite-size confluent corrections. The dependence in the deviation
of the crossing point for $L$ and $sL$ size lattices was estimated by 
Binder \cite{BINDER1}
\begin{equation}
\beta_{\rm c}(L,sL) - \beta_{\rm c} 
\sim \frac{1 - s^{-\omega}}{s^{1/\nu} - 1}L^{-\omega - 1/\nu}\ ,
\label{SHIFT}
\end{equation}
where $\omega$ is the universal exponent for the
corrections-to-scaling.

The infinite volume critical point the value
\begin{equation}
\beta_{\rm c}=-0.7065(5)[+2][-2]\ ,
\label{BETAC}
\end{equation}
where the  errors in brackets correspond to the variations in the
extrapolation when we use the values $\omega=0.5$ and $\omega=2$
respectively. In Figure 5 we plot eq. (\ref{SHIFT}) for $\omega=1$.

\begin{figure}[t]
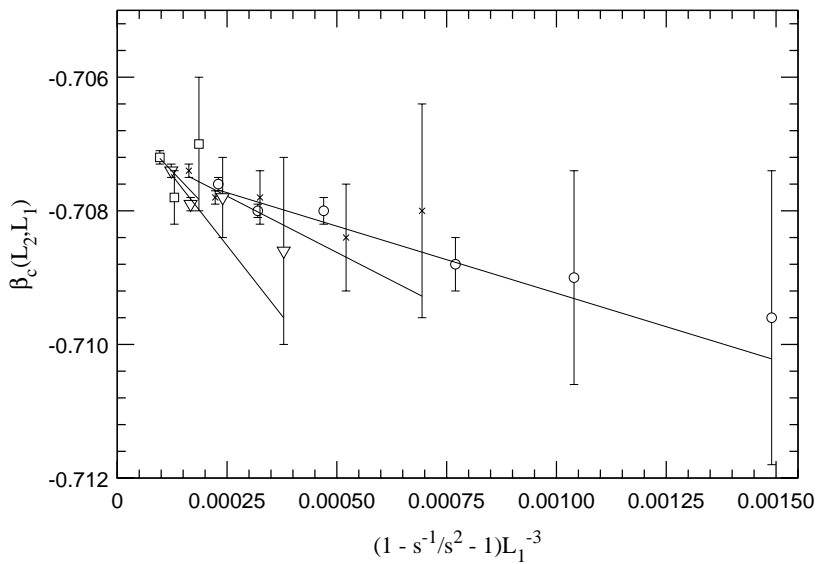

\centerline{
\fpsxsize=8cm
\def\fpsangle{270}
\fpsbox[70 90 579 760]{figura5.ps }}
\caption{\small
Extrapolation to $\beta_{\rm c} (\infty)$ for $L_1 = 6, 8, 10, 12$
(circles, crosses, triangles and squares symbol respectively)} 
\label{BINFI}
\end{figure}

Using the previous value of $\beta_{\rm c}$ we can compute the Binder
cumulant at this point. In table
\ref{TABLE_BINDER} we quote the obtained values. The result points to
that the Binder cumulant stays constant in the critical region. This
result would be compatible with a non zero value of the renormalized
coupling when $L$ increases.

\begin{table}[h]
{
\begin{center}
{
\begin{tabular}{|c|c|c|c|}\hline
Lattice sizes  &$U_L(\beta_{\rm c}(\omega=0.5))$ &$U_L(\beta_{\rm c}
		(\omega=1))$ 
               &$U_L(\beta_{\rm c}(\omega=2))$        \\ \hline
$6$       &0.4435(15) &0.4437(12)   &0.4438(12)           \\   \hline
$8$       &0.4406(15) &0.4409(12)   &0.4413(12)            \\   \hline
$10$      &0.4407(14) &0.4411(16)   &0.4414(14)            \\  \hline
$12$      &0.436(4) &0.437(3)   &0.438(3)            \\  \hline
$16$      &0.435(3) &0.436(3)   &0.437(3)            \\   \hline
$20$      &0.429(5) &0.431(5)   &0.433(5)            \\  \hline
$24$      &0.428(6) &0.430(7)   &0.433(7)            \\  \hline
\end{tabular}
}
\end{center}
}
\caption[a]{ Binder cumulant for various lattices sizes at the 
extrapolated critical point for $\omega = 0.5,1,2$.}
\protect\label{TABLE_BINDER}

\end{table}

Concerning the possibility of having logarithmic corrections in the
determination of the critical coupling, from the numerical point of
view, it is not possible
to discern between the $\omega$ effect, and a logarithmic correction.

\subsubsection{Thermal Critical exponents: $\alpha$, $\nu$}

The critical exponent associated to correlation length can be obtained
from the scaling of:
\begin{equation}
\kappa = \frac{\partial \log{M}}{\partial \beta}\ ,
\end{equation}
where $M$ is an order parameter for the transition, $M_{\rm PAF}$
for our purposes. In the critical region $\kappa \sim L^{1/\nu}$. As
$\kappa$ is a flat function of $\beta$, is not crucial the point where
we actually measure. The results displayed in table \ref{TABLE_EXPO}
have been obtained measuring at the crossing point of the Binder
parameters for lattice sizes $L$ and $2L$ using (\ref{QUOTIENT}).

For measuring $\alpha / \nu$ we study the scaling of the specific heat
\begin{equation}
C = \frac{\partial \langle E_2 \rangle}{\partial \beta_2}\ ,
\end{equation}
We expect that $C$ scales as $A+BL^{\alpha/\nu}$, where $A$ is usually
non-negligible.  In Figure 6 we plot the specific heat
measuring at (\ref{BETAC}), as well as at the peak of the specific
heat, as a function of $L$. We observe a linear behavior for intermediate
lattices. For the largest lattice the slope decreases. The weak first
order behavior~\cite{POTTS1} ($\alpha/\nu=1$ for small lattices that
becomes $d$ for large enough sizes) seem hardly compatible with our 
data. If we neglect the $A$ term (what is asymptotically correct), 
and compute the exponent using eq. (\ref{QUOTIENT}) we obtain 
$\alpha/\nu \approx 0.3$ for intermediate lattices that reduces to
$\alpha/\nu=0.15(2)$ for the (20,24) pair.

\begin{figure}[t]
\centerline{
\fpsxsize=8cm
\def\fpsangle{90}
\fpsbox[70 90 579 760]{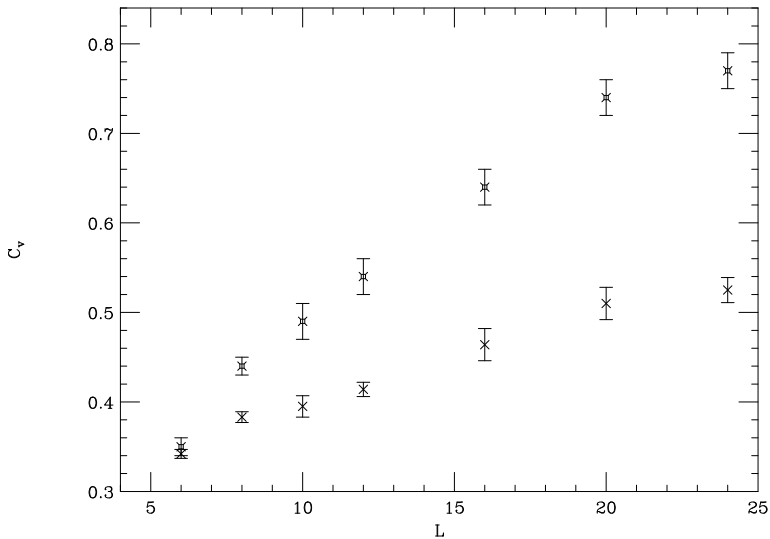 }}
\caption{\small Specific heat at the peak (triangle symbols) and at $\beta=-0.7068$ (cross symbols) as a function of the lattice size.  } 
\label{HEAT}
\end{figure}

However, to give a conclusive answer for the value of $\alpha$ 
statistics on larger lattices are mandatory.

\subsubsection{Magnetic Critical exponents: $\gamma$, $\beta$}

The exponents $\gamma$ and $\beta$ can be obtained respectively from
the scaling of susceptibility and magnetization:
\begin{equation}
\chi \equiv V \langle M^2 \rangle \sim L^{\gamma / \nu} 
\end{equation}
\begin{equation}
M \sim L^{-\beta / \nu}
\end{equation}

Where $M$ is an order parameter for the phase transition.  In Figure 7
upper part, we plot the quotient between $M_{\rm PAF}$
for lattices $L$ and $2L$ as a function of the quotient between the
Binder cumulants for both lattice sizes.

\begin{figure}[t]
\centerline{
\fpsxsize=8cm
\def\fpsangle{90}
\fpsbox[70 90 579 760]{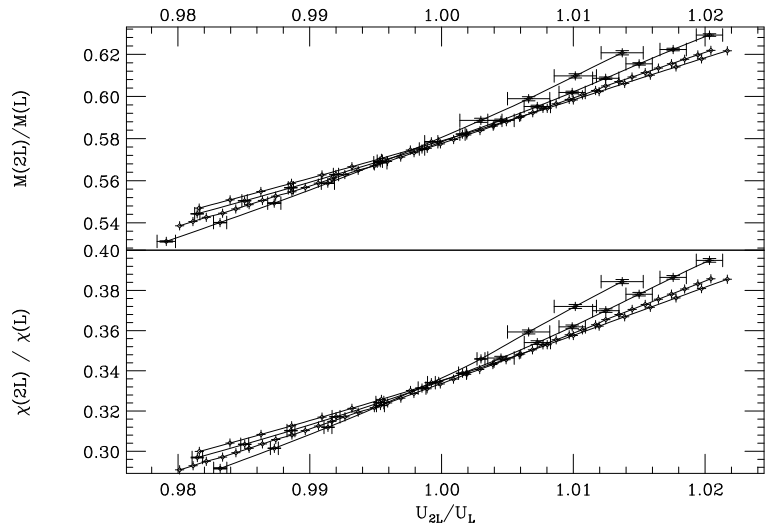} }
\caption{\small Quotients to obtain $\beta/\nu$ and $\gamma/\nu$  } 
\label{l2l.ps}
\end{figure}

For large $L$ in the critical region we should obtain a single curve,
the deviations corresponding to corrections to scaling.
In the lower part of Figure 7 we plot the same function for
susceptibility.

The values for $\gamma$ and $\beta$ are summarized in Table \ref{TABLE_EXPO}.

\begin{table}[h]
{
\begin{center}
{
\begin{tabular}{|c|c|c|c|}\hline
Lattice sizes  &$\gamma / \nu$  &$\beta / \nu$   &$\nu$      \\  \hline
\multicolumn{4}{|c|}{(without logarithmic corrections)} \\ \hline
$6/12$                    &2.417(3)    &0.791(4)  &0.474(10) \\  \hline
$8/16$                    &2.403(3)    &0.792(6)  &0.483(8)  \\  \hline
$10/20$                   &2.410(2)    &0.790(4)  &0.471(6)  \\  \hline
$12/24$                   &2.403(5)    &0.797(5)  &0.483(7)  \\  \hline
$20/24$                   &2.398(5)    &0.802(4)  &0.487(6)  \\  \hline
\multicolumn{4}{|c|}{(with logarithmic corrections)} \\ \hline
$6/12$                    &2.301(3)    &0.849(4)    &0.484(9)   \\    \hline
$8/16$                    &2.300(3)    &0.850(5)    &0.489(7)   \\   \hline
$10/20$                   &2.315(2)    &0.843(3)    &0.488(5)   \\   \hline
$12/24$                   &2.314(2)    &0.842(5)    &0.487(5)     \\   \hline
$20/24$                   &2.317(5)    &0.839(4)    &0.498(5)  \\   \hline
\end{tabular}
}
\end{center}
}
\caption[a]{ Critical Exponents for the PM-PAF phase transition in the 
$\rm{F}_4$ lattice}
\protect\label{TABLE_EXPO}

\end{table}

\subsection{Logarithmic corrections}

 We now address the question of the possibility of logarithmic corrections
in the AF O(4) model. For the thermal critical exponents, the situation
seems clear, they are compatible with the classical exponent $\nu = 0.5$.
For the magnetic exponents, the situation is more involved. In principle,
one can think that they disagree from MFT due to logarithmic corrections.
We have no perturbative predictions about the form in which these
corrections would affect $\xi_L$ for the AF case. However, one
expects that such corrections slightly modify the critical exponents,
as occurs in the FM case.
It could be possible that logarithmic corrections modify largely the
previous critical exponents and drift them to the FM ones. To sort this
out, we have considered the possibility of
a behavior FM like, so that $\xi_L \sim L(\ln L)^{1/4}$. 
In the lower part of Table \ref{TABLE_EXPO} we quote the values of the
critical exponents for the PAF phase transition when logarithmic
corrections are included (formula (\ref{QUOLOG})). We see how in effect
the magnetic critical exponents are too far from the classical ones
for being the result of a logarithmic correction to the MFT predictions.
It is interesting to compare this situation with that in the RP$^2$
model in $d=4$ \cite{RP241} where small deviations from MFT exponents
can be explained as logarithmic corrections.

\section{Conclusions and open problems}

\begin{figure}[t]
\centerline{
\fpsxsize=8cm
\def\fpsangle{90}
\fpsbox[70 90 579 760]{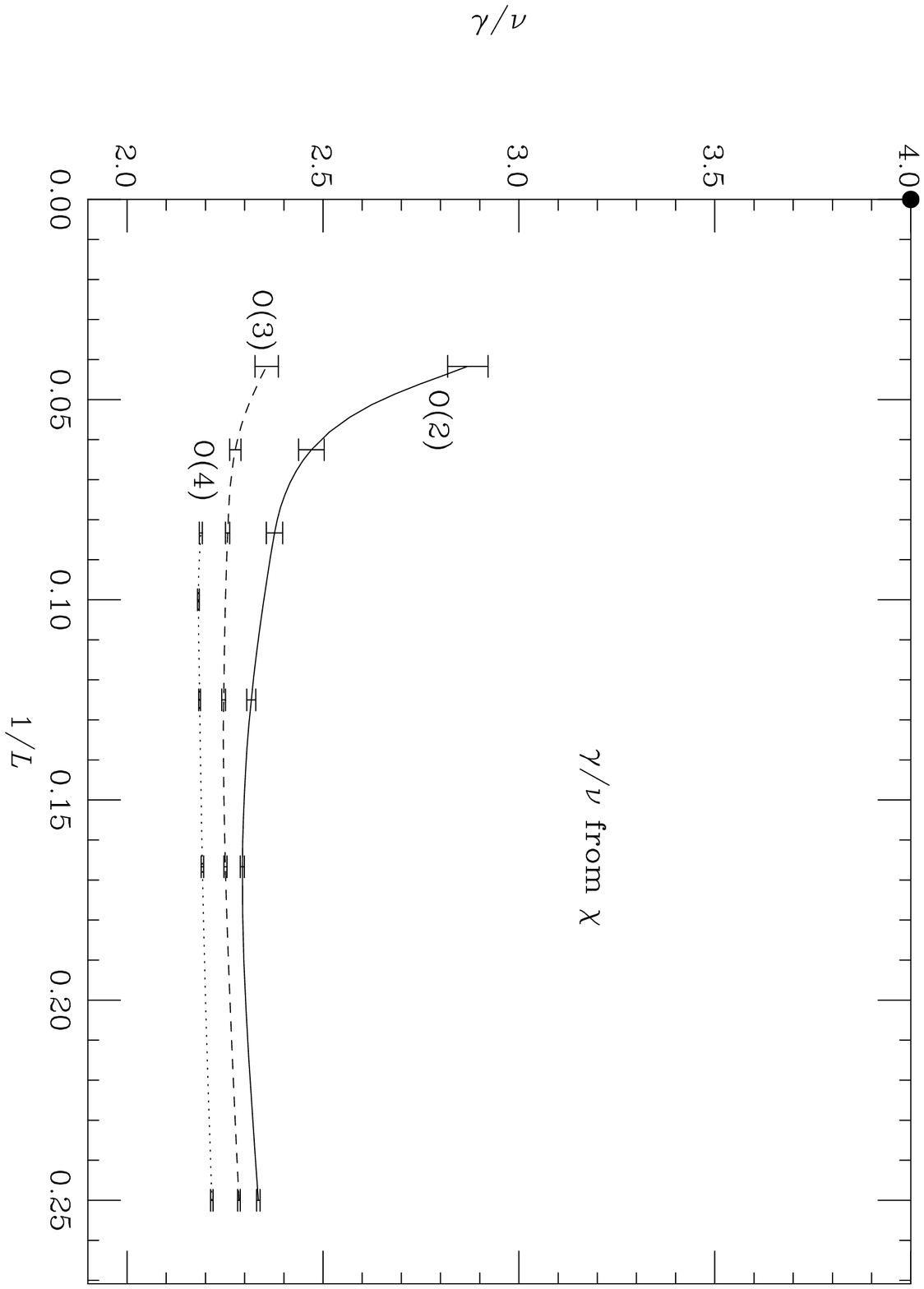} }
\caption{\small Effective $\gamma/\nu$ exponents in $O(N)$ with N=2,3,4.  } 
\label{ON}
\end{figure}

We have studied the phase diagram of the four dimensional O(4)
model with first and second neighbors couplings. We have found that
the presence of couplings with opposite signs makes frustration to
manifest in form of antiferromagnetism in different dimensionalities.
Whether or not these vacua have interest from the point of view
of Quantum Field Theory is an open question.
The transition between the paramagnetic phase, and the AF plane ordered
one is compatible with very weak first order and with higher order.
To shed some light on this question we have studied the evolution
of the effective critical exponents in the limit $\beta_1=0$. We found
that up to $L=24$ the exponents are in disagreement with the Mean Field
predictions. Specifically, from our $\gamma/\nu$ estimation
(or $\beta/\nu$ using hyper-scaling relation) the exponent $\eta$
asociated with the anomalous dimension of the field is $\eta \approx -0.4$.
This fact itself would imply the non-triviality of the theory
because Green functions would not factorize anymore. We have also 
measured the Binder cumulant at the critical point, finding that it stays
almost constant when increasing the lattice size.

However, before claiming about the important issue of a non-trivial
phase transition in four dimensions we have tried to acomodate the data
in two possible scenarios: 1) logarithmic triviality; 2) weak first
order scenario.

The mere observation of critical exponents different from the Mean Field
ones is in principle still compatible with a logarithmic trivial theory.
We have contrasted our data with the perturbative predictions derived
in the FM case, which do not fit the results. It would
be possible to obtain triviality also with a logarithmic exponent in equation 
(\ref{QUOLOG}) different from 1/4. We can fix the critical exponents
to its mean field value and compute this parameter from the numerical
data. The results obtained shown a non-asymptotic behavior with values
ranging from 0.8 to 1.2 for the lattices used.
In short, this model encounters the typical problems observed by other
triviality studies, which make the numerical approach to this issue
so difficult: one would need much stronger analytic insight to rule out
logarithmic triviality.

In the other hand, the possibility of being behind a very weak first
order phase transition is a possibility that cannot be discarded a priori.
Our data do not evidence any two state signature. Moreover,
the stability of our measure of $\gamma/\nu$ for lattices ranging from
$L=6$ to $L=24$, which are more than a hundred standard deviations apart
the MF value, makes this hypothesis rather unlikely.
However the question is wether or not this hypothesis can be ruled
out beyond any doubt.
The phase transition in this region for the Ising case, O(1), has been shown
to be first order \cite{ISING1}. Lately, the models O(2) and O(3)
have been considered with the purpose of checking wether or not the
order of the phase transition changes with $N$ \cite{ON1} in lattice
sizes as larger as $L=48$.
Evidence is given for a first order character of the transition
in O(2). In O(3) the transition is weaker, and despite no double peak
structure is observed, the behavior of the effective exponents leaves
little doubts about the first order character of O(3) as well (see Figure
\ref{ON}). In the O(4) case we cannot observe any particular trend
in the effective $\gamma / \nu$ . Indeed, since the transition seems
to be continuous in the limit $N \rightarrow \infty$, $N=4$ might be
a sort of critical $N$ value from which on the transition is second order.

However, from the observations in O(2) and O(3) we would be prone
to consider the weak first order scenario as the most plausible.
If the transition turns out to be
first order, it would be weaker than the observed in O(3), and hence
lattices larger than $L=48$ would be needed to observe such asymptotic
behavior in the critical exponents.

\newpage

\chapter{La transici\'on de fase en el modelo SU(2)-Higgs \label{capSU2}}
\thispagestyle{empty}
\markboth{\protect\small CAP\'ITULO \protect\ref{capSU2}}
{\protect\small {\sl La transici\'on de fase en el modelo SU(2)-Higgs}}

\cleardoublepage        

\section{Introducci\'on}

Los modelos Higgs describen la autointeracci\'on del campo de
Higgs y su acoplamiento con los campos gauge SU(2)$_L$
y U(1)$_Y$. 

En la aproximaci\'on en la que se desprecian los acoplos de Yukawa 
entre el campo Higgs y los fermiones, 
estos modelos est\'an considerados como una razonable descripci\'on
del sector electrod\'ebil del Modelo Est\'andar. En particular
se persigue mediante su formulaci\'on en la red dar cuenta
de los aspectos no perturbativos de la teor\'{\i}a electrod\'ebil,
como por ejemplo la generaci\'on de masa mediante el mecanismo de
ruptura espont\'anea de la simetr\'{\i}a.

Los acoplamientos gauge de U(1)$_Y$ y SU(2)$_L$ se denotan respectivamente
por $g^{\prime}/2$ y $g$. La relaci\'on entre ellos es
a trav\'es del \'angulo de Weinberg, concretamente la relaci\'on es 
$g^{\prime}/g = 
\tan \theta_{\rm W}$, puesto que este \'angulo es peque\~no 
($\sin^2 \theta_{\rm W} \approx 0.23$),
podemos despreciar en un primer paso el grado de libertad
U(1)$_Y$. En este sentido SU(2)-Higgs puede considerarse como
el modelo Higgs m\'as simple.

El inter\'es del estudio de estos modelos no es tanto la necesidad
de determinar par\'ametros medibles en la naturaleza, sino m\'as
bien profundizar en el entendimiento te\'orico de la formulaci\'on de la
teor\'{\i}a electrod\'ebil. A diferencia de la QCD, esta
teor\'{\i}a no es asint\'oticamente libre, el punto fijo gaussiano
a acoplamiento cero es inestable ultravioleta, y la existencia de un punto 
fijo no trivial se hace necesaria para la formulaci\'on de la teor\'{\i}a 
en el continuo.

La acci\'on del modelo SU(2)-Higgs consiste en una parte pura gauge
m\'as un t\'ermino de interacci\'on escalar-gauge:

\be
S[U,{\bf {\Phi}}]_{\lambda} = S_{\rm g} [U] + S_{{\bf {\Phi}}} [U,{\bf {\Phi}}] \ .
\ee

La interacci\'on entre la parte escalar y la gauge 
est\'a controlada por el {\sl hopping parameter}, $\kappa$.
Una normalizaci\'on adecuada cuando se est\'a interesado en trabajar
en el l\'{\i}mite de autoacoplamiento cu\'artico infinito 
(regimen no perturbativo) es la siguiente:
\be
\begin{array}{lll}
S_{{\bf {\Phi}}} [U,{\bf {\Phi}}]_{\lambda}  & = &\displaystyle \sum_{x} {\bf {\Phi}}^{\dagger}(x) {\bf {\Phi}}(x) - \frac{1}{2} \kappa \sum_{x, \mu} 
\tr {\bf {\Phi}}^{\dagger}(x) U_{\mu}(x) {\bf {\Phi}}(x + \mu)  \\ 
& \displaystyle  + & \lambda \sum_{x} [{\bf {\Phi}}^{\dagger}(x) 
{\bf {\Phi}}(x) - 1]^2 \ ,
\end{array}
\ee

con esta normalizaci\'on
el l\'{\i}mite $\lambda \rightarrow \infty$ equivale a fijar
el modulo del campo Higgs a la unidad.

La parte pura gauge, $S_{\rm g}[U]$, se expresa como la suma sobre
las plaquetas orientadas positivamente:

\be
S_{\rm g} = \beta \sum_{P} (1 - \frac{1}{2} \tr U_{\rm P}) \ ,
\ee
 
donde $\beta = 4/g^2$. En el l\'{\i}mite $\beta \rightarrow \infty$ 
el campo gauge se desacopla y se recupera el modelo ${\bf {\Phi}}^4$, el cual
presenta una transici\'on de fase
entre una regi\'on sim\'etrica (peque\~no $\kappa$) y una regi\'on 
en la que la simetr\'{\i}a se ha roto espontaneamente ($\kappa$ grande)
\cite{ITZYKSON2}.

\begin{figure}[!t]
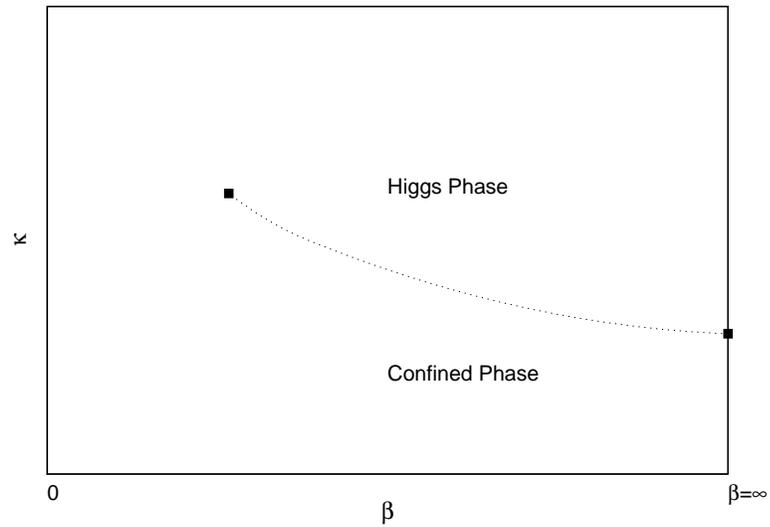

\centerline{
\fpsxsize=8cm
\def\fpsangle{270}
\fpsbox[70 90 579 760]{figure_1.ps} }
\caption{\small {Gr\'afico esquem\'atico del
diagrama de fases del modelo SU(2)-Higgs con modulo fijo 
($\lambda = \infty$)}. } 
\label{SU2}
\end{figure}

Cuando $\beta$ es finita ($g \neq 0$) esta transici\'on de fase se prolonga
en una l\'{\i}nea de transiciones de fase en el interior del espacio 
de acoplos \cite{DROU2,FRAD2,CREU2}, como se muestra en la figura \ref{SU2}. 

La continuaci\'on de la fase sim\'etrica de ${\bf {\Phi}}^4$ cuando
el acoplo gauge es distinto de cero es una
regi\'on {\sl confinante} en el sentido de que los campos gauge SU(2)
confinan las part\'{\i}culas escalares de materia. En este sentido
la fase sim\'etrica de SU(2)-Higgs es an\'aloga a QCD.

La continuaci\'on de la fase de simetr\'{\i}a rota
cuando el coupling gauge es distinto de cero es la fase {\sl Higgs}.
En esta fase,
mediante el mecanismo de ruptura espont\'anea de la simetr\'{\i}a,
el campo de Higgs adquiere un valor esperado en el vac\'{\i}o
distinto de cero. Los bosones vectoriales
adquieren su masa a trav\'es de su acoplamiento con el campo
escalar. Los tres bosones de Goldstone del modelo
${\bf {\Phi}}^4$ se hacen masivos al mezclarse con los bosones gauge
vectoriales. El resultado es un isovector con spin 1, el boson $W$.

Esta l\'{\i}nea de transiciones de fase termina en un valor finito
del espacio de acoplamientos, estando
las fases {\sl Higgs} y {\sl confinante} conectadas analiticamente.

En cuanto al orden de la transici\'on de fase, para peque\~nos acoplamientos
existen argumentos perturbativos
debidos a Coleman y Weinberg \cite{COLE-WEI2} apuntando que
la transici\'on deber\'{\i}a ser de primer orden.
El estudio num\'erico de la transici\'on de fase 
para peque\~nos e intermedios valores de $\lambda$ parece confirmar 
esta conjetura \cite{LANG22,MONT2,XU2}.

Sin embargo se observa que la
transici\'on se debilita (las discontinuidades son m\'as peque\~nas)
conforme $\lambda$ aumenta. En el l\'{\i}mite
$\lambda = \infty$ la opini\'on generalizada es que la transici\'on
sigue siendo de primer orden aunque extremadamente d\'ebil, 
si bien no existen evidencias que justifiquen esta creencia.
Los \'ultimos resultados num\'ericos fueron obtenidos
por Montvay en 1985, en redes no demasiado grandes y a pesar
de observar metaestabili\-dades en los ciclos de hist\'eresis no se
pudieron obtener resultados concluyentes \cite{LM2}.

Tratando de arrojar algo de luz sobre este punto hemos enfocado el problema
desde un punto de vista distinto, formulando el modelo SU(2)-Higgs
en un espacio de parametros extendido. Sin embargo la motivaci\'on
no es s\'olo hacer un estudio con alta estad\'{\i}stica del modelo,
sino tratar de extraer las propiedades generales que presentan
las transiciones de primer orden d\'ebiles en la red en modelos
gauge acoplados a campos escalares.
Con este prop\'osito se ha a\~nadido a la acci\'on est\'andard un acoplo extra
entre el campo gauge y el campo escalar conectando segundos vecinos
en la red. Este acoplamiento nuevo ser\'a usado como un par\'ametro
para reforzar o debilitar la transici\'on, permitiendo
as\'{\i} estudiar su mecanismo de debilitamiento cuando este 
acoplamiento se hace tender a cero.

\newpage

\begin{center}
{\bf Abstract}
\end{center}

{\sl

The properties of the Confinement-Higgs phase transition in the 
SU(2)-Higgs model with fixed modulus are investigated. 
We show that the system exhibits a transient behavior up to {\it L}=24 
along which,  
the order of the phase transition cannot be discerned.
To get stronger conclusions about this point, without going to
prohibitive large lattice sizes, we have introduced a second 
(next-to-nearest neighbors) gauge-Higgs
coupling ($\kappa_2$). On this extended parameter
space we find a line of phase transitions which become
increasely weaker as $\kappa_2 \rightarrow 0$.
The results point to a first order character
for the transition with the standard action ($\kappa_2 = 0$). 
}

\section{The Model}

The SU(2) lattice gauge model coupled to an scalar field, in the
fundamental representation of the gauge field can be described by the
action

\ba
S_{\lambda} & = \displaystyle \beta \sum_{p} [1 - 
\frac{1}{2} \tr U_p] - \frac{1}{2} \kappa_1 \sum_{x, \mu} 
\tr {\bf {\Phi}}^{\dagger}(x) U_{\mu}(x) {\bf {\Phi}}(x + \mu) + \nonumber \\ 
& \displaystyle  + \lambda \sum_{x} [{\bf {\Phi}}^{\dagger}(x) {\bf {\Phi}}(x) - 1]^2 + \sum_{x} {\bf {\Phi}}^{\dagger}(x) {\bf {\Phi}}(x) \ .
\ea

Where U$_{\mu}(x)$ represents the link variables, and $U_p$ are their products
along all the positive oriented plaquettes of a four-dimensional
lattice of side {\it L}.

The scalar field at the site $x$ is denoted by ${\bf {\Phi}}(x)$, being
$\lambda$ the parameter controlling its radial mode. 
In the limit ($\lambda = \infty$, $\beta = \infty$) the model becomes a pure
O(4)-symmetric scalar model.

As we pointed out in the previous section, the PT line ends at some
finite value of the parameters ($\beta$, $\kappa_1)$
The endpoint moves towards larger $\beta$ values as $\lambda$ increases.
For $\lambda \geq 0.1$ the endpoint crosses to the $\beta > 0$ region,
and in the limit $\lambda = \infty$ the phase transition 
ends at ($\kappa_1 \approx 0.6$, $\beta \approx 1.6$) \cite{LANG22}.
It is commonly believed that the transition at this point is second order with
classical critical exponents \cite{ROB2}, however a careful numerical
study would be necessary.

In the scaling region, and for finite $\lambda$, the phase transition 
turns out to be first order. Also,
the transition becomes weaker as $\beta$ or $\lambda$ increases. In
particular in the limit $\beta = \infty$ (spin model)
the transition is second order with classical critical exponents
\cite{ITZYKSON2}.

In the limit $\lambda = \infty$ the situation is less transparent,
and it is not clear whether the transition is weak first order or higher
order.

The study of the model with the parameter $\lambda = \infty$ is equivalent
to fixing the modulus of the Higgs field, ${\bf {\Phi}}^{\dagger} \cdot {\bf {\Phi}} = 1$. 
In the pioneer work, \cite{LRV2}, the PT was considered first order for finite
$\lambda$ and second order for $\lambda = \infty$. Later,
larger statistics, and the hypothesis of a universal behavior of
the PT for all values of $\lambda$, seem to point to a first order 
character of the PT in the scaling region. Nowadays,
though it is generally believed that the transition is still first
order in this limit, the numerical proofs \cite{LM2,MONT2} 
on which rely these statements are not conclusive, as the
authors safely conclude, because the statistic and lattice sizes
are not enough for excluding the possibility of a higher 
order phase transition in the edge $\lambda = \infty$.

The model with fixed modulus is described by the action
\be
S_{\infty} = \beta \sum_{p} [1 - \frac{1}{2} \tr U_p]
- \frac{1}{2}\kappa_1 \sum_{x, \mu} \tr {\bf {\Phi}}^{\dagger}(x) U_{\mu}(x) {\bf {\Phi}}(x + \mu)  \ .
\ee

As we shall show below, 
we have simulated this model up to lattices {\it L}=24, and the result
is still compatible with a second or higher order PT, however,
we have no indications of asymptoticity
in the behavior of the observables, and the results are compatible with
a very weak first order PT too. This means that larger lattices
are needed to overcome these transient effects,  
but the added difficulty here is that such lattices would suffer
of termalization problems, and severe autocorrelation times.
Altogether makes this approach too CPU expensive for nowadays computers.

In order to get a more conclusive answer without going to prohibitive
large lattice sizes, we have studied the model in an extended parameter space. 
For this purpose
we have introduced a second coupling between the gauge and the scalar
field connecting next-to-nearest neighbors on the lattice,
in such a way that the new action reads
\be
S = S_{\infty}
- \frac{1}{4} \kappa_2 \sum_{x, \mu<\nu}
Tr {\bf {\Phi}}^{\dagger}(x) [U_{\mu}(x) U_{\nu}(x+\mu) +
U_{\nu}(x) U_{\mu}(x+\nu)] {\bf {\Phi}}(x + \mu + \nu)     
\label{ACTION}
\ee

Within this parameter space we expect to get a global vision on what
the properties of the PT are, and also, to give a stronger conclusion 
about the order. In the region of
$\kappa_2$ positive, (competing interaction effects appear if $\kappa_2 < 0$) 
this extended model is expected to belong
to the same universality class than the {\it standard} one ($\kappa_2$ = 0) 
since both models posses the same symmetries.
The effect of the new coupling $\kappa_2 > 0$ is to reinforce the
transition but should not change the order if the system has not
tricritical points. An example of such behavior appears in the 
O(4)-symmetric $\sigma$ model with second neighbors coupling. This model 
presents a ($\kappa_1^c,\kappa_2^c$) line of phase
transitions which is second order, since the model with $\kappa_2=0$
shows a second order PT too \cite{YO2}. 

The action (\ref{ACTION}) has the following symmetries:
\begin{itemize}
\item{} $\kappa_1 = \kappa_2 = 0$.
\ba
\beta  & \rightarrow & -\beta \nonumber \\
U_{\mu}(x) & \displaystyle \rightarrow & (-1)^{\sum_{\rho \neq \mu} x_{\rho}} U_{\mu}(x) 
\ea
\item{} $\kappa_1 \rightarrow -\kappa_1$, ${\bf {\Phi}}(x)$ fixed
\be
U_{\mu}(x) \rightarrow - U_{\mu}(x) 
\ee
\item{} $\kappa_1 \rightarrow -\kappa_1$, $U_{\mu}(x)$ fixed
\be
\displaystyle {\bf {\Phi}}(x) \rightarrow \displaystyle (-1)^{\sum_{\mu=0}^{d-1} x_{\mu}} {\bf {\Phi}}(x) 
\label{STAG}
\ee
\end{itemize}

The action is not symmetric under the
change $\kappa_2 \rightarrow -\kappa_2$. The existence of couplings
$\kappa_1$ and $\kappa_2$  
with opposite signs would make frustration to appear, and
very different vacua are possible \cite{YO2}. 
In this work we are interested in the regions free of frustration
effects. Taking into account the symmetry properties of the action,
the phase diagram in the region $\kappa_2 >$ 0 will be symmetric
with respect to the axis $\kappa_1$ = 0, and hence, we can restrict
the study to the quadrant $\kappa_1 >$ 0.

We define the normalized energy associated to the plaquette term
\be
E_0 = \frac{1}{N_{l_0}} \sum_{p} (1 - \frac{1}{2}\tr U_p) \ ,
\ee 
and also the energies associated to the links
\be
E_1  = \frac{1}{N_{l_1}} \sum_{x,\mu} \frac{1}{2} \tr {\bf {\Phi}}^{\dagger}(x) U_{\mu}(x) {\bf {\Phi}}(x + \mu)  \ ,
\ee
\be
E_2  =  \frac{1}{N_{l_2}} \sum_{x, \mu<\nu} \frac{1}{4}
\tr {\bf {\Phi}}^{\dagger}(x) [U_{\mu}(x) U_{\nu}(x+\mu) +
U_{\nu}(x) U_{\mu}(x+\nu)] {\bf {\Phi}}(x + \mu + \nu)   
\ee

where $N_{l_0}$ = 6V, $N_{l_2}$ = 12V and $N_{l_1}$ = 4V.

With these definitions $E_0 \rightarrow 0$ when $\beta \rightarrow \infty$
and $E_i \rightarrow 1$ when $\kappa_i \rightarrow \infty$.

On the three-dimensional ($\beta$, $\kappa_1$, $\kappa_2$)
parameter space we consider the plane $\beta=2.3$. 
On this plane there is a PT line ($\kappa_1^c, \kappa_2^c$).
We expect to learn on the properties of this PT on the region
($\kappa_1^c \neq 0, \kappa_2^c \neq 0$), where the signals are
clearer, with regards to applying what we learn to the $\emph standard$
case $\kappa_2=0$. 

To monitorize the strength of the phase transition, we measure
the existence of latent heat, and the behavior of the specific heat.
As we shall see, 
for $\kappa_1 = 0$ the transition is first order, with a
clearly measurable latent heat. We will see how this transition weakens
along the PT line for increasing values of $\kappa_1$.

\begin{figure}[!t]
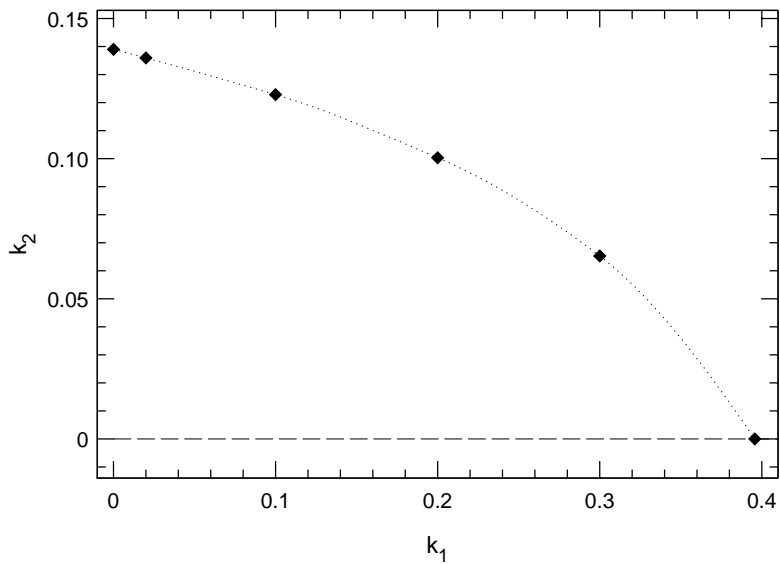

\centerline{
\fpsxsize=8cm
\def\fpsangle{270}
\fpsbox[70 90 579 760]{figure_2.ps} }
\caption{\small {Phase diagram obtained from the MC simulation.} } 
\label{PHASESU2}
\end{figure}

\section{Numerical study}

\begin{figure}[!t]
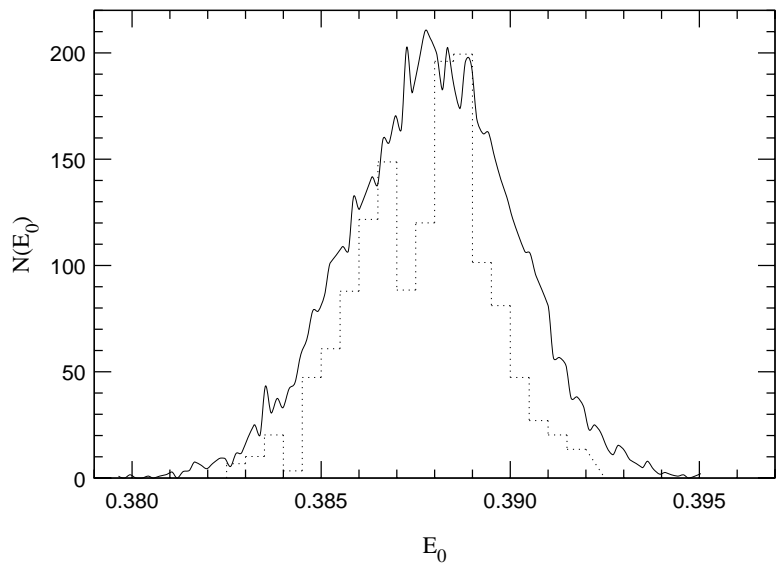

\centerline{
\fpsxsize=8cm
\def\fpsangle{270}
\fpsbox[70 90 579 760]{figure_3.ps} }
\caption{\small {Normalized energy distribution for L=12, 
$\lambda=\infty$, $\kappa_1$=0.395, $\beta$=2.3 from \protect\cite{LM2} 
(dotted line), compared with the distribution we obtain at the same couplings 
(solid line) in $L$=12 too, when statistics is increased by 
one order of magnitude.}}
\label{LMf}
\end{figure}

We have simulated the model in a $L^4$ lattice with periodic boundary
conditions. For the update we have employed a combination of
heat-bath and over-relaxation algorithms (ten over-relax sweeps
followed by a heat-bath sweep). For the simulation we used the RTNN
machine, consisting of a network of 32 PentiumPro 200MHz. processor.
The total CPU time employed has been the equivalent of 3 years of PentiumPro.

Monte Carlo methods provide information about the thermodynamic
quantities at a particular value of the couplings. We have used
the Spectral Density Method (SDM) \cite{FERRENBERG2} to extract information
on the values of the observables in a finite region around the
simulation point. In particular it is useful to have a precise
location of the coupling where some observables have a maximum,
as well as an accurate measure of the value of that maximum.

From the Monte Carlo simulation at some coupling $\kappa$, 
we got the histogram $H(E)$ which is an 
approximation to the density of states. Using the SDM approximation
the probability of finding the system with
an energy E at a different coupling $\kappa^{\prime}$ can
be written as:
\be
P_{\kappa^{\prime}}(E) \propto 
H(E) e^{(\kappa^{\prime} - \kappa) V E} \ .
\label{SD}
\ee

The region of validity of the SDM approximation
is $\Delta \kappa \sim 1/(V\sigma)$, being $\sigma$ the width
of the distribution $H(E)$. 
Although $\sigma$ gets the maximum values in the critical region, 
the approximation has been very useful, specially for tuning the
couplings where to measure.

Concerning the lattice sizes, we have used lattices ranging 
from $L$=6 to $L$=24. For the small lattices ($L$=6, 8 and 12) we have 
done 4$\times$10$^5\tau$ iterations, being
$\tau$ the largest autocorrelation time for the energy, 
which ranges from $\tau \approx 10$
in $L$=8 to $\tau \approx 35$ in $L$=24.
For the largest lattices, $L$=20 and 24, we run up to 10$^5\tau$
MC iterations.

The statistical errors are computed with the jackknife method.

\section{Results}

We shall make the discussion with the first-neighbors link
energy, E$_1$, but as far as the critical behavior is concerned, 
we could carry out the analysis with any of the energies. We remark
that an appropriate linear combination of E$_1$, E$_2$ and E$_0$
could give slightly more accurate results \cite{U1H2}.

We have considered fixed values of $\kappa_1$ (0, 0.02,
0.1, 0.2 and 0.3) and sought the $\kappa_2$ critical for
every line, $\kappa_2^c (\kappa_1)$.
We have also studied the case $\kappa_2$= 0 varying $\kappa_1$
which corresponds to the usual SU(2)-Higgs model.

The SDM has been used to locate the apparent critical point,
defined through the specific heat behavior. From the specific
heat matrix:
\be
C_v^{i,j}(L) = \frac{\partial E_i}{\partial \kappa_j} \ ,
\label{MAT}
\ee
we obtain the best signal for C$_v^{1,2}(L)$, which can
be calculated as (we shall omit the superscript from now on)
\be
C_v(L) = 4 L^d (\langle E_1 E_2 \rangle - \langle E_1 \rangle \langle E_2 \rangle)  \ .
\ee

In a first order phase transition, C$_v^{\rm {max}} (L)$ behaves, 
asymptotically, proportional to the volume,  $L^d$.
If the PT is second order, the dominant behavior for
C$_v(L)^{\rm{max}}$ 
is $L^{\alpha /\nu}$ which diverges too provided that $\alpha >$ 0.
At the upper critical dimension $\alpha$ = 0,  and one has to go further
the leading order, appearing logarithmic divergences \cite{LOG2}.

As a consequence of this divergent behavior, in a finite lattice C$_v(L)$ 
shows a peak at some value of the
coupling which will be taken as apparent critical point,  
$\kappa_2^{\ast}(L)$.

In Figure \ref{PHASESU2} we plot the critical line ($\kappa_1^c, \kappa_2^c$).
This line is obtained by extrapolation to the thermodynamic 
limit according to $\kappa_2^c (\infty) = \kappa_2^{\ast} (L) - A L^{-d}$.
We point out that this extrapolation is valid only in first order
phase transitions. If the PT is second order the power in $L$
is ($-1/\nu$).

In continuous PT scale invariance holds, and the
thermodynamic magnitudes, such as the specific heat, or susceptibilities
do scale.
However if the transition is first order the correlation
length remains finite and hence there is not scaling properties,
and no critical exponents can be defined.

Nevertheless in a first order PT we can ask how large is the lattice
we need in order to observe the asymptotic behavior of C$_v$. In particular, 
with abuse of language one can measure a $\emph pseudo$ $\alpha/\nu$
exponent to get insight on the nature of the PT: the larger is the lattice
we need to measure $\nu = 1/d$, the weaker the PT is. Following this,
first order PT can be classified according to their degree of weakness.

\begin{figure}[!t]
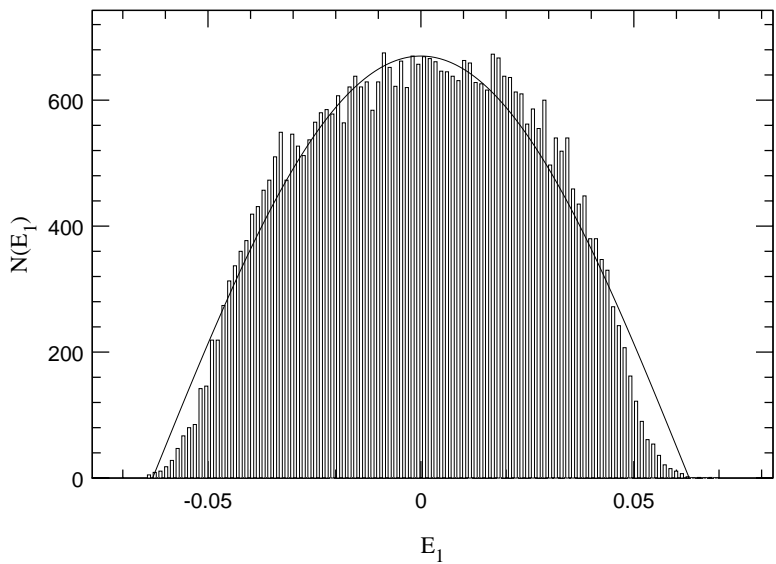

\centerline{
\fpsxsize=8cm
\def\fpsangle{270}
\fpsbox[70 90 579 760]{figure_4.ps} }
\caption{\small {E$_1$ distribution on the axis $\kappa_1 = 0$ 
on a L=8 lattice
at ($\kappa_2 = 0.15, \beta = 2.3$). The cosine fit is very accurate
in spite of the finite $\beta$ value, in this region of parameters
the pure gauge term couples slightly both sub-lattices.} }
\label{EJE}
\end{figure}

\begin{figure}[!t]
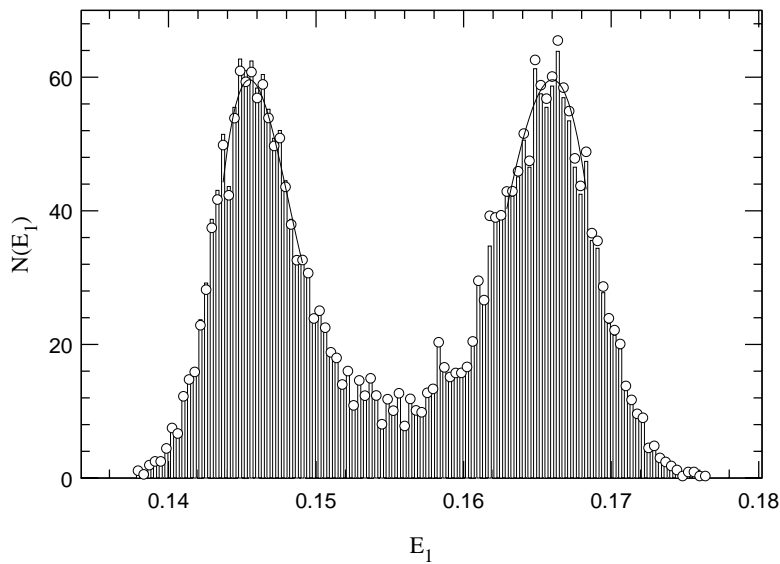

\centerline{
\fpsxsize=8cm
\def\fpsangle{270}
\fpsbox[70 90 579 760]{figure_5.ps} }
\caption{\small {Normalized distribution of E$_1$ 
at ($\kappa_1=0.2$, $\kappa_2=0.10036$) in
a L=16 lattice. The cubic fit at the maxima to get
$\Delta E$ is superimposed.}}
\label{FIT}
\end{figure}

\begin{figure}[!t]
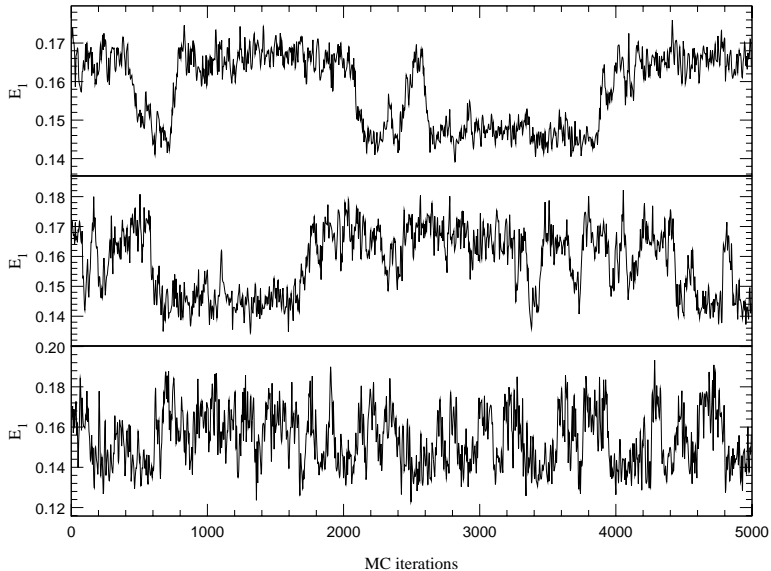

\centerline{
\fpsxsize=8cm
\def\fpsangle{270}
\fpsbox[70 90 579 760]{figure_6.ps} }
\caption{\small {MC evolution of E$_1$ at ($\kappa_1=0.2, \kappa_2^{\ast}(L)$)
for $L$=8 (lower part), $L$=12 (middle) and $L$=16 (upper part).} }
\label{MET}
\end{figure}   
      
\begin{figure}[!t]
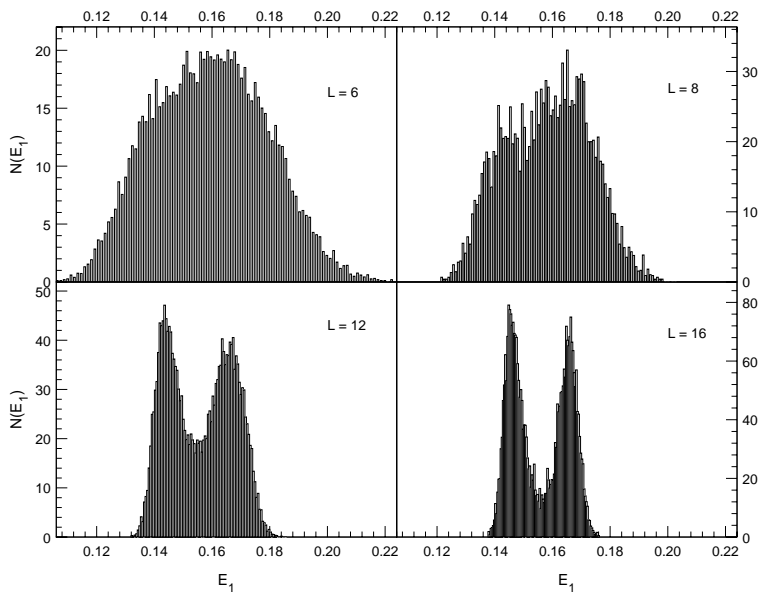

\centerline{
\fpsxsize=8cm
\def\fpsangle{270}
\fpsbox[70 90 579 760]{figure_7.ps} }
\caption{\small {Normalized distribution of E$_1$ at 
($\kappa_1=0.2, \kappa_2^{\ast}(L)$)
for L=6, 8, 12 and 16.}} 
\label{HK02}
\end{figure}

However, the so called
weak first-order PT appear often in literature (see \cite{JUAN2,LAFPOTS2}
and references there in) as
PT characterized by a transient behavior with a non-measurable latent
heat. Let be $\xi_c$  the correlation length of the system
at the critical point in the thermodynamic limit. In a finite
lattice of size $L$, the first order behavior will be evidenced
if $L \geq \xi_c$. For lattice sizes much smaller than $\xi_c$ the system
will behave like in a second-order PT, since the correlation
length is effectively infinite. 
As an example, in Figure \ref{LMf} we plot the energy distribution
in a L=12 lattice at ($\beta=2.3, \kappa_1=0.395$) obtained in \cite{LM2},
compared with the one we obtain at the same couplings and in the same L,
when the statistics increases by one order of magnitude. 
We observe that the order of the PT can not be discerned at this volume, 
even when the statistics is enough. Termalization effects can also
contribute to mistake the histogram structure.

The entire line ($\kappa_1^c, \kappa_2^c$) is first order, but the weak
character increases as $\kappa_2 \rightarrow 0$. We will make a quantitative
description of the weakening phenomenon
by studying the specific heat, and the latent heat.

But before going on, we shall make a remark concerning the behavior
on $\kappa_1 = 0$.
As we pointed out, the system is symmetric under the change $\kappa_1
\rightarrow -\kappa_1$. The transformation (\ref{STAG}) maps
the positive $\kappa_1$ semi-plane with energy E$_1$, onto the
negative $\kappa_1$ semi-plane with energy -E$_1$. The transition
across this axis is first order because the energy is discontinuous. 
In the limit $\beta \rightarrow \infty$, and in $\kappa_1 = 0$, the system
decouples in two independent sublattices, each one constituted by the first 
neighbors of the other. The first neighbors energy for this system 
is proportional
to $\cos \theta$, being $\theta$ the angle between the symmetry breaking
direction of the scalar field in both sub-lattices. In Figure \ref{EJE} we plot
the E$_1$ distribution for L=8 at $(\kappa_1 =0,\kappa_2 = 0.15)$ and
$\beta = 2.3$. We see that the agreement with a cosine distribution
is quite good in spite of the finite $\beta$ value.

\begin{figure}[!t]
\centerline{
\fpsxsize=8cm
\def\fpsangle{270}
\fpsbox[70 90 579 760]{figure_8.ps} }
\caption{\small {Normalized distributions of E$_1$ for 
$\kappa_1$ = 0.02, 0.1, 0.2 and 0.3 
at $\kappa_2^{\ast}(12)$.}} 
\label{H12}
\end{figure}

\begin{figure}[!t]
\centerline{
\fpsxsize=8cm
\def\fpsangle{90}
\fpsbox[70 90 579 760]{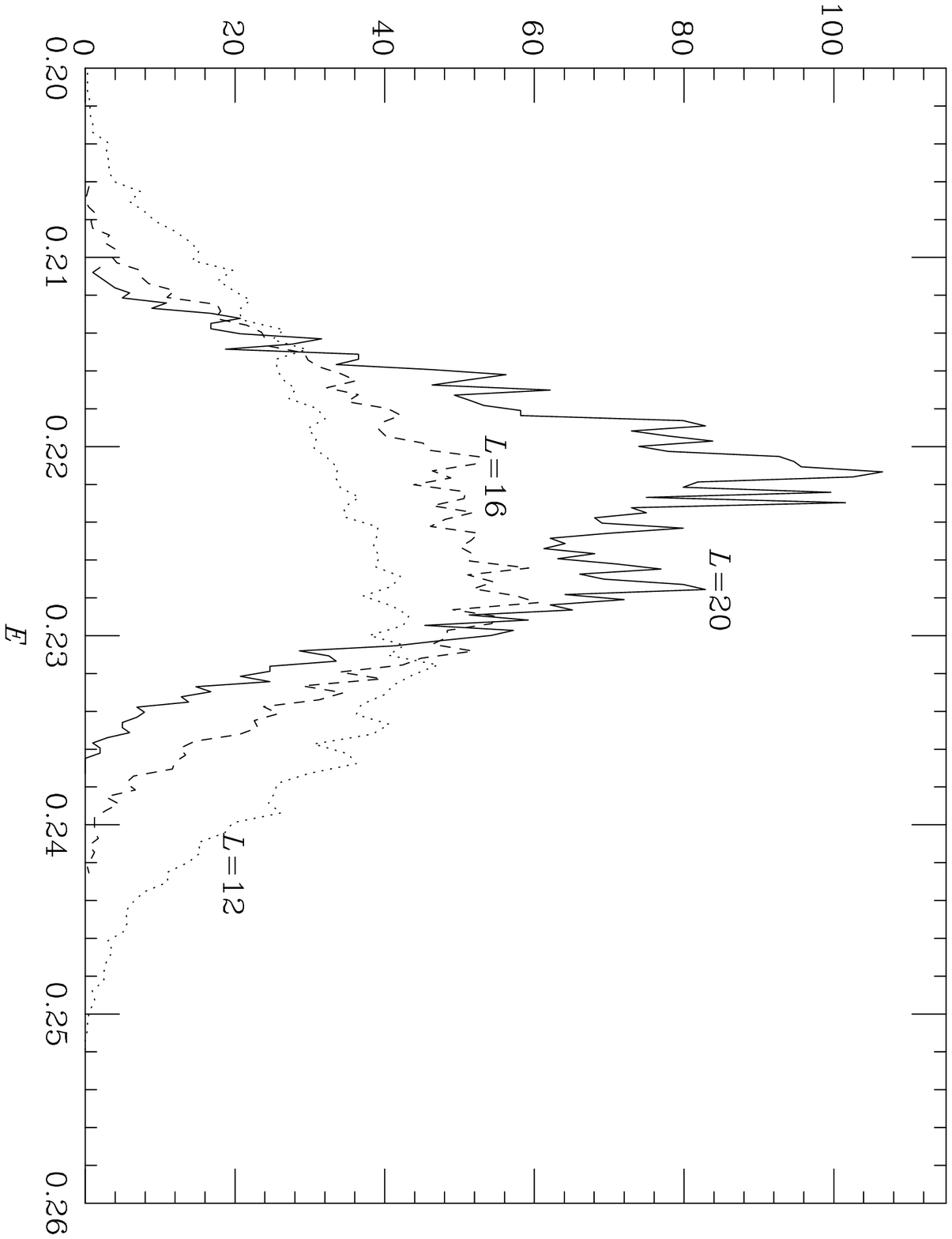} }
\caption{\small {Normalized distribution of 
E$_1$ at $\kappa_1$ = 0.3 for L=12,16 and 20}} 
\label{1620}
\end{figure}

\subsection{Latent Heat}

Along the apparent critical line we have done simulations for different
lattice sizes and stored the plaquette and links energies to construct the 
histograms for the energy distributions.
In a first-order phase transition the energy has a discontinuity 
which manifest in the appearance of latent heat, $\Delta E$. 
This quantity is not well
defined in a finite lattice, so we measure the distance between the
two maximum of the energy distribution, and extrapolate to
the thermodynamic limit. The drawback of this approximation is that
the maxima of the energy distribution are difficult to discern, since
this function at the apparent critical point is very noisy 
We have used a cubic spline at the maxima in order to get a 
more reliable estimation (Figure 5).

In Figure \ref{MET} we show the MC evolution
of E$_1$ for $L$=8, 12 and 16 at $\kappa_1 = 0.2$. 
In $L$=8 the latent heat is not clearly measurable. We observe in
the MC evolution how the two-state signal becomes cleaner as
the lattice size increases (see Figure \ref{HK02}).

In Figure \ref{HK02} we plot the distribution of E$_1$  
at $\kappa_1 = 0.2$ at the apparent
critical point $\kappa_2^{\ast} (L)$ for $L$=6, 8, 12 and 16.
A remarkable stability of $\Delta E_1$ with the volume is observed. This
is a common feature  for all values of $\kappa_1$.

As we have already pointed out, the transition weakens when increasing
$\kappa_1$ and larger lattices are needed in order to observe a
measurable latent heat. We give a quantitative
description of this fact in Figure \ref{H12}, where
the distribution of $E_1$ for several values 
of ($\kappa_1,\kappa_2^{\ast}(12)$) is displayed. 
The two-state signal is no longer measurable in $L$=12 at $\kappa_1 = 0.3$. 
The first evidences of two-states appear in $L$=20 (see Figure \ref{1620}).
but from its energy distribution we can only give an approximate value
for $\Delta E_1(L=20)$ since the two-peaks appear too close to each other.

In Figure \ref{DELTA} we plot $\Delta E_1(L)$ and $\Delta E_2(L)$  
as a function of 1/L$^4$, in order to get $\Delta E_i$ in the
thermodynamic limit with a linear fit.
Finally we quote these values in Table \ref{TAB}, together 
with the change in the action (\ref{ACTION}) between the two
phases.

\begin{table}[t]
{

\begin{center}
{
\begin{tabular}{|c|c|c|c|}\hline
coupling  &$\Delta E_1 (\infty)$ &$\Delta E_2 (\infty)$  &$\Delta S$ \\ \hline
$\kappa_1$ = 0             & -          &0.0366(8)  &0.0134(12) \\ \hline
$\kappa_1$ = 0.02          &0.0162(6)   &0.0347(5)  &0.0137(13) \\ \hline
$\kappa_1$ = 0.1           &0.0162(7)   &0.0345(9)  &0.0094(10) \\ \hline
$\kappa_1$ = 0.2           &0.0179(7)   &0.0201(8)  &0.0078(12)  \\ \hline
$\kappa_1$ = 0.3 & $\approx$0.006 &$\approx$ 0.012 &$\approx$ 0.0026 \\ \hline   
\end{tabular}
}
\end{center}
}
\caption[a]{\small {$\Delta E(\infty)$ for E$_1$ and E$_2$, and variation
of the action.}}
\protect\label{TAB}

\end{table}

\begin{figure}[!t]
\centerline{
\fpsxsize=8cm
\def\fpsangle{270}
\fpsbox[70 90 579 760]{figure_10.ps} }
\caption{\small {$\Delta E_1$ (upper plane)
and $\Delta E_2$ (lower plane) as a function of 1/L$^4$. The two-peak
structure is not clearly observed in L=6 at any $\kappa_1$ value.
The values quoted for this lattice size are upper bounds.}} 
\label{DELTA}
\end{figure}

\begin{figure}[!t]
\centerline{
\fpsxsize=8cm
\def\fpsangle{90}
\fpsbox[70 90 579 760]{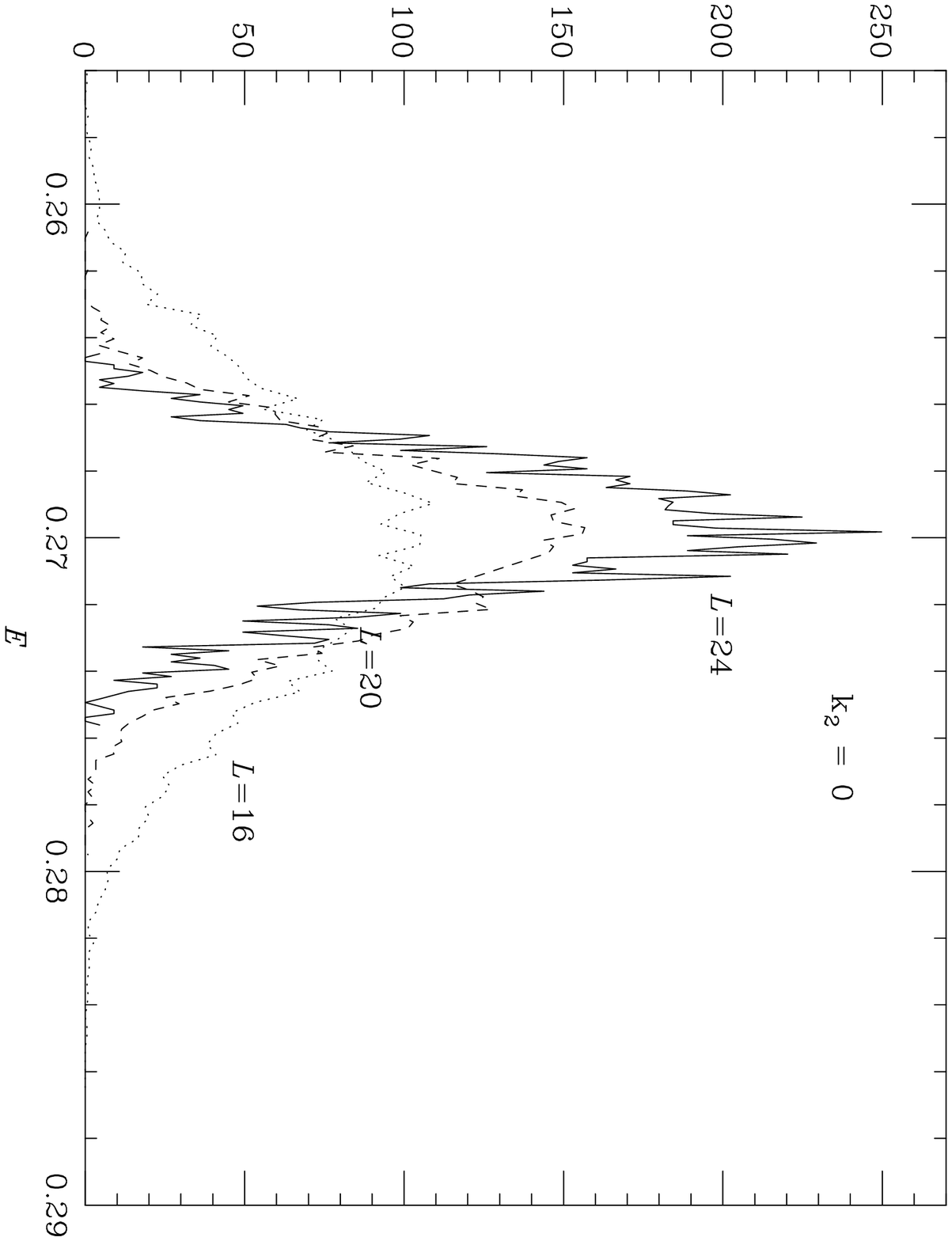} }
\caption{\small {Normalized distribution of E$_1$ at $\kappa_2$=0,
in $L$=16, 20 and 24 at the peak of the specific heat} }
\label{2024}
\end{figure}

From the energy distributions at $\kappa_2 = 0$, see Figure \ref{2024}
we have no direct
evidences of the existence of latent heat. However, on the larger
lattices one can observe non-gaussianities in the energy distributions. 
Such asymmetries  could precede 
the onset of clear two-peak structures in larger lattices, however
this is just a guess. We conclude that no information concerning the
order of the PT can be obtained from the energy distributions
up to $L$=24.

\subsection{Specific Heat}

We have done MC simulations at the points predicted by SDM, 
($\kappa_1, \kappa_2^{\ast}(L)$), in order to measure accurately
the peak of $C_v(L)$.

\begin{figure}[!t]
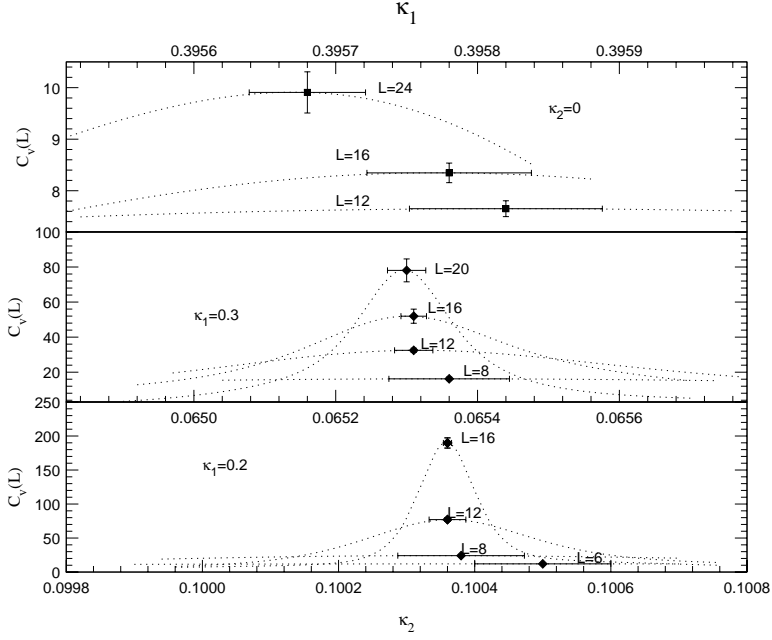

\centerline{
\fpsxsize=8cm
\def\fpsangle{270}
\fpsbox[70 90 579 760]{figure_12.ps} }
\caption{\small {C$_v^{\rm {max}} (L)$ at $\kappa_1=0.2$ (lower plane),  at 
$\kappa_1=0.3$ (middle) and at $\kappa_2=0$ (upper plane).
The dotted line is the SDM extrapolation. } }
\label{PEAKS}
\end{figure}

As an example we show in Figure \ref{PEAKS} the value of $C_v(L)$ 
around its maximum for various lattice sizes at
$\kappa_1$ = 0.2 (upper plane) and at $\kappa_1$ = 0.3 (lower plane).
We observe that the maximum of the specific heat grows slower when
increasing $\kappa_1$, indicating a weakening in the PT.

\begin{figure}[!t]
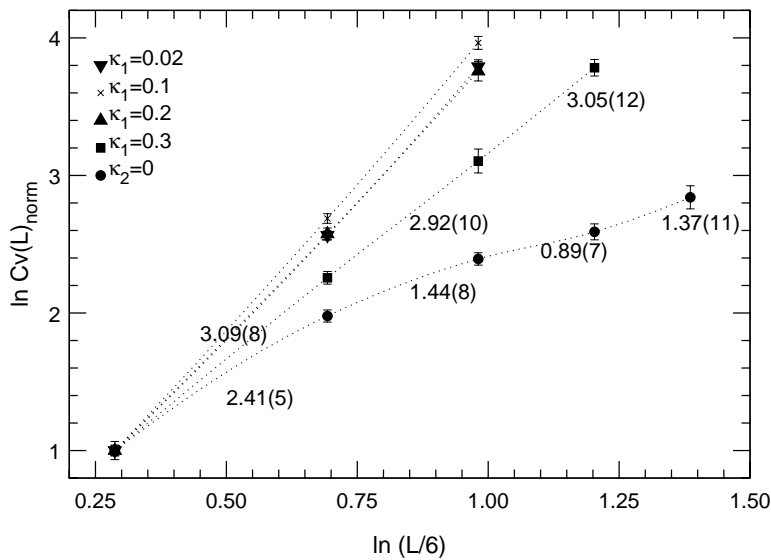

\centerline{
\fpsxsize=8cm
\def\fpsangle{270}
\fpsbox[70 90 579 760]{figure_13.ps} }
\caption{\small {C$_v^{\rm {max}}$ for the 
various $\kappa_1$ values and $\kappa_2=0$.
We have normalized the values with respect to 
C$_v^{\rm {max}}$(8)/C$_v^{\rm {max}}$(6). 
The slope of the segments is indicated when smaller than 4.}} 
\label{CV1N}
\end{figure}

In Figure \ref{CV1N} we draw C$_v^{\rm {max}} (L)$ relative to 
C$_v^{\rm {max}}(6)$ as a function of the lattice size. 
The values have been normalized to 
C$_v^{\rm {max}}(8)$/C$_v^{\rm {max}}(6)$
in order to compare distinctly the behaviors for different $\kappa_1$.
The slope of the segment joining
the values of C$_v^{\rm {max}}$ in consecutive
lattices gives the $pseudo$ $\alpha/\nu$ exponent.
We observe that such slope is approximately 4 at $\kappa_1$ = 0.02, 
0.1 and 0.2 for all the volumes we compare.  However the transition
at $\kappa_1$ = 0.3 evidences much more weakness. We do not
have evidences of asymptoticity in C$_v^{\rm {max}}$ till L=20, as could
be expected from the energy distributions (Figure \ref{1620}).
The slope of the segment joining C$_v^{\rm {max}}(16)$ with 
C$_v^{\rm {max}}(20)$
is 3.05(12) which is almost the asymptotic value expected for
a first order PT.

As expected, only if the two-peak structure is
observed in the energy distributions and $\Delta E$ is stable,
the maximum of the specific heat will grow up like the volume, $L^d$.

At $\kappa_2$ = 0 we are within the transient region even for L=24. 
We remark that in this case C$_v$ is defined by the element C$^{1,1}$
of the specific heat matrix (\ref{MAT}) 
since this is the most natural choice
at this point, and also is the best signal we measure.
As we observe in Figure \ref{CV1N}, at $\kappa_2$ = 0 C$_v^{\rm {max}}(L)$
seems to tend to a constant value as V$\rightarrow \infty$ up to
$L$=20. From our previous discussions we should conclude that
either the correlation length at the transition point $\xi_c$
is much larger than the lattice size up to L=20, or the transition
is second order with $\alpha = 0$ in the thermodynamic limit. 
However, in $L$ = 24 things are changing, C$_v^{\rm max}(L=24)$
starts to run away of this quasi-plateau, and the $pseudo$ 
$\alpha/\nu$ exponent grows again.
As can be observed in Figure \ref{CV1N}, at $\kappa_1$ = 0.3 the $pseudo$ 
$\alpha/\nu$ exponent decreases for the segment $L$ = 12-16,
with respect to the value in the segment $L$ = 8-12. The lattice 
$L$ = 20 is enough to overcome the transient
region, but the behavior is qualitatively the same that in $\kappa_2$ = 0,
though the transition is stronger.

We believe that this behavior is general for weak first
order transitions in four dimensions. There exists a transient region in
which the correlation length is effectively infinite 
compared with the lattice size, and
the system behaves like suffering a second order PT with
thermal index $\alpha \approx 0$ in the thermodynamic limit.

\subsection{Binder Cumulant}

In order to check the consistency of our results 
we have also considered the behavior of the Binder cumulant

\be
V_L = 1 - \frac{\langle E_1^4 \rangle_L}{3\langle E_1^2 \rangle_L^2} \ .
\label{CUMUL}
\ee

This quantity behaves differently depending on the order
of the PT. If the transition is second order the minimum
of the cumulant,
$V_L^{\rm {min}}$ approach 2/3 in the thermodynamic limit.
However if the transition is first order, $V_L^{\rm {min}}$ 
tends a value smaller
than 2/3 reflecting the non-gaussianity of the energy
distribution at the transition point.

\begin{figure}[!t]
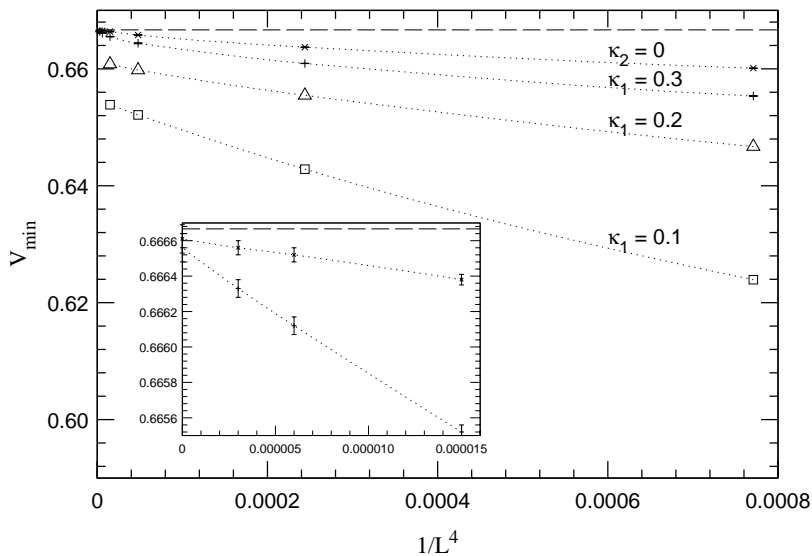

\centerline{
\fpsxsize=8cm
\def\fpsangle{270}
\fpsbox[70 90 579 760]{figure_14.ps} }
\caption{\small {$V_L^{\rm {min}}$ for $\kappa_1$ = 0.1, 0.2,0.3 
and $\kappa_2$ = 0,
as a function of 1/L$^4$. The dashed line represents the value 2/3.}} 
\label{CUM}
\end{figure}

In Figure \ref{CUM} we plot $V_L^{\rm {min}}$ for several
$\kappa_1$ values.

For those values of $\kappa_1$ in which the PT is distinctly
first order, the minimum of the Binder cumulant stays
safely away from 2/3, as
we observe $V_L^{\rm {min}}$ extrapolated to L$\rightarrow \infty$ is
0.65401(4) at $\kappa_1$ = 0.1. However, this value reachs
0.66637(5) at $\kappa_1$ = 0.3, and 0.66657(8) at $\kappa_2=0$.
Again we find a tight difference between a very weak first order PT
and a continuous one.

For the sake of discussing quantitatively the 
order of magnitude of the latent heat in the limit $\kappa_2=0$,
from the energy distributions we find that at $\kappa_2=0$
one can approximately locate one the peaks of the energy 
at E$_a \approx$ 0.273. The other should be at certain 
E$_b$ = E$_a - \Delta$, being $\Delta$ the latent heat.

In the thermodynamic limit the energy distributions are
two delta functions situated at E$_a$ and E$_b$, then

\be
V_{\infty}^{min} = 1 - \frac{2(E_a^4 + E_b^4)}{3(E_a^2 + E_b^2)^2}  \ .
\label{EQ}
\ee

If we use V$_{\infty}^{\rm {min}}$ = 0.66657 and E$_a$ in (\ref{EQ})
the value we got for the latent heat is $\Delta \approx 0.006$
which is of the same order as the one expected from the histograms.

\section{Conclusions}

The order of the Confinement-Higgs phase transition in the
SU(2)-Higgs model with fixed modulus is a highly non trivial issue.
We have used an extended parameter space, in order to get
a global vision on the problem. On this extended parameter
space we have found a line of first order phase transitions 
which get weaker
as $\kappa_2 \rightarrow$ 0. We have also observed that, because
of the computer resources needed, it is too ambitious trying
to measure two-peak energy distributions in the limit $\kappa_2$ = 0.
However, on this point, we can extract conclusions 
from the behavior of the specific heat. 

As we have discussed along the paper, a fake second order PT
seems to exists for a range of $L$ in very weak first order
phase transitions.
We have applied Finite Size Scaling properties along this 
transient region to compute a $pseudo$ $\alpha/\nu$ critical index.
We want to be extremely careful at this point, this computation
is completely meaningless when the transition is first order, 
since Scaling does not hold, but it can be used
as a technical tool to catalogue the PT when there is no direct
evidences, as in this case.
Using the relation $\alpha = 2 - \nu d$, we got $\nu$ varying
in the interval (0.36, 0.41) in the range $L = 8, \dots, 20$. 
Calculated from $L$=20 and $L$=24, $\nu\approx0.35$.
We expect this behavior to be transitory, and when going to larger
lattices sizes, if the transition is second order,
$\nu$ should reach its mean field value $\nu=1/2$.
If the transition is first
order this value should go to 1/d, indicating that the specific 
heat maximum grows like the volume $L^d$. We believe that this is
the case, since the $pseudo$ $\nu$ exponent
in $L$=24, instead of approaching 1/2, starts to decrease.
An example of weak first order PT showing a similar
behavior is described in \cite{ISAF2}.

In what concerning the motivation of introducing a second coupling,
we pointed out that $\kappa_2$ should not change 
the order of the PT because do not change the symmetry properties.
This argument is heuristic, but the phase
diagram we found supports this assertion.
As far as the order of the PT is concerned, we think that this approach
can be useful when dealing with PT of
questionable order in the sense that it is not clear whether
the transition is weakly first order or higher order. The hope
is that it could be applied to other more controversial models.

\newpage

\chapter{El modelo U(1) gauge en $D$=4 con
topolog\'{\i}as toroidal \-y esf\'erica \label{capU1}}
\thispagestyle{empty}
\markboth{\protect\small CAP\'ITULO \protect\ref{capU1}}
{\protect\small {\sl U(1) gauge con topolog\'{\i}a toroidal y esf\'erica}}

\cleardoublepage        

\section{Introducci\'on}

La transici\'on de la teor\'{\i}a libre de Dirac a la
electrodin\'amica cu\'antica (QED) implica reemplazar
una simetr\'{\i}a global por una local es decir, 
hacer gauge la teor\'{\i}a. 
Concretamente de la acci\'on correspondiente a la teor\'{\i}a libre 
de Dirac:
\be
S_F^{(0)} = \int dx^4 {\bar {\psi}} (x) 
(i \gamma^{\mu} \partial_{\mu} - M ) \psi(x)  \ ,
\ee
invariante bajo transformaciones globales de U(1), es decir
del tipo $G= e^{i \Lambda}$, se pasa  a la electrodin\'amica
cu\'antica (QED) haciendo que la invarianza global sea tambien
local, esto es, invariante bajo cambios locales de fase 
pertenecientes a U(1), es decir $G(x) = e^{i \Lambda (x)}$.
Esta transformaci\'on implica introducir un potencial gauge $A_{\mu} (x)$
y reemplazar las derivadas ordinarias por derivadas covariantes:
$D_{\mu} = \partial_{\mu} + ie A_{\mu}(x)$. 

Sin embargo para que el campo gauge juegue un papel en la din\'amica
hay que a\~nadir a la acci\'on original un t\'ermino cin\'etico
que permita su propagaci\'on. Para ello se introduce el tensor
${\mathcal F}_{\mu \nu}$:
\be
{\mathcal F}_{\mu \nu}(x) = - \frac{i}{e} [D_{\mu},D_{\nu}] = 
\partial_{\mu} A_{\nu}(x) - \partial_{\nu} A_{\mu}(x)  \ ,
\ee

de manera que la acci\'on que describe la QED en el espacio eucl\'{\i}deo
queda:
\ba
S_{QED}^{\rm cont} & = & S_G + S_F \nonumber \\
 & = &\frac{1}{4} \int dx^4 {\mathcal F}_{\mu \nu}(x) 
{\mathcal F}_{\mu \nu}(x) + \int dx^4 {\bar {\psi}} (x) 
(i \gamma^{\mu} D_{\mu} + M ) \psi(x)  \ .
\ea

En lo que sigue estaremos interesados en el estudio de
la parte pura gauge en la red, donde una posible formulaci\'on
de la teor\'{\i}a del continuo est\'a dada por la acci\'on de Wilson
\cite{WILSON}:
\be
S_{\mathcal W} = \beta \sum_{\rm n} \sum_{\mu < \nu}
[ 1 - \frac{1}{2}(U_{\mu \nu}(n) + U_{\mu \nu}^{\dagger}(n)) ] \ ,
\ee

donde $U_{\mu \nu}(n)$ es el usual producto ordenado de las links
que forman cada plaqueta
\be
U_{\mu \nu}(n) = U_{\mu}(n) U_{\nu}(n+\mu) U_{\mu}^{\dagger}(n+\nu) 
U_{\nu}^{\dagger}(n) \ .
\ee

En el l\'{\i}mite en el que el espaciado de la red se hace tender
a cero se recupera la acci\'on del continuo, 
$S_G$, con la siguiente correspondencia:

\ba
U_{\mu} (n) &=& e^{iea A_{\mu}(n)} \ ,  \\
\beta & = & \frac{1}{e^2} \ .
\ea

En efecto, usando la versi\'on discretizada de ${\mathcal F}_{\mu \nu}$:
\be
{\mathcal F}_{\mu \nu}(n) = \frac{1}{a} [ (A_{\nu} (n+\mu) - A_{\nu}(n)) -
(A_{\mu} (n+\nu) - A_{\mu}(n)) ]  \ ,
\ee

se tiene que la variable plaqueta se puede escribir como:
\be
U_{\mu \nu} (n) \approx e^{iea^2 {\mathcal F}_{\mu \nu}(n)} \ ,
\ee

y por tanto la acci\'on en el l\'{\i}mite $a \rightarrow 0$ queda:
\be
S_{\mathcal W} 
\approx \frac{1}{4} \sum_{\rm n} \sum_{\mu,\nu} a^4 
{\mathcal F}_{\mu \nu} {\mathcal F}_{\mu \nu} \ .
\ee

El modelo U(1) compacto en $d=4$
posee dos fases. Una es la regi\'on de acoplamiento fuerte
en la cual los fotones est\'an confinados, y la otra 
es la regi\'on de acoplamiento
d\'ebil en la que los fotones son portadores de un potencial tipo
culombiano. La transici\'on de fase que separa ambas regiones
tiene lugar en $\beta \sim 1$ y
ha sido ampliamente estudiada debido a la relevancia del modelo
U(1) para estudiar QED. Sin embargo 
el estudio de esta transici\'on de fase
ha resultado ser m\'as complicado de lo que se pod\'{\i}a
esperar en un principio, de su al menos aparente simplicidad. 
En la actualidad, las cuestiones relacionadas con el orden de la 
transici\'on de fase, y con los mecanismos que la provocan, siguen
siendo temas controvertidos.

La mayor\'{\i}a de los estudios num\'ericos han sido realizados
en redes c\'ubicas en $4D$ con condiciones de contorno peri\'odicas, es
decir en un hipertoro ($\HT$). Los estudios iniciales de este modelo
apuntaban a que la transici\'on era continua, puesto que las
simulaciones num\'ericas no evidenciaron ninguna se\~nal de metaestabilidad
\cite{CRE,LAU,BHA}.
Las redes m\'as grandes utilizadas en aquellos trabajos fueron
$L=4,5$. Sin embargo, cuando los recursos de los ordenadores permitieron
simular en redes m\'as grandes, se empezaron a observar metaestabilidades
e histogramas con doble pico de $L=6$ en adelante, revelando que la
transici\'on ten\'{\i}a calor latente y por lo tanto ser\'{\i}a de
primer orden \cite{VIC,VIC2,UBE}.
Esta idea fue corroborada por trabajos que aproximaron
el problema desde el punto de vista del Grupo de Renormalizaci\'on
\cite{HAS,ALF}.

Actualmente, quedan un cierto n\'umero de problemas
abiertos.
\begin{enumerate}
\item
Las simulaciones desde $L=6$ hasta $L=12$ revelan calor latente, 
pero su valor decrece con $L$.
\item
El exponente cr\'{\i}tico efectivo $\nu$ est\'a en el intervalo
(0.29,0.32) que es distinto de lo que se espera en primer
orden, $\nu = 0.25$, y tambien del valor correspondiente al
segundo orden trivial, $\nu=0.5$.
\end{enumerate}

Estos hechos llevaron a considerar la posibilidad de que el
calor latente fuese a cero en el l\'{\i}mite termodin\'amico.
Es decir, la transici\'on de fase en U(1) podr\'{\i}a a pesar de todo ser
continua si los dos picos que aparecen en las simulaciones 
fuesen un efecto de tama\~no finito \cite{TCP}.

Aparte de la acci\'on de Wilson, se consideraron acciones extendidas
para tratar de clarificar el problema mediante su formulaci\'on
en un espacio de fases extendido. En este contexto, una acci\'on
que incluye un t\'ermino proporcional a las plaquetas al cuadrado
fue propuesta por G.~Bhanot a comienzos de los '80 \cite{BHA2}:

\be
S_{\mathcal {EW}} = \beta \sum_{\rm n} \sum_{\mu < \nu}
[ 1 - \frac{1}{2}(U_{\mu \nu} + U_{\mu \nu}^{\dagger}) ] +
 \gamma \sum_{\rm n} \sum_{\mu < \nu}
[ 1 - \frac{1}{2}(U_{\mu \nu}^2 + U_{\mu \nu}^{\dagger 2}) ] \ .
\ee

Esta acci\'on posee el mismo l\'{\i}mite continuo que la acci\'on
de Wilson con la identificaci\'on $\beta + 4\gamma = 1/g^2$.

En este trabajo se estudi\'o el espacio de par\'ametros extendido
($\beta,\gamma$) en $L=4,5$. Se encontr\'o que la linea transici\'on
de fase desconfinante se extiende para todo $\gamma$. En estas
redes s\'olo se observaron metaestabilidades en $\gamma \geq +0.2$,
sugiriendo que la linea de transiciones de fase $\beta_{\rm c}(\gamma)$
se hace de primer orden para $\gamma \geq +0.2$, mientras que
en $\gamma < +0.2$ la transici\'on aparec\'{\i}a como de segundo
orden.

Simulaciones posteriores, siempre en redes $\HT$, han mostrado la
existencia de calor latente para todo valor de $\gamma$, tanto positivo
como negativo, para ret\'{\i}culos suficientemente grandes.
Por tanto, el trabajo de Bhanot no es exacto cuantitativamente, pero
cualitativamente sus conclusiones si son correctas: los valores de $\gamma$
positivos refuerzan la transici\'on en el sentido que el calor latente
crece cuanto m\'as positivo es este par\'ametro, 
mientras que los valores negativos de $\gamma$ debilitan la transici\'on
y calor latente decrece.

Actualmente se acepta con generalidad que la transici\'on es de
primer orden para valores positivos de $\gamma$. Sin embargo para
$\gamma \leq 0$ son posibles dos escenarios: 1) la linea de transiciones
$\beta_{\rm c}(\gamma)$ es de primer orden para todo $\gamma$ finito,
positivo o negativo; 2) Existe un $\gamma_{\rm TCP} \leq 0$ punto
tricr\'{\i}tico a partir del cual la transici\'on es de segundo orden
\cite{TCP}.

Este problema fue estudiado desde el punto de vista del Grupo de
Renormalizaci\'on \cite{ANA}.
Los resultados dan soporte al primer escenario descrito, en concreto
se encuentra un debilitamiento de la 
transici\'on al hacer $\gamma$ m\'as ne\-gativo,
pero ello sin cambiar el caracter de primer orden, sino simplemente 
debilit\'andola.

Hasta aqu\'{\i} se han descrito los resultados existentes en 
la topolog\'{\i}a $\HT$.

Sin embargo a lo largo del estudio de esta transici\'on de fase
se han usado otras topolog\'{\i}as aparte de la toroidal.
El motivo para ello es la idea debatida hace algunos a\~nos sobre la
influencia de los monopolos en la transici\'on de fase de U(1).
Concretamente se argument\'o que debido a las condiciones de contorno
periodicas existen monopolos no triviales, en el sentido
de que no son contra\'{\i}bles a un punto \cite{GUP}. 
Al ser los monopolos configuraciones asociadas a m\'{\i}nimos
locales de la acci\'on, su existencia hace crecer la energ\'{\i}a,
y se sugiri\'o que el salto en la energ\'{\i}a observado en la
transici\'on de fase se debe a la desaparici\'on de estos monopolos,
que en la regi\'on desconfinada no existen. 

Estas hip\'otesis indujeron a trabajar en redes con condiciones
de contorno esf\'ericas, puesto que en la esfera
todos los monopolos son contra\'{\i}bles a un punto \cite{CN,BAIG}.
En concreto ha sido sugerido que los dos estados metaestables
desaparecen cuando se trabaja en una red constru\'{\i}da considerando
la superficie de un cubo en $5D$ \cite{CN,JCN}, 
que tiene la misma topolog\'{\i}a que
la esfera. La ausencia de doble pico, junto la estimaci\'on del
exponente cr\'{\i}tico $\nu \sim 0.37$ medido en $\gamma = 0, -0.2, -0.5$
ha inducido a los autores en \cite{JCN} a decir que los dos picos
observados en el toro son un efecto de tama\~no finito, y que trabajando
en la esfera se ve que la linea $\beta_{\rm c}(\gamma)$ 
es cr\'{\i}tica para $\gamma \leq 0$ y est\'a caracterizada por un
exponente cr\'{\i}tico $\nu \sim 1/3$.

El papel de los monopolos no triviales mencionados anteriormente
parece estar descartado en cuanto a su influencia en el orden
de la transici\'on de fase \cite{REBBI}. Por lo tanto
a primera vista la situaci\'on es bastante extra\~na. 
Puesto que que los monopolos no triviales no parecen jugar ning\'un papel,
uno esperar\'{\i}a obtener antes el comportamiento asint\'otico en una red 
homog\'enea e invariante por traslaciones, es decir la toroidal, que en una red
manifiestamente no homogenea, como la superficie de un cubo en $5D$,
y en la cual la invariancia traslacional se recupera s\'olo en el
l\'{\i}mite de volumen infinito.

Como se ha venido exponiendo a lo largo de esta memoria, la ausencia
de doble pico no prueba que una transici\'on sea de segundo orden.
Adem\'as la aparici\'on de un exponente $\nu \sim 0.37$, intermedio entre
el valor de primer orden y el de segundo orden trivial, ha sido encontrado
en otros modelos como caracterizador del regimen transitorio de las
transiciones de primer orden d\'ebiles \cite{LAF,APE}.

Para tratar de arrojar algo de luz sobre este problema se ha hecho
un estudio sistem\'atico de la linea de transiciones de fase
$\beta_{\rm c}(\gamma)$ en el toro y en la red con topolog\'{\i}a
esf\'erica empleada en \cite{CN,JCN} y que denotaremos por $\HS$.
Para esta red hemos empleado tama\~nos m\'as grandes de los medidos
anteriormente para ver si la se\~nal de doble pico aparece.
En la red $\HT$ hemos mejorado la estad\'{\i}stica que hab\'{\i}a
hasta la fecha, midiendo en redes hasta $L=24$ para algunos valores de 
$\gamma$.

\newpage

\begin{center}
{\bf Abstract}
\end{center}

{\sl 
We have performed a
systematic study of the phase transition in the pure 
compact U(1) lattice gauge theory in the extended coupling 
parameter space ($\beta , \gamma$) on toroidal and spherical lattices. 
The observation of 
a non-zero latent heat in both topologies for all  
investigated $\gamma \in [+0.2,-0.4]$,  
together with an exponent $\nu_{\rm {eff}} \sim 1/d$ when large enough
lattices are considered, lead us to conclude that the phase transition
is first order. For negative $\gamma$, our results point to an
increasingly weak first order transition as $\gamma$ is made more negative.
}

\section{Description of the model and observables}

We shall consider the parameter space described by the extended
Wilson action, which can be expressed in terms of the plaquette
angle in the following way:

\begin{equation}
S = - \beta \sum_{\rm p} \cos \theta_{\rm p} - \gamma \sum_{\rm p} 
\cos 2\theta_{\rm p} \ .
\end{equation}

We use for the simulations the conventional $4D$ hypercube with
toroidal boundary conditions (hypertorus), and, for
comparison we also consider the surface of a $5D$ cube which is
topologically equivalent to a $4D$ sphere. Contrarily to the hypertorus,
the surface of a $5D$ cube is not homogeneous. There are a number of sites
which do not have the maximum connectivity, namely 8 neighbors. Due to this
fact, uncontrolled finite size effects are expected to turn up. 
Their influence can be somehow alleviated by the introduction of 
appropriate weight factors in those inhomogeneous sites \cite{JCN}.
Since the topology remains unchanged and we do not expect the order
of the phase transition to be affected by the rounding, we do not
use weight factors. 

As a notational remark, we label with $L$ and $N$ the side length
for the $\HT$ topology and for the $\HS$ topology respectively.

Next we define the energies associated to each term in the action
\begin{eqnarray}
E_{\rm p} = \frac{1}{N_{\rm p}} \langle \sum_{\rm p} \cos \theta_{\rm p} 
\rangle \ , \\
E_{\rm {2p}} = \frac{1}{N_{\rm p}} \langle \sum_{\rm p} \cos 2 \theta_{\rm p} 
\rangle \ ,
\end{eqnarray}

where $N_{\rm p}$ stands for the number of plaquettes.
On the hyper-torus this number is simply proportional to the volume,
namely the forward plaquettes are
$N_{\rm p} = 6 L^4$. On the sphere the number of plaquettes has a
less simple expression, and can be computed as a function of $N$
as $N_{\rm p}= 60(N-1)^4+20(N-1)^2$. 
In this case, the system is
not homogeneous and $N_{\rm p}$ is not proportional to the number of points
on the four dimensional surface, which is $N^5-(N-2)^5$, some points having a 
number of surrounding plaquettes less than the possible maximum $12$, 
as opposed to what happens on the torus .
 In order to allow comparison, we 
define $L_{\rm {eff}}=(\frac{N_{\rm p}}{6})^{1/4}$, in such a way that a 
hypertorus of $L=L_{\rm {eff}}$ has the same number of plaquettes as
the corresponding hypersphere.

The specific heat and the Binder cumulant are useful quantities
to monitorize the properties of a phase transition. 
At the critical point, they are known to posses different thermodynamical 
limits depending on whether the transition is first order or higher order,
and hence, their behavior with increasing lattice size can give
some clues as to the order of the phase transition.
We have studied the Finite Size Scaling (FSS) of these two energy cumulants.

The specific heat is defined for both energies as:

\begin{equation}
C_{\rm v} = \frac{\partial}{\partial\beta}E_{\rm p} = 
N_{\rm p} ( \langle E_{\rm p}^2 \rangle - \langle E_{\rm p} \rangle^2 ) \ ,
\label{CSPE}
\end{equation}

\begin{equation}
C_{\rm {2v}} = \frac{\partial}{\partial \gamma}E_{\rm {2p}} = 
N_{\rm p} ( \langle E_{\rm {2p}}^2 \rangle - \langle E_{\rm {2p}} \rangle^2 ) \ .
\label{CSPE2}
\end{equation}

As we have observed a high correlation between both energies, the
results being qualitatively the same, we shall only report
on the observable defined for
the plaquette energy $E_{\rm p}$.

In a second order phase transition, scaling theory predicts the \-specific
heat maximum to diverge as $L^{\alpha/\nu}$.
If the transition is first order it is expected
to diverge like the volume $L^d$ (or strictly like the number of
plaquettes $N_{\rm p}$) reflecting that the maximum of the energy fluctuation
has the size of the volume. This is expected to hold in the asymptotic
region of the transition $L \gg \xi_c$. In the transient region,
namely $L < \xi_c$, the specific heat is expected to grow more slowly than
the volume. 

We shall use the number of plaquettes $N_{\rm p}$ to study the specific heat
behavior in order to have a single parameter for both the $\HT$ and 
the $\HS$ topologies. This is of course equivalent to doing the discussion
as a function of $L$ on the $\HT$ and $L_{\rm {eff}}$ on the $\HS$.
We have in this case $C_v \sim (N_{\rm p})^{\alpha/d\nu}$.

We have also studied the behavior of the fourth cumulant of the energy
\begin{equation}
V_L = 1 - \frac{\langle E_{\rm p}^4 \rangle_L}{\langle E_{\rm p}^2 
\rangle^2_L} \ .
\end{equation}

When the energy distribution describing the system is gaussian $V_L 
\rightarrow 2/3$ in the thermodynamic limit. This is the case for a second
order phase transition. If the transition is first order, far from
the critical coupling $\beta_{\rm c}$, $V_L$ also tends to 2/3, reflecting the
gaussianity of the energy distribution. However, at $\beta_{\rm c}$ the distribution 
can be described by two gaussians centered about the energy of each
metastable state $E_1$ and $E_2$, and hence this quantity has the
non-trivial thermodynamic limit:
\begin{equation}
V_{L \rightarrow \infty} \rightarrow 1 - \frac{2(E_1^4 + E_2^4)}{3(E_1^2 + E_2^2)^2} < 2/3 \ .
\end{equation}

Another interesting quantity measurable from the density of states
is the distribution of the partition function zeroes, or Lee-Yang
zeroes \cite{LEE}. To clarify the role of the partition function zeroes
let us first stress a well known fact: there are no phase transitions
on a finite volume, phase transitions arise in the thermodynamical limit
and are signaled by non-analyticity in the free energy:

\begin{equation}
F(\beta,V) = - \frac{1}{\beta V} \log Z(\beta) \ .
\end{equation}

$Z(\beta)$ is a linear combination of exponentials, and hence an analytical
function. This implies that the free energy can be singular only where
$Z(\beta) = 0$. However if the coupling $\beta$ is real, and the volume
is finite, that linear combination is a sum of positive terms and hence, the
zeroes of $Z(\beta)$ are located in the complex plane of the coupling $\beta$.
The onset of the phase transition in the limit of infinite volume
is signaled by a clustering of the zeroes on the real axis at $\beta_{\rm c}$.

Here follows the description of our procedure, for a general description
of the method see \cite{ENZO}. 

As mentioned previously we work only with $E_{\rm p}$, and then we 
consider only the spectral density method \cite{FALCI} for this variable.
Since we work at fixed $\gamma$ we consider only the $\beta$ coupling
derivatives.
 
From the MC simulation at $\beta$ we obtain an approximation to the density
of states  which allows us to compute the normalized energy distribution
$P_{\beta} (E)$. The energy distribution can be expressed as

\begin{equation}
P_{\beta}(E) = \frac{1}{Z} W(E) e^{-\beta E}  \ .
\label{PROB}
\end{equation}

We use the standard reweighting technique to obtain from the distribution
measured at $\beta$ the distribution 
at another coupling $\omega$, which is complex in the more general
case $\omega = \eta + i \xi$. The standard reweighting formula is:

\begin{equation}
P_{\omega}(E) = \frac{P_{\beta}(E) e^{-(\omega - \beta)E} }
{\sum_E P_{\beta}(E) e^{-(\omega - \beta)E} }  \ ,
\end{equation}

using (\ref{PROB}) and the normalization condition $\sum_E P_{\beta}(E) =1$
one obtains

\begin{equation}
\frac{Z(\omega)}{Z(\beta)} = 
\sum_E P_{\beta}(E) e^{-(\omega - \beta)E}  \ .
\end{equation}

We can factorize the contributions from the real and 
imaginary part:

\begin{equation}
\frac{Z(\omega)}{Z(\beta)} = 
\sum_E P_{\beta}(E) (\cos(E\xi) + i\sin(E\xi)) e^{-(\eta - \beta)E} \ .
\end{equation}

This is the standard reweighting formula extended to the complex parameter
space of couplings. As a first observation we have a pure oscillating factor
due to $Im(\omega) \neq 0$. Since $E$ is $O(V)$ this is a rapidly
oscillating function which makes it impossible to locate zeroes with
large imaginary part.

The real part of the coupling contributes to the well known
exponential damping, $e^{-(\eta - \beta)E}$, which is telling us that we can
trust the extrapolation only in a small neighborhood $\beta \pm \eta$.

Since $Z(\Re \omega)$ has no zeroes in a finite volume, 
an easy way to locate the zeroes numerically is looking at the minima
of the function  $|G(\omega)|^2$, where:

\begin{equation}
G(\omega) = \frac{Z(\omega)}{Z(\Re \omega)} = \frac
{\sum_E P_{\beta}(E) (\cos(E\xi) + i\sin(E\xi)) e^{-(\eta - \beta)E}}
{\sum_E P_{\beta}(E) e^{-(\eta - \beta)E}} \ .
\end{equation}

The function to minimize is:

\begin{equation}
|G(\omega)|^2 = 
\frac{(\sum_E P_{\beta}(E) e^{-(\eta - \beta)E} \cos(E \xi))^2
 + (\sum_E P_{\beta}(E) e^{-(\eta - \beta)E} \sin(E \xi))^2 }
{(\sum_E P_{\beta}(E) e^{-(\eta - \beta)E})^2} 
\label{F2}
\end{equation}

One interesting property of the partition function zeroes concerns
the estimation of critical exponents. Denoting by $\omega_0$
the coupling where the first zero is located, the distance to the real axis
scales with the $\nu$ exponent:

\begin{equation}
Im(\omega_0) \sim L^{-1/\nu} \ .
\label{NU}
\end{equation}

We shall use this property to compute the effective $\nu$ exponent.

From $P_{E} (\beta)_L$, we can measure the free energy gap, $\Delta F(L)$,  
which is the difference between 
the minima and the local maximum of the free energy \cite{FREE}.
We use the spectral density method to get, from the measured histograms,
a new histogram where both peaks have equal height. We take
the logarithm of those histograms and measure the energy gap.
A growing $\Delta F(L)$ in the asymptotic region of the transition
implies a first order phase transition. An increase proportional
to $L^{d-1}$ is expected \cite{FREE}.
For the transition to be
second order, $\Delta F(L)$ must stay constant with increasing lattice
sizes.

\begin{figure}[!t]
\centerline{
\fpsxsize=7cm
\def\fpsangle{90}
\fpsbox[70 90 579 760]{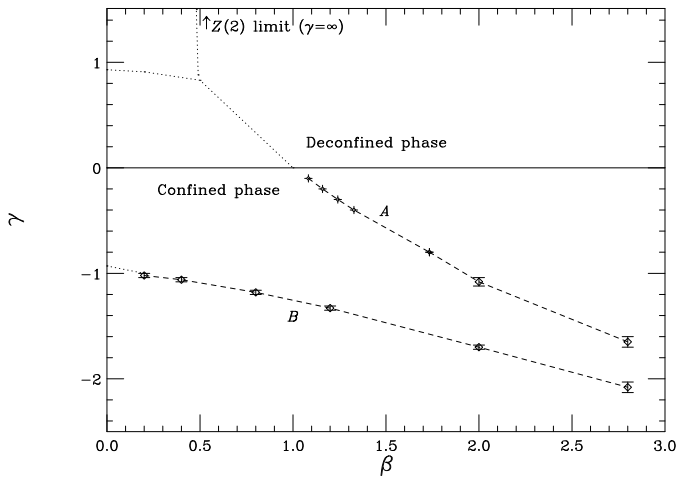} }
\caption{\small {Phase diagram of U(1) in the 
extended $(\beta, \gamma)$ parameter
space. The dotted lines have been taken from \protect\cite{BHA2}. 
The symbols correspond to our simulations on the torus. The crosses
correspond to the peak of $C_{\rm v}(L=16)$, the errors are not visible
in this scale. The diamonds have been obtained with hysteresis
cycles in $L=8$.}}
\label{PHD}
\end{figure}

\subsection{Schwinger-Dyson Equations}

As a further check we have implemented the Schwinger-Dyson equations (SDE)
\cite{FALCI} which allow one to recover the simulated couplings 
from the Montecarlo data.

Let {\rm A($\theta$)} be an operator with null expectation value:
\begin{equation}
\langle A(\theta) \rangle = 
Z^{-1} \int [d\theta] A(\theta) e^{-S[\theta]} \equiv 0 \ .
\end{equation}

Derivating with respect to $\theta$ this equation trivially yields

\begin{equation}
\langle \frac{\partial A(\theta)}{\partial \theta} \rangle =
\langle A(\theta) \frac{\partial S[\theta]}{\partial \theta} \rangle \ ,
\label{SDEQ}
\end{equation}

which is the equation of movement of the operator {\rm A($\theta$)}
or Schwinger-Dyson equation. When the action depends on several 
couplings the equation (\ref{SDEQ}) can be expressed as:

\begin{equation}
\langle \frac{\partial A(\theta)}{\partial \theta} \rangle =
\sum_i \beta_i \langle A(\theta) 
\frac{\partial S_i[\theta]}{\partial \theta} \rangle  \ .
\label{SDD}
\end{equation}

This equation relates the values of the couplings with the expectation
values we measure from the MC simulation. We need as many independent equations
as couplings we have to determine in the action. In our case, in order
to measure both $\beta$ and $\gamma$, we need two operators with null
expectation value in order to have two independent tests.
At each lattice site $n$ and for every direction $\mu$
we consider the operators:
\begin{eqnarray}
A(\theta) = \sin \theta_p = \sin(\theta_{n,\mu} - \theta_{stap}) \ , \\ 
B(\theta) = \sin 2\theta_p = \sin2(\theta_{n,\mu} - \theta_{stap}) \ ,
\end{eqnarray}

where $\theta_{stap}$ is the staple of the link labeled by ($n$,$\mu$).

Applying equation (\ref{SDD}) to those operators we get:
{\small {
\begin{eqnarray}
\langle \sum_{\rm p} \cos(\theta_{n,\mu} - \theta_{stap}) \rangle = 
\beta \langle \sum_{\rm p} \sin(\theta_{n,\mu} - \theta_{stap})
\sum_{\rm p} \sin(\theta_{n,\mu} - \theta_{stap}) \rangle + \nonumber  \\ 
 + 2\gamma \langle \sum_{\rm p} \sin(\theta_{n,\mu} - \theta_{stap}) 
\sum_{\rm p} \sin2(\theta_{n,\mu} - \theta_{stap}) \rangle 
\end{eqnarray}

\begin{eqnarray}
\langle \sum_{\rm p} 2\cos2(\theta_{n,\mu} - \theta_{stap}) \rangle = 
\beta \langle \sum_{\rm p} \sin2(\theta_{n,\mu} - \theta_{stap})
\sum_{\rm p} \sin(\theta_{n,\mu} - \theta_{stap}) \rangle + \nonumber \\ 
 + 2\gamma \langle \sum_{\rm p} \sin2(\theta_{n,\mu} - \theta_{stap})
\sum_{\rm p} \sin2(\theta_{n,\mu} - \theta_{stap}) \rangle 
\end{eqnarray}
}
}
On the hypertorus $\sum_{\rm p}$ means the sum over the plaquettes
in positive and negative directions ($\pm \mu$)
bordering the link (12 plaquettes). On the sphere one has to be careful,
since not all the links have 12 surrounding plaquettes, and the sum
has to be understood as extended to the existing plaquettes.

These equations hold for all $n$ 
in such a way that we can sum up the equations
for every single link and average over the number of links.
We quote in Table \ref{TABLA_SD} for the tests done at $\gamma = -0.2$
using both topologies. They show a perfect agreement between the simulated
couplings and the ones recovered from the MC expectation values.

\begin{table}[h]
{

\begin{center}
{
\small{
\begin{tabular}{|c||c|c||c||c|c|c|} \hline
\multicolumn{3}{|c||}{$\HT$ topology} &  \multicolumn{3}{|c|}{$\HS$ topology} \\ \hline  \hline
$L$ & $\beta_{\rm {sim}}$ & $(\beta,\gamma)_{\rm {SD}}$ 
&N &$\beta_{\rm {sim}}$ &$(\beta,\gamma)_{\rm {SD}}$   \\ \hline
$6$ &1.1460 &1.1466(44),-0.2009(26)        
&6 &1.1587 &1.1592(14),-0.1998(12)                     \\ \hline
$8$ &1.1535  &1.1527(22) -0.1997(15)       
&8  &1.1600  &1.1599(12) -0.1999(8)               \\ \hline
$12$ &1.1582 &1.1581(12),-0.1999(8)       
&10  &1.1602 &1.1602(7),-0.2001(9)                   \\ \hline
\hline

\end{tabular}
}
}
\end{center}

}
\caption[a]{\small{Couplings obtained from the MC simulations at $\gamma$=-0.2
using the Schwinger-Dyson equations.}}
\protect\label{TABLA_SD}

\end{table}

\section{Numerical Simulation}

\begin{table}[h]
{

\begin{center}
{
\small{
\begin{tabular}{|c|c||c|c|c|c|c|c|c|} \hline
\multicolumn{7}{|c|}{$\gamma = +0.2$} \\  \hline
$N$  &$L_{\rm {eff}}$ &$\beta_{\rm {sim}}$ &$\tau$ &$N_\tau$  
&$\beta^{\ast} (L)$ &${\rm Im}(\omega_0)$  \\ \hline
$4$  &$5.383$    &0.8910  &70   &29142  &0.8906(2) &0.0169(1)       \\ \hline
$5$  &$7.150$    &0.8855  &210   &8000  &0.8855(2) &0.00604(6)   \\ \hline
$6$  &$8.921$    &0.8834  &520   &3800  &0.88330(5) &0.00244(6)   \\ \hline
$7$  &$10.694$   &0.8818  &790   &2300  &0.88182(3) &0.00090(3)    \\ \hline
$8$  &$12.469$   &0.88095 &930   &1930  &0.88097(3) &0.00040(2)   \\ \hline
\hline \hline
\multicolumn{7}{|c|}{$\gamma = 0$} \\  \hline
$N$  &$L_{\rm {eff}}$ &$\beta_{\rm {sim}}$ &$\tau$ &$N_\tau$ 
&$\beta^{\ast} (L)$ &${\rm Im}(\omega_0)$  \\ \hline
$6$  &$8.921$    &1.0128 &240   &8300 &1.01340(5)  &0.00616(8)      \\ \hline
$8$  &$12.469$   &1.0125 &400   &2300 &1.0127(2)   &0.00196(4)     \\ \hline
$10$ &$16.021$   &1.0120 &780   &1200 &1.01212(3)  &0.00075(3)     \\ \hline
$12$ &$19.574$   &1.0119 &850   &1150 &1.01194(2)  &0.00031(2)     \\ \hline
$14$ &$23.123$   &1.0117 &920   &1900 &1.01168(2)  &0.00013(1)    \\ \hline
\hline \hline
\multicolumn{7}{|c|}{$\gamma = -0.2$} \\  \hline
$N$  &$L_{\rm {eff}}$ &$\beta_{\rm {sim}}$ &$\tau$&$N_\tau$   
&$\beta^{\ast} (L)$ &${\rm Im}(\omega_0)$  \\ \hline
$6$  &$8.921 $   &1.1587 &160   &2000  &1.1587(4)  &0.0107(3)   \\ \hline
$7$  &$10.694$   &1.1600 &320	&2800  &1.1597(1)  &0.00550(7)  \\ \hline
$8$  &$12.469$   &1.1597 &510   &1000  &1.1603(2)  &0.00321(4)   \\ \hline
$10$ &$16.021$   &1.1602 &680   &1500  &1.1604(2)  &0.00171(2)   \\ \hline
$12$ &$19.574$   &1.1604 &820   &1200  &1.1602(1)  &0.00083(1)   \\ \hline
$14$ &$23.123$   &1.1605 &900   &1100  &1.16048(5) &0.00044(1)   \\ \hline
$16$ &$26.684$   &1.1604 &1150  &1200  &1.16038(2) &0.00023(1)  \\ \hline
\end{tabular}
}
}
\end{center}

}
\caption[a]{\footnotesize{Statistics of the data obtained 
for the $\HS$ topology}}
\protect\label{TABLA_ESFERA}

\end{table}

\begin{table}[!t]
{

\begin{center}
{
\small{
\begin{tabular}{|c|c|c|c|c|c|} \hline
\multicolumn{6}{|c|}{$\gamma = -0.1$}   \\ \hline
$L$ & $\beta_{\rm {sim}}$ & $\tau$ & $N_{\tau}$ &$\beta^{\ast} (L)$ 
&$Im(\omega_0)$  \\ \hline \hline
$6$ &1.0720 &350   &1900 &1.0716(2) &0.0097(1)       \\ \hline
$8$ &1.0784  &640  &1400  &1.0786(2)  &1.1539(2)    \\ \hline
$12$ &1.0820 &820   &750  &1.0818(1) &0.00114(2)   \\ \hline
$16$ &1.08278 &930    &900   &1.0827(1)  &0.00040(2)   \\ \hline
$20$ &1.0833 &1150   &1100  &1.0833(1) &0.00020(1)    \\ \hline
\hline \hline
\multicolumn{6}{|c|}{$\gamma = -0.2$}   \\ \hline
$L$ & $\beta_{\rm {sim}}$ & $\tau$ & $N_{\tau}$ &$\beta^{\ast} (L)$ 
&$Im(\omega_0)$  \\ \hline \hline
$6$   &1.1460 &380   &2000     &1.1452(2)  &0.0115(1)            \\ \hline
$8$   &1.1535  &620   &1900    &1.1539(2)  &0.00506(6)           \\ \hline
$12$  &1.1582 &840   &1200     &1.1582(2)  &0.00152(3)           \\ \hline
$16$  &1.15935 &920   &900     &1.1593(1)  &0.00060(2)           \\ \hline
$20$  &1.1599 &1150 &900       &1.1599(1)  &0.00028(1)           \\ \hline
\hline \hline
\multicolumn{6}{|c|}{$\gamma = -0.3$} \\ \hline  
$L$&$\beta_{\rm {sim}}$ &$\tau$  &$N_{\tau}$ &$\beta^{\ast} (L)$ 
&$Im(\omega_0)$  \\  \hline \hline
$6$&1.2255   &340  &2000  &1.2237(4) &0.0138(1)     \\   \hline
$8$&1.2344   &560   &1800 &1.2340(1) &0.00623(4)     \\   \hline
$12$&1.2395   &770  &1000 &1.2395(2) &0.00190(3)    \\  \hline
$16$ &1.2410  &900  &900  &1.2410(1) &0.00084(2)    \\  \hline
$20$ &1.2416  &1100  &1200  &1.24156(5)  &0.00041(2)     \\  \hline
$24$ &1.2417  &1200   &1100 &1.24162(5)  &0.00022(1)     \\  \hline
\hline \hline
\multicolumn{6}{|c|}{$\gamma = -0.4$} \\ \hline
$L$&$\beta_{\rm {sim}}$ &$\tau$  &$N_{\tau}$ &$\beta^{\ast} (L)$ 
&$Im(\omega_0)$  \\ \hline \hline
$6$   &1.3090  &400  &1800   &1.3082(4)  &0.0152(1)          \\ \hline
$8$   &1.3192 &600  &1600    &1.3194(3)  &0.00704(5)           \\ \hline
$12$   &1.3258 &710  &1400   &1.3259(1)  &0.00238(3)            \\ \hline
$16$   &1.32775 &840  &900   &1.3278(1)  &0.00108(2)            \\ \hline
$20$   &1.3285 &930  &1600   &1.3284(1)  &0.00054(2)            \\ \hline
$24$   &1.3286 &1150  &1500  &1.3286(1)  &0.00029(1)             \\ \hline
\end{tabular}
}
}
\end{center}

}
\caption[a]{\footnotesize{Statistics of the data obtained
in the $\HT$ topology.}}
\protect\label{TABLA_TORO}

\end{table}

Most of the work has been done by simulating the subgroup 
$Z(1024) \subset U(1)$, since the phase transition associated to the discrete
group lies safely far away. The overrelax algorithm can be applied only
in the simulations with $\gamma = 0$ so that the gain in statistical
quality due to the overrelax effect is limited to the Wilson 
action, where we have simulated both, the full group $U(1)$
and the discrete one for the sake of comparison.
For $\gamma \neq 0$ we have simulated the discrete group since so the
simulation is considerably speeded up.

For every lattice size we consider, we perform trial runs to locate
the coupling $\beta^{\ast}(L)$ where the specific heat
shows a peak. We use the standard reweighting techniques to extrapolate
in a neighborhood of the simulated coupling.
Once the peak is located within an error in the fourth
digit of the coupling, we perform an intensive simulation there
to get $P_{E}(\beta^{\ast})_L$. The statistics performed at these
pseudo-critical
couplings are reported in Table \ref{TABLA_ESFERA} and Table \ref{TABLA_TORO}
for the $\HS$ and the $\HT$ topology respectively.

Typically we measure the energies every 10 $MC$ sweeps in order
to construct the energy histogram (\ref{PROB})
From this energy distribution we 
obtain the cumulants of the energy we are interested in, and the critical
exponents using Finite Size Scaling techniques.
We remark that one of the main sources of systematic error 
when measuring critical exponents, is the
indetermination in the coupling $\beta^{\ast}(L)$ where to measure.

The simulations have been done in the {\sl RTNN} machine consisting of 32
Pentium Pro 200MHz processors. The total CPU time used is the equivalent
of 6 Pentium Pro years.

We have updated using a standard Metropolis algorithm with 2 hits. The
acceptance has always been between 65\% and 75\%. 

In order to consider the statistical quality of the simulation,
following \cite{SOKAL} we define the unnormalized autocorrelation 
function for the energy
\be
C(t) = \frac{1}{N-t} \sum_{i=1}^{N-t} E_i E_{i+t} - \langle E \rangle^2 \ ,
\ee
as well as the normalized one
\be
\rho(t) = \frac{C(t)}{C(0)} \ .
\ee

The integrated autocorrelation time for the energy, $\tau^{int}$, can be 
measured using the window method
\be
\tau^{int}(t) = \frac{1}{2} + \sum_{t^{\prime} = 1}^t \rho(t^{\prime}) \ ,
\ee

for large enough $t$, which is in practice selected self-consistently.
We use $t$ in the range 5$\tau^{int}$, 10$\tau^{int}$, and we check
that the obtained $\tau^{int}$ remains stable as the window in $t$
is increased. 

We have always started from hot and cold configurations in order to make
sure that the system does not remain in a long living metastable state,
which could be interpreted as a Dirac sheet. The results coming from
both types of starts have always been indistinguishable.

\section{The phase transition line $\beta_{\rm c}(\gamma)$}

We have studied the deconfinement-confinement phase transition line
at several values of $\gamma$ (Figure \ref{PHD}).

In the region of negative $\gamma$, apart from the deconfinement
transition line, there is another transition line provoked by the
competing interaction between the couplings ($\beta,\gamma$).
In the limit $\gamma = -\infty$ the model is not dynamical since for
all finite $\beta$, $\cos \theta_{\rm p} = 0$. So we expect this
line to end at the corner $(\gamma = - \infty, \beta=~+\infty)$.
We have not performed a deep study
of this transition line, however the simulations at $L=8$ revealed
double-peak structures pointing to a first order character.
The limit $\gamma = + \infty$ is equivalent to a $Z(2)$ theory and
the critical point can be calculated exactly by self-duality.

We have studied carefully the region
between lines $A$ and $B$
finding no signatures of the existence of additional lines.

We have focused on the transition line $A$,
at several values of $\gamma$, on the hyper-torus and on the hyper-sphere.

The structure of this section is the following:

First, we describe the results at $\gamma = +0.2$ on the spherical topology.
Our purpose is to observe how does the $\HS$ topology behave in a
non controversial region, and comparing with the known results on the
hypertorus.

Second, we study the point $\gamma = 0$ (Wilson action) on the spherical 
lattice. This coupling has been 
recently claimed to be the starting point of a critical line
(infinite correlation length at the critical point) which extends in
the range $\gamma \leq 0$, with an associated critical exponent
$\nu \sim 0.37$ \cite{JCN}. Since this assertion relies on the
absence of two-state signals on spherical lattices up to $N=10$,
we shall check whether or not double peak structures set in when
larger spheres are considered.

Third, we go to the region of negative $\gamma$ values. 
We are aware of no previous systematic study on the hypertorus 
in this region, 
so we run simulations up to $L=24$ at $\gamma = -0.1,-0.2,-0.3,-0.4$.
Motivated by the 
results on the hyper-torus we have just run a single $\gamma$
negative value on the hyper-sphere. For the sake of comparison with \cite{JCN}
we choose this value to be $\gamma = -0.2$.

Finally, we discuss the Finite Size Scaling behavior 
exhibited by both topologies.

\begin{figure}[!b]
\centerline{
\fpsxsize=7cm
\def\fpsangle{90}
\fpsbox[70 90 579 760]{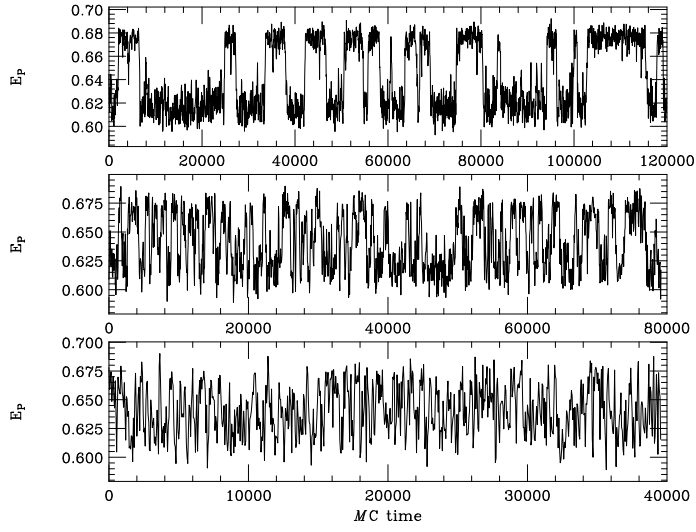} }
\caption{\small {MC evolution of $E_{\rm p}$ at $\gamma=+0.2$ on the $\HS$
topology for $N=6$ (lower window),$N=7$ (middle) and $N=8$ (top).}}
\label{EVOL02}
\end{figure}

\begin{figure}[!b]
\centerline{
\fpsxsize=7cm
\def\fpsangle{90}
\fpsbox[70 90 579 760]{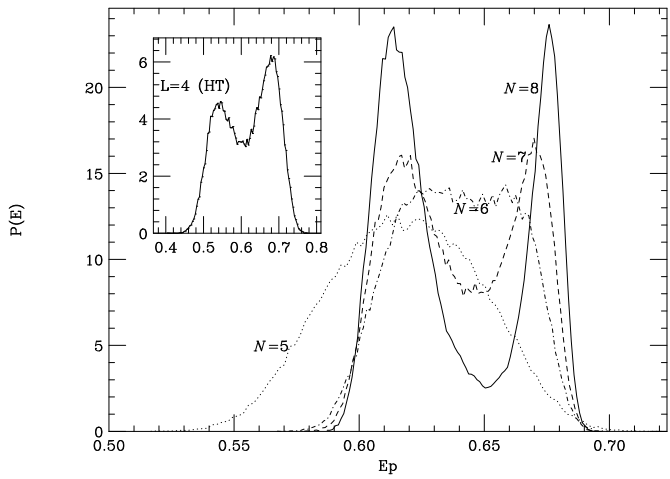} }
\caption{\small {$E_{\rm p}$ distribution at $\gamma=+0.2$ on the $\HS$
topology. The small window is the distribution we obtained on the
$\HT$ topology in $L=4$ at $\beta =0.8595$ }}
\label{HISTO02}
\end{figure}

\begin{figure}[!]
\centerline{
\fpsxsize=7cm
\def\fpsangle{90}
\fpsbox[70 90 579 760]{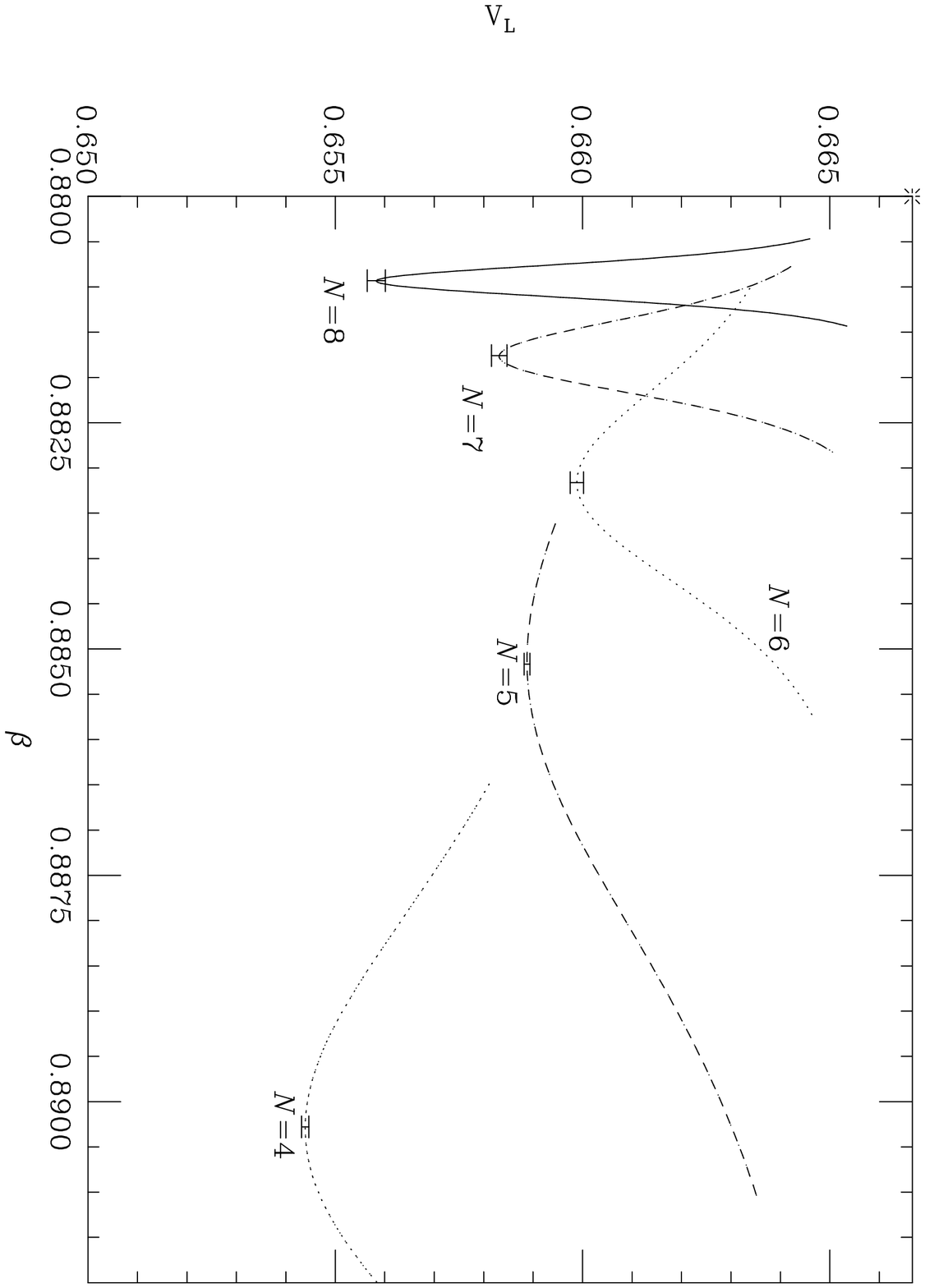} }
\caption{\small {Binder cumulant at $\gamma = +0.2$
on the $\HS$ topology in $N=4,5,6,7,8$. The cross in the upper corner 
signals the second order value $2/3$.}}
\label{BINDER02}
\end{figure}

\begin{figure}[!]
\centerline{
\fpsxsize=7cm
\def\fpsangle{90}
\fpsbox[70 90 579 760]{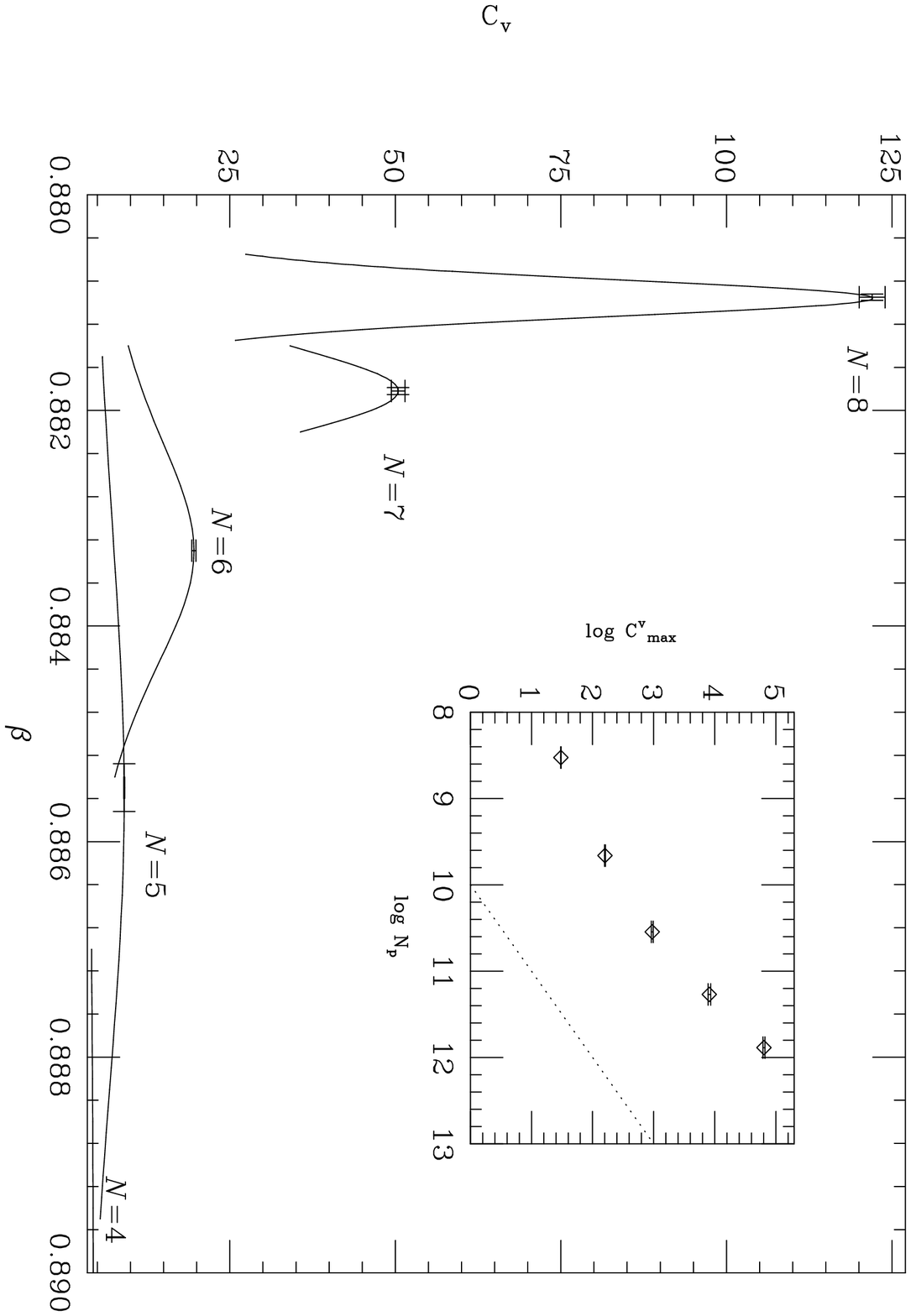} }
\caption{\small {Specific heat maximum, and Ferrenberg-Swendsen extrapolation 
(solid line) at $\gamma = +0.2$ on the $\HS$ topology.
The small window represents $C_{\rm v}^{\rm {max}} (N_{\rm p})$ 
The dotted line corresponds to the slope expected in a
first order phase transition.}}
\label{PEAKS02}
\end{figure}

\begin{figure}[!]
\centerline{
\fpsxsize=7cm
\def\fpsangle{90}
\fpsbox[70 90 579 760]{h01.ps} }
\caption{\small {$E_{\rm p}$ distribution at $\gamma=+0.1$ on the $\HS$
topology.}}
\label{HISTO01}
\end{figure}

\subsection{Results at $\gamma=+0.2$ on the spherical topology}

As we have pointed out,
there is a general agreement on considering the phase transition 
first order in the region of positive $\gamma$.

The lattice sizes used range from $N=4$ to $N=8$, which correspond
to $L_{\rm {eff}} \sim 5$ and $L_{\rm {eff}} \sim 12$ respectively
see (Table \ref{TABLA_ESFERA}). 

In Figure \ref{EVOL02}, the MC evolution for $N=6,7,8$ is shown. We remark that
no multicanonical update is needed to obtain a very high rate of flip-flops
up to $N=~8$. However, on the torus, for $L > 6$ the probability of tunneling
between both metastable states is so tiny that a reasonable rate
of flip-flops is not accessible to ordinary algorithms.
Probably the inhomogeneity of the sphere
is increasing the number of configurations with energies which 
correspond neither to the confined nor to the deconfined phase, 
but in between.
These configurations make the free energy gap to decrease and hence
the tunneling is easier on the $\HS$ topology. In short, the sites without
maximum connectivity act as catalysts of the tunneling.

In Figure \ref{HISTO02} 
the energy distributions are plotted. The distribution at
$N=6$ is distinctly non-gaussian, but a blatant two-peak structure is
observed only from $N=7$ ($L_{\rm {eff}} \sim 11$) on. So, 
when comparing with the result at $L=4$ on the torus, (see 
small window in Figure \ref{HISTO02})
a first observation is that the 
onset of a two-state signal is particularly spoiled
by the $\HS$ topology, at least in 4$D$ pure compact U(1) gauge theory.

Concerning the latent heat,
we remark its stability already at $N=7$, or, if anything, its increase
from $N=7$ to $N=8$.
One would even say that the same happens
between the positions of the would-be two states in $N=6$ and the position
observed in $N=7$. The behavior of the Binder cumulant reflects this fact
(see Figure \ref{BINDER02}). We observe a rapid growth of $V_L^{\rm {min}}$
for small lattices sizes, apparently towards $2/3$ (second order value).
This growth stops when the splitting of the two peaks is observed.
The splitting of the two peaks is reflected by smaller values in
$V_L^{\rm {min}}$. We remark that the errors quoted for $V_L^{\rm {min}}$
are calculated taking into account the indetermination in the
value of the minimum, but not the displacement in the position of the
coupling where the minimum appears.

In our opinion, mainly two reasons can give account of this behavior. 
The first one comes from general grounds: at the very
asymptotic region of a first order phase transition, the 
energy jump gets sharper and sharper, and a slightly increase of the
latent heat could be expected. The second one would be
the increasing restoration of homogeneity in the hypersphere when
increasing the lattice size. The latent heat observed in small $N$
might be affected by the inhomogeneity of the hyper-sphere. 

At this point we cannot give a single reason for this to happen.
We postpone a stronger conclusion to the section devoted to the Finite
Size Scaling discussion.

The value of the latent heat when obtained with
a cubic spline fit to the peaks in $N=8$ is $C_{\rm {lat}} = 0.064(2)$.
Results obtained with mixed hot-cold starts in $N=9$ seems to give
an energy jump around 0.067. However the reliability of this
method is very limited and we do not dare to extract strong conclusions 
from it.
Taking into account that he cubic spline at $N=7$ gives 0.053(2) a possible
scenario would be a slowly increasing  latent heat towards its
asymptotic value $C_{\rm {lat}}(\infty)$.

The peak of the specific heat for the different lattice sizes is displayed
in Figure \ref{PEAKS02}.
The continuous line represents the $FS$ extrapolation.
The plot of the peak value, $C_v^{\rm {max}}(N_{\rm p})$ as a function of the
plaquette number reveals a linear relationship, and hence a first order
character.

We have also run simulations at $\gamma = +0.1$ in order to get
an estimation for the latent heat. The results, being qualitatively
similar to those encountered at $\gamma=+0.2$ are not reported in detail.
We plot the energy distribution in Figure \ref{HISTO01}.

\subsection{Results at $\gamma =0$ on the spherical topology}

Let us first situate the status of the studies using the toroidal topology
with the Wilson action.
As we have pointed out in the introduction, the transition was believed
to be continuous till the simulation of a $L=6$ lattice revealed
the first signs of the existence of two metastable states \cite{JER}. 
Numerical simulations up to $L=10$ showed a rapid decrease of the latent
heat when increasing the lattice size, suggesting that the two metastable
states might superimpose in the limit $L \rightarrow \infty$ \cite{VIC}.
However, in \cite{VIC2}, Azcoiti et al. suggest that the latent heat
starts to stabilize at $L \sim 12,14$. 

When one tries to simulate large lattices the tunneling becomes scarce and
the use of multicanonical simulations for those lattice sizes seems to be
in order. \-However, in spite of its usefulness in spin models, 
as far as we know, there are no
multicanonical simulation showing a substancial flip-flop
rate improvement for this model.

Lattice sizes up to $L=16$ have also been studied using multihistogramming
techniques \cite{UBE} and RG approaches \cite{HAS,ALF}. The results
support the idea of a quasi stable latent heat for $L > 12$.

Topological considerations stressed already
in the introduction led some authors to use lattices homotopic 
to the sphere.
In the case of the Wilson action ($\gamma = 0$) the two-state signal
is absent up to $N=10$ \cite{JCN}. Following this, we have simulated
on the spherical topology in lattices ranging from $N=6$ to $N=14$,
finding that the two-state signal sets in from $N=12$ on. 

\begin{figure}[!t]
\centerline{
\fpsxsize=7cm
\def\fpsangle{90}
\fpsbox[70 90 579 760]{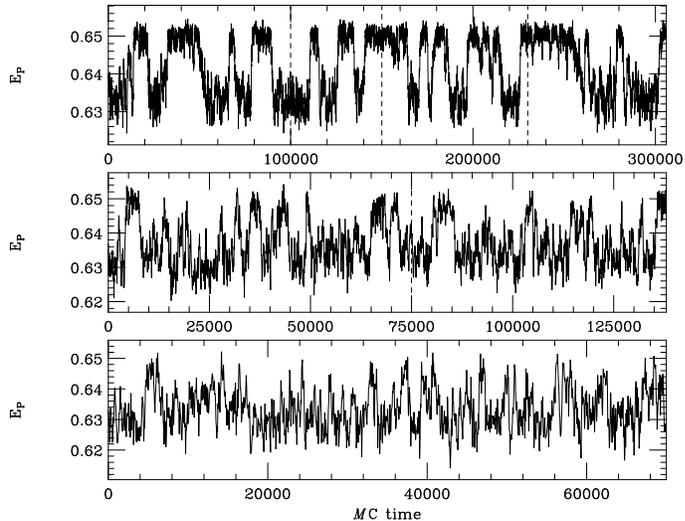} }
\caption{\small {MC evolution of $E_{\rm p}$ at $\gamma=0$ on the $\HS$
topology for $N=10$ (lower window), $N=12$ (middle) and $N=14$ (top).
The vertical dashed lines separate different runs}}
\label{EVOL0}
\end{figure}

The MC evolution for $N=10,12,14$ is plotted in Figure \ref{EVOL0}.
For the $N=14$ lattice we have four independent runs, signaled
by the dashed lines in the figure, all of them giving the same 
predictions.

\begin{figure}[!b]
\centerline{
\fpsxsize=7cm
\def\fpsangle{90}
\fpsbox[70 90 579 760]{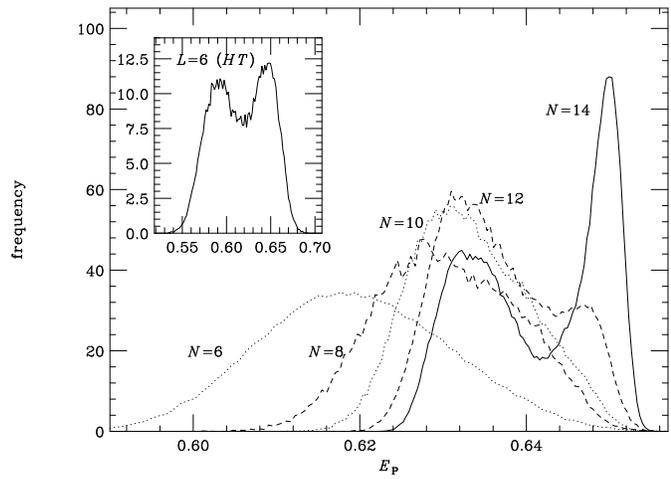} }
\caption{\small {$E_{\rm p}$ distribution at $\gamma=0$ on the $\HS$
topology. The small window corresponds to our simulation in the $\HT$
topology in L=6 at $\beta = 1.0020$.}}
\label{HISTO0}
\end{figure}

In Figure \ref{HISTO0} the distribution of $E_{\rm p}$ is plotted in lattices
ranging from $N=6$ to $N=14$. We observe that a two-peak structure is
revealed first time by the histogram in $N=12$, which has a 
$L_{\rm {eff}} \sim 19$. On the toroidal topology the equivalent 
signal is observed already at $L=6$ (see small window in Figure \ref{HISTO0}).

\begin{figure}[!b]
\centerline{
\fpsxsize=7cm
\def\fpsangle{90}
\fpsbox[70 90 579 760]{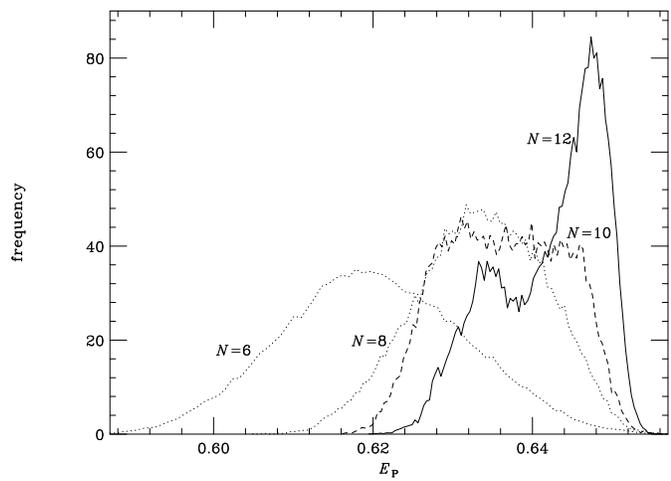} }
\caption{\small {$E_{\rm p}$ distribution at $\gamma=0$ on the $\HS$
topology simulating the full group U(1).}}
\label{HISTO0O}
\end{figure}

\begin{figure}[!t]
\centerline{
\fpsxsize=7cm
\def\fpsangle{90}
\fpsbox[70 90 579 760]{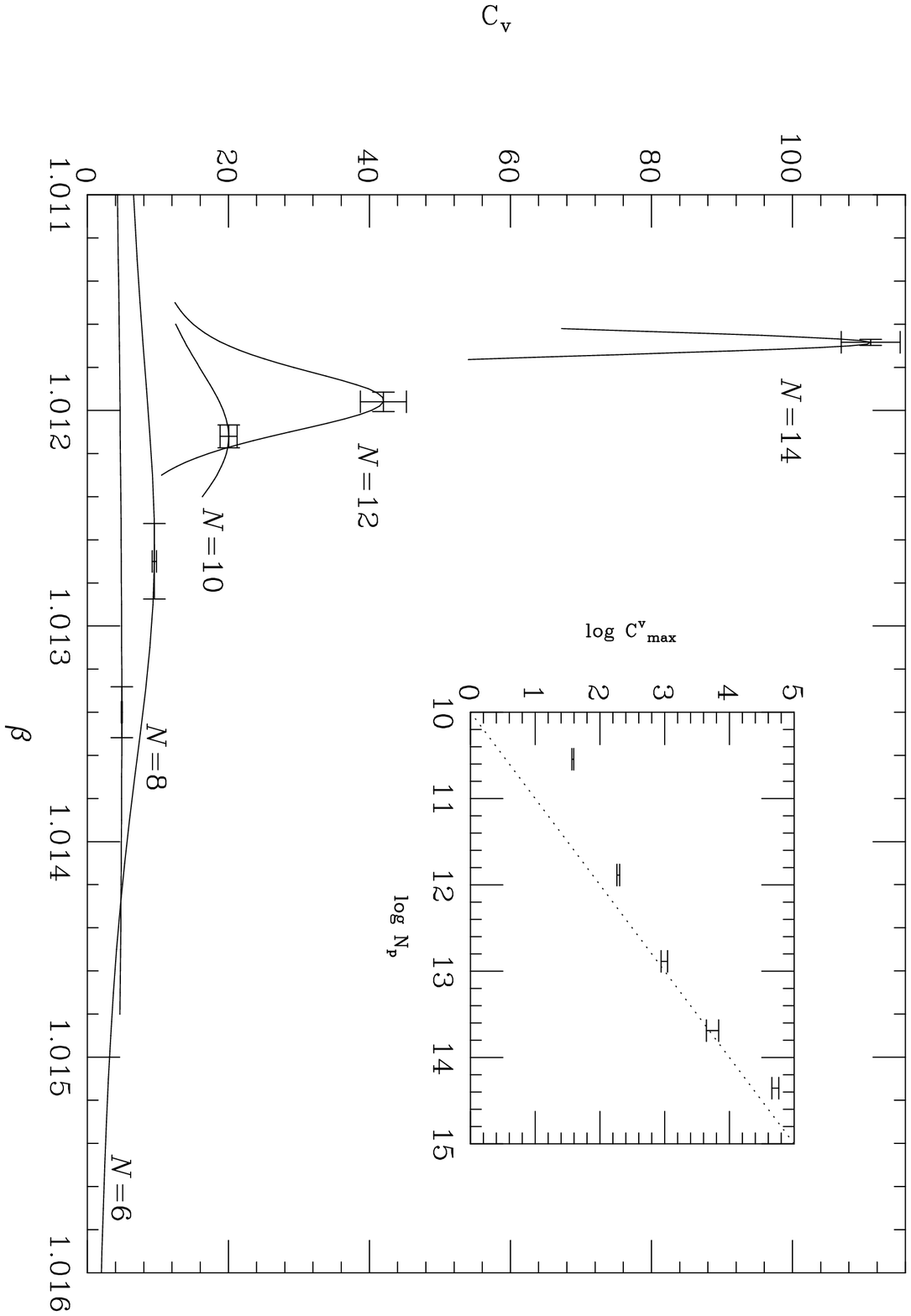} }
\caption{\small {Specific heat maximum, and Ferrenberg-Swendsen extrapolation 
(solid line) at $\gamma = 0$ on the $\HS$ topology. 
The small window represents $C_{\rm v}^{\rm {max}} (N_{\rm p})$ 
The dotted line corresponds to the slope expected in a
first order phase transition. }}
\label{PEAKS0}
\end{figure}

\begin{figure}[!]
\centerline{
\fpsxsize=7cm
\def\fpsangle{90}
\fpsbox[70 90 579 760]{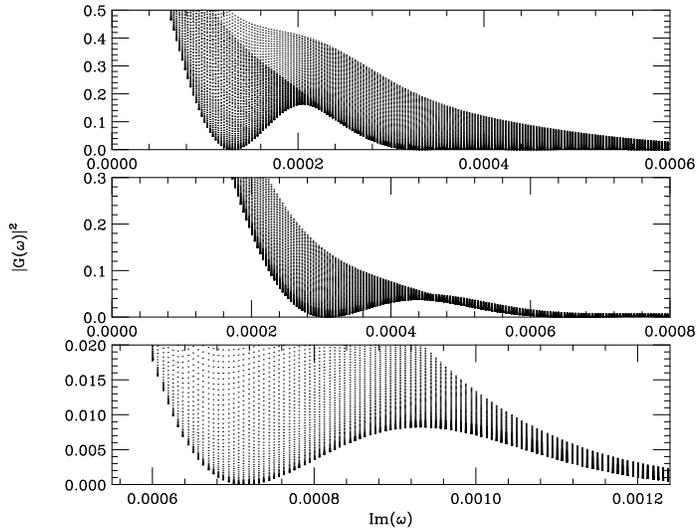} }
\caption{\small {Plot of (\protect\ref{F2}) for $N=10$ (lower window), 
$N=12$ (middle) and $N=14$ (top) at $\gamma = 0$ in the $\HS$ topology.}}
\label{FISHER0}
\end{figure}

\begin{figure}[!]
\centerline{
\fpsxsize=7cm
\def\fpsangle{90}
\fpsbox[70 90 579 760]{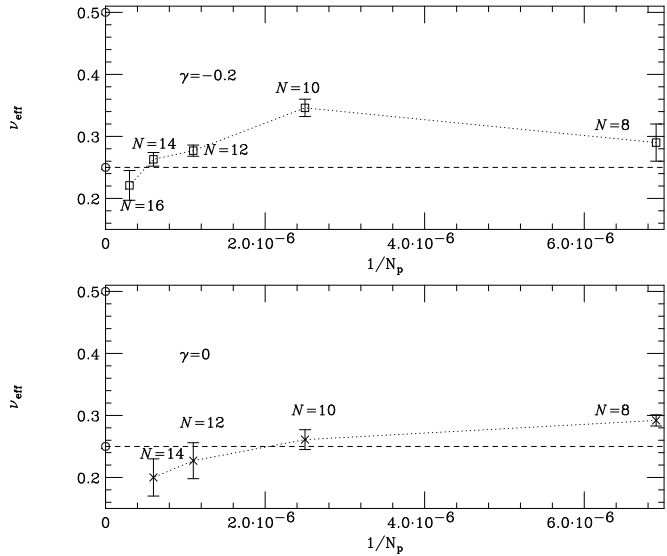} }
\caption{\small {Effective exponent $\nu$ at $\gamma = 0$ 
(lower window), and at
$\gamma = -0.2$ (upper window) on the spherical topology. In the smaller
lattices an $\nu_{\rm {eff}} \sim 1/3$ is observed which becomes $1/d$
when large enough lattices are considered.}}
\label{NUESFERA}
\end{figure}

\begin{figure}[!]
\centerline{
\fpsxsize=7cm
\def\fpsangle{90}
\fpsbox[70 90 579 760]{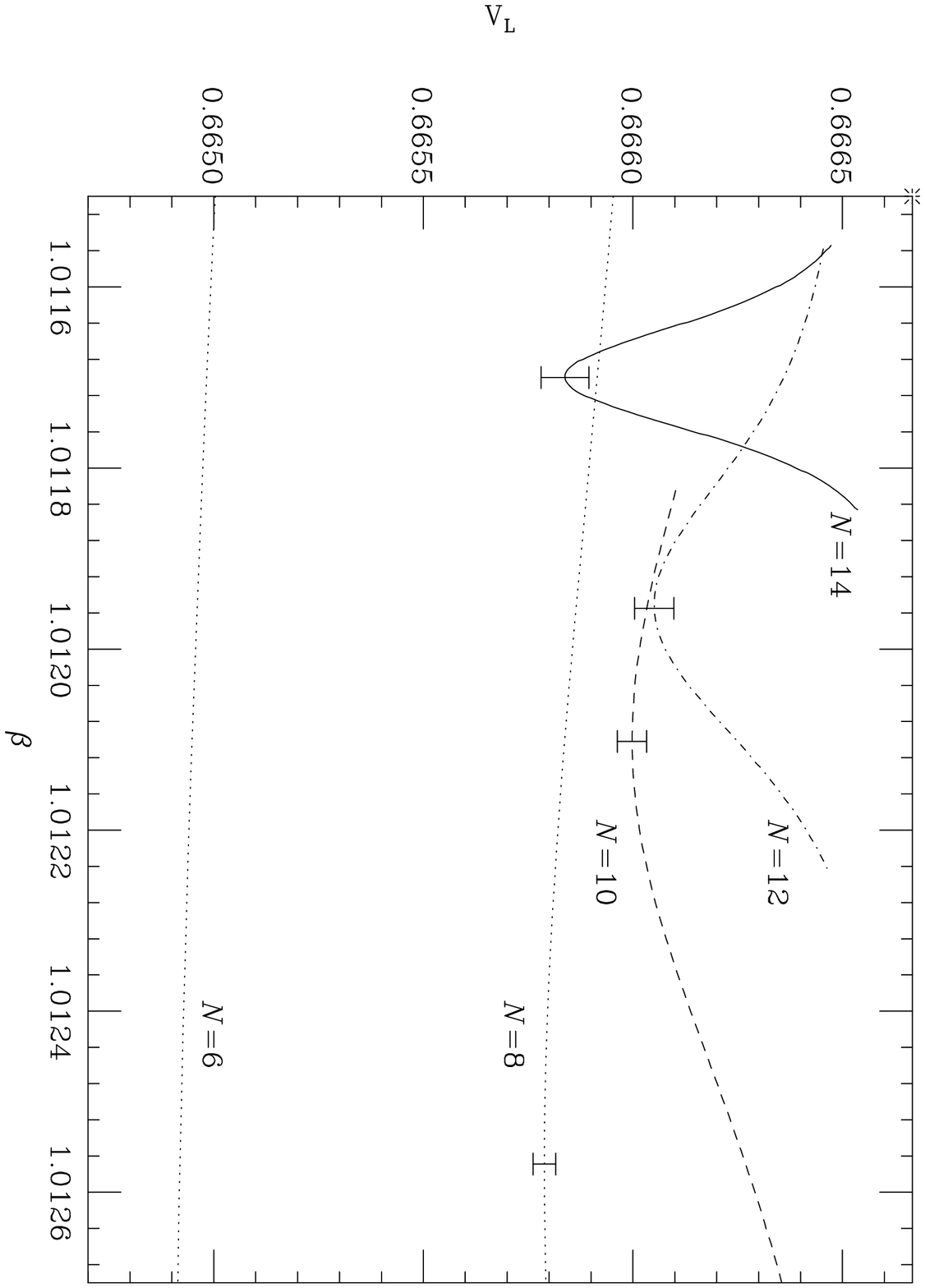} }
\caption{\small {Binder cumulant at $\gamma = 0$ (Wilson action)
on the $\HS$ topology in $N=7,8,10,12,14$. The cross in the upper corner 
signals the second order value $2/3$.}}
\label{BINDER0}
\end{figure}

Having in mind the results of the previous section at $\gamma = +0.2$,
it seems that the minimum $L_{\rm {eff}}$ required 
to observe a two-peak structure
in the spherical topology is around three times the minimum lattice size $L$
needed in the torus to observe two peaks. 

The lattice sizes used in \cite{JCN} at $\gamma =0 $
ranged from $N=4$ to $N=10$, so it is not surprising that they did not see
any two-peak structure. Our results are in that sense compatible with
theirs, though ours show a faster divergence
for $C_{\rm v}^{\rm {max}}(N)$ with increasing $N$ than the one
observed in \cite{JCN} in the lattice sizes we share (N=6,8,10).
However we are particularly confident on this point because our
simulation has been performed closer to the peak of the specific
heat $\beta^{\ast}(L)$, and so we expect the Ferrenberg-Swendsen
extrapolation to be more precise.

For the sake of comparison with the full group U(1),
simulations with the Wilson action have been performed at $\gamma=0$.
At this coupling an implementation of overrelax is possible,
and the global decorrelating effect should manifest itself in a better
statistical quality.

In Figure \ref{HISTO0O} we show the histograms of $E_{\rm p}$ from
$N=6$ to $N=12$ simulating the full group. The results are
fully compatible with those obtained from simulations with $Z(1024)$.

This test being performed, we go back to the description of
the results obtained for the discrete group.

In Figure \ref{PEAKS0} we plot $C_{\rm v}(N)$ for different $N^{\prime}$s,
together with the Ferrenberg-Swendsen extrapolation. The small
window is a $\log - \log$ plot of $C_v^{\rm {max}}(N)$ as a function
of $N_{\rm p}$. 
A linear behavior is observed from $N=10$ on, which is even faster
than linear when $N=14$ is taken into account. This fact by itself implies
the first order character of the transition since it means that the
maximum in the energy fluctuation has the size of the volume.
As a further check we have measured the $\nu$ exponent from the scaling
of the Fisher zeroes (see Table \ref{TABLA_ESFERA}).

In Figure \ref{FISHER0} we plot $|G(\omega)|^2$ for several lattice
sizes on the spherical topology at $\gamma=0$. The different curves
stand for the different $\omega$ we extrapolate.
For small lattice sizes $Im(\omega_0)$ is larger, 
and the damping is more severe
than for the larger lattices since in the later the imaginary part
contributes with a faster oscillating function. Actually, we observe
that for $N=14$ even a second minimum could be measured before the signal
is damped, while for $N=10$ one can measure accurately only the first
one.

The results for the
effective $\nu$ are plotted in Figure \ref{NUESFERA} lower window.
As could be expected from the behavior of the specific heat, for small
lattices the effective exponent is somewhat larger than $0.25$. It
gets compatible with the first order value from $N=10$ on. In the largest
lattices the distance between the two peaks slightly increases, and we
measure a $\nu_{\rm {eff}}$ slightly smaller than the first order value.
As a $\nu < 0.25$ is strictly impossible we expect this to be a transient
effect due to finite size effects associated to the observed splitting 
up of the two peaks.

Concerning the latent heat, we observe a behavior completely analogous
to the one observed at $\gamma = +0.2$. We do observe a two
peak structure quite stable when comparing $N=12$ and $N=14$,
but it slightly increases in $N=14$. The plot of the Binder cumulant
reflects again this fact (see Figure \ref{BINDER0}). The fast growth
of $V_L^{\rm {min}}$ towards $2/3$ is preempted by the onset of
double peaked histograms from $N=12$ on, and it even decreases in $N=14$.

The cubic spline fit in $N=14$ at the peaks gives 
a latent heat $C_{\rm {lat}} \approx 0.018(2)$.
which is compatible with the results suggested by extrapolating
to infinite volume the values obtained on the torus up to $L=14$. 
Our results are hence supporting the conjecture stressed in \cite{VIC2}
about a quasi stabilization of the latent heat on the torus
from $L=12$ on.

\begin{figure}[!t]
\centerline{
\fpsxsize=7cm
\def\fpsangle{90}
\fpsbox[70 90 579 760]{evoltoro02.ps} }
\caption{\small {MC evolutions at $\gamma = -0.2$ in L=12,16,20 on the
$\HT$ topology.}}
\label{EVOLTORO02}
\end{figure}

\begin{figure}[!]
\centerline{
\fpsxsize=7cm
\def\fpsangle{90}
\fpsbox[70 90 579 760]{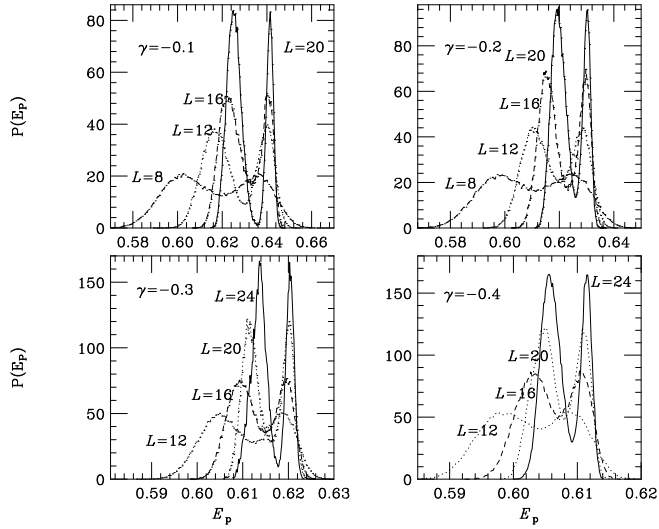} }
\caption{\small {$E_{\rm p}$ distributions at the different 
negative $\gamma$ on the
$\HT$ topology. We have run up to $L=20$ at $\gamma=-0.1,-0.2$ and up to
$L=24$ at $\gamma=-0.3,-0.4$.}}
\label{HISTO_TORO}
\end{figure}

\begin{figure}[!]
\centerline{
\fpsxsize=7cm
\def\fpsangle{270}
\fpsbox[70 90 579 760]{figure14.ps} }
\caption{\small {Free energy gap $\Delta F(L)$ (lower window) 
and latent heat (upper window) for the different negative $\gamma$ in 
the $\HT$ topology.}}
\label{GAP_LAT}
\end{figure}

\begin{figure}[!]
\centerline{
\fpsxsize=7cm
\def\fpsangle{90}
\fpsbox[70 90 579 760]{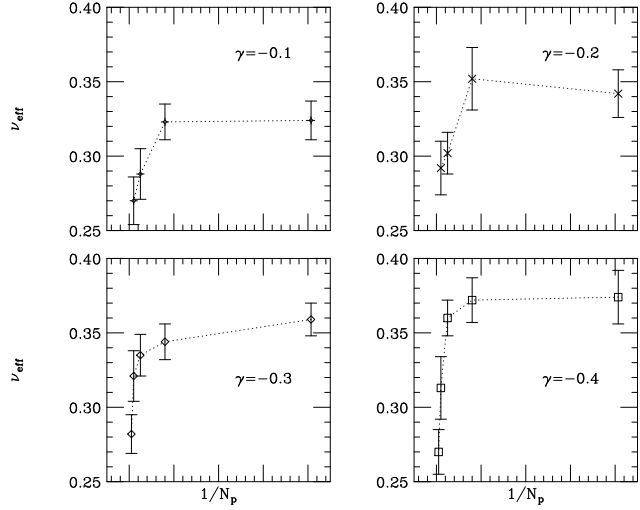} }
\caption{\small {Effective $\nu$ exponent for the different 
negative $\gamma$ in the $\HT$ topology.}}
\label{NU_TORO}
\end{figure}

\begin{figure}[!t]
\centerline{
\fpsxsize=7cm
\def\fpsangle{90}
\fpsbox[70 90 579 760]{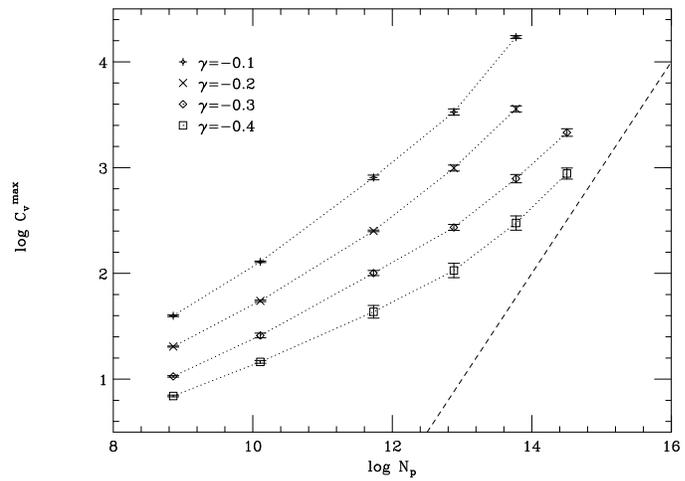} }
\caption{\small {$C_v^{\rm {max}}(L)$ as a function of $N_{\rm p}$ 
for the different negative $\gamma$ values on the $\HT$ topology. 
We have used $\log - \log$ scale 
for the clarity of the
graphic sake. The dotted line corresponds to the first order slope.}}
\label{PEAKS_TORO}
\end{figure}

\subsection{Results for $\gamma < 0$}

\subsubsection{Toroidal topology}

We have performed a systematic study of the transition at several
negative values of $\gamma$ on the toroidal topology.  

We find that the two-state signal persists for all $\gamma$ values
we consider. 

The MC evolution at $\gamma=-0.2$ is plotted 
for several lattices sizes in figure \ref{EVOLTORO02}.
As a general observation the more negative is $\gamma$,
the flip rate becomes scarce for increasing
lattice size. 
As an example, at $\gamma = -0.4$ the flip rate
in $L=20$ is comparable to the one observed in $L=16$ at $\gamma=-0.2$.

The energy histograms reveal an increasing weakness
of the transition when going to more negative $\gamma$ values
(see Figure \ref{HISTO_TORO}). A double peak structure is clearly
visible at $\gamma = -0.1$ in $L=8$, while at $\gamma = -0.4$ one has
to go to $L=12$ to observe an equivalent signal.

In what concerns the behavior of the free energy gap $\Delta F(L)$,
it grows for all investigated lattice sizes at all negative $\gamma$ values
(see Figure \ref{GAP_LAT} lower window) signaling the first order character
of the transition.
The value of $L$ at which $\Delta F(L)$
starts growing is certainly larger as the value of $\gamma$ is more negative.
This is another test of the increasing weakness of the transition as
$\gamma$ gets more negative. 

The statistics performed on the torus are reported in 
Table \ref{TABLA_TORO}. We also quote for the different 
negative $\gamma$, 
the value of $\beta^{\ast}(L)$ and the position of partition function
zero closest to the real axis. We have computed
from the imaginary part of the zeroes the effective
$\nu$ exponent between consecutive lattice sizes following (\ref{NU}).

In Figure \ref{NU_TORO} we plot for the different $\gamma$ values
the $\nu_{\rm {eff}}$ we measure. In all cases a $\nu_{\rm {eff}} \sim 1/3$
is observed for small lattice sizes, which gets closer to 0.25 when
the lattice is large enough. From this figure the trend of
$\nu_{\rm {eff}}$ seems rather clear towards the first order value.

From the energy distributions we measure the latent heat through a
cubic spline at the peaks.
Taking into account the value $\nu_{\rm {eff}}=0.25$ we measure,
we plotted it as a function of the inverse of the
volume $L^{-4}$, which is also the expected behavior of the latent
heat when the transition is first order.
The latent heat can be extrapolated to a value which
is safely far from zero (see Figure \ref{GAP_LAT} upper window).

In Figure \ref{PEAKS_TORO} we plot $C_v^{\rm {max}}(L)$ as a function
of $N_{\rm p}$. As could be expected from the behavior of the effective
exponent $\nu$, the maximum of the specific heat for small lattices diverges
slower than the volume. For the smaller lattices the effective 
exponent is $\alpha/\nu \sim 1.4$. This value increases monotonically
with the lattice size. In the largest ones we observe $\alpha/\nu \sim 3.5$.

\subsubsection{Spherical topology}

On the toroidal topology we have found that the minimum lattice
size required to observe a two state signal is obtained through
an apropriate combination $(\gamma,L_{\rm {min}})$, 
with increasing $L_{\rm {min}}$ for increasingly negative $\gamma$,
the behavior being qualitatively similar for all the $\gamma$ values
we have investigated. In view of this, we have studied a single
$\gamma < 0$ value on the spherical lattice to check if the
two state signal is absent on this topology. For the sake of comparison
with \cite{CN,JCN} we choose this value to be $\gamma = -0.2$.

\begin{figure}[!]
\centerline{
\fpsxsize=7cm
\def\fpsangle{90}
\fpsbox[70 90 579 760]{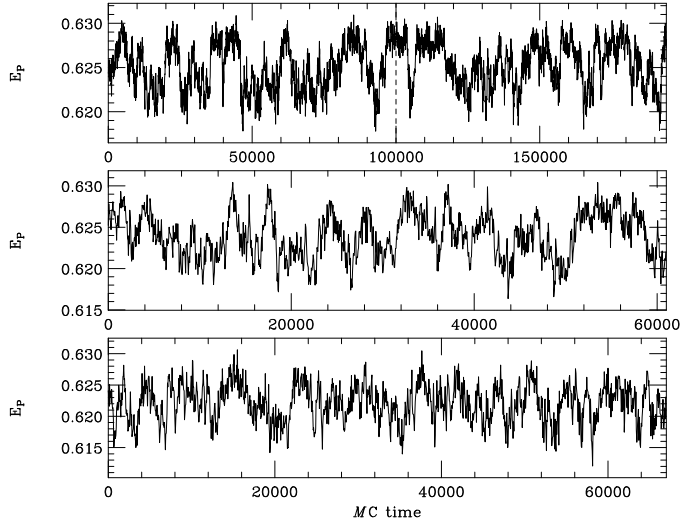} }
\caption{\small {MC evolution of $E_{\rm p}$ in $N=12$ (lower window), $N=14$ 
(middle window) and $N=16$ (upper window) at $\gamma = -0.2$ on 
the $\HS$ topology. The two different runs in $N=16$ are separated by
a vertical dashed line.}}
\label{EVOL_02}
\end{figure}

We have run simulations on spheres ranging from $N=6$ ($L_{\rm {eff}} \sim 8$)
to $N=16$ ($L_{\rm {eff}} \sim 26$). The MC evolution is plotted in 
Figure \ref{EVOL_02}.

\begin{figure}[!]
\centerline{
\fpsxsize=7cm
\def\fpsangle{90}
\fpsbox[70 90 579 760]{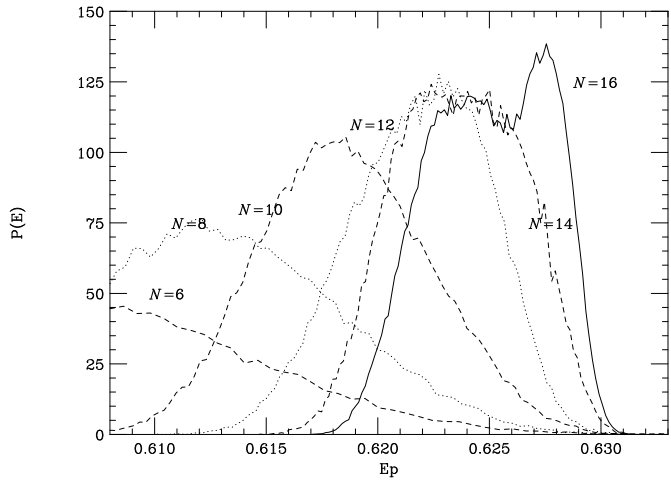} }
\caption{\small {$E_{\rm p}$ distributions at $\gamma = -0.2$ on 
the $\HS$ topology.}}
\label{HISTO_02}
\end{figure}

The distribution in $N=14$ is distinctly
non gaussian and the splitting of the peaks occurs in $N=16$ (see Figure
\ref{HISTO_02}). We remark that the simulations in $N=~16$ are extremely
expensive in CPU. In order to alleviate thermalization we have parallelized
the code using shared memory in two PPro processors. In this lattice we
have run two independent simulations starting from different configurations
the results being fully compatible.

\begin{figure}[!]
\centerline{
\fpsxsize=7cm
\def\fpsangle{90}
\fpsbox[70 90 579 760]{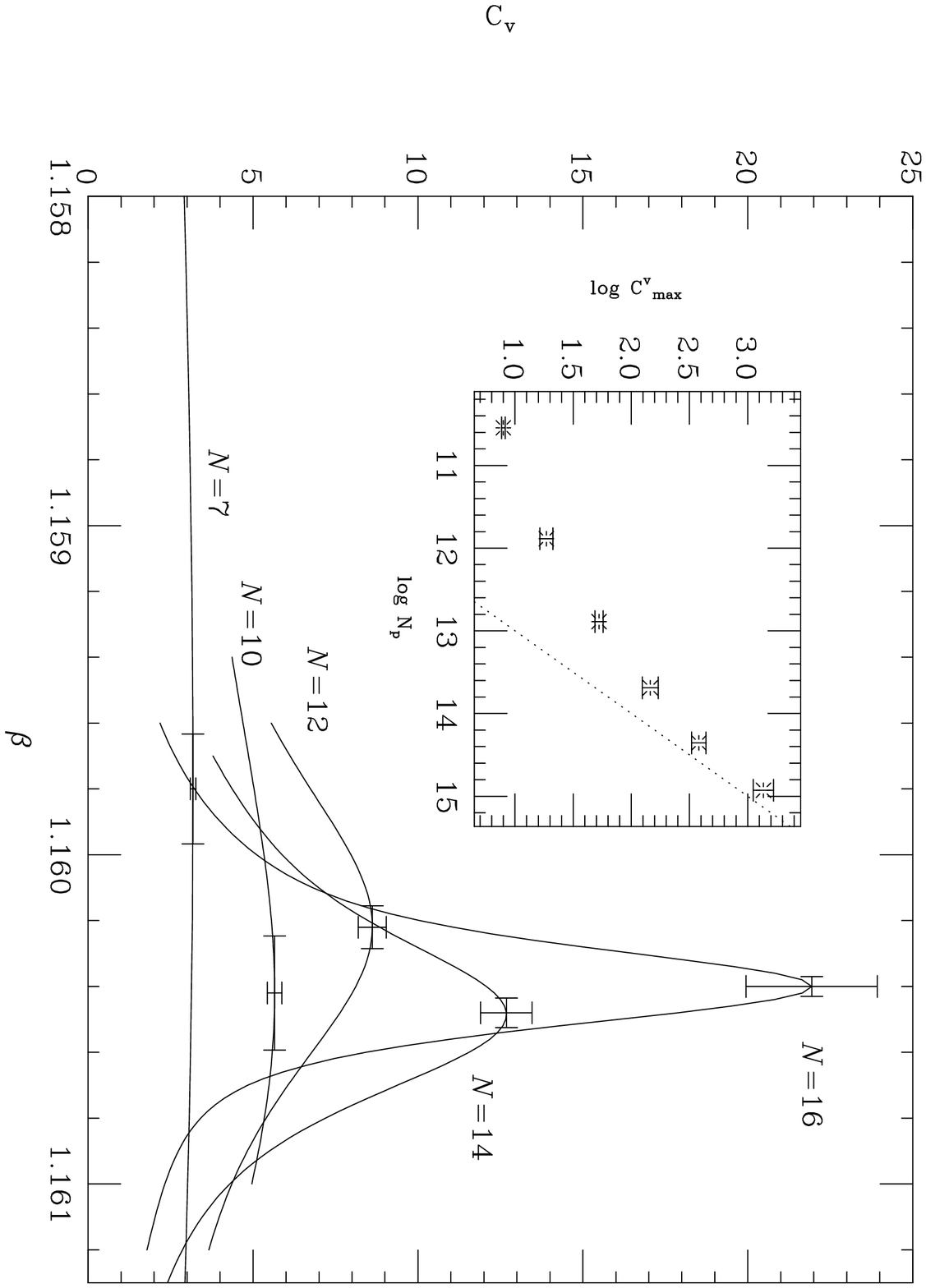} }
\caption{\small {Specific heat maximum, and Ferrenberg-Swendsen extrapolation 
(solid line) at $\gamma = -0.2$. 
The small window represents $C_{\rm v}^{\rm {max}} (N_{\rm p})$ 
The dotted line correspond to the slope expected in a
first order phase transition.}}
\label{PEAKS_02}
\end{figure}

\begin{figure}[!]
\centerline{
\fpsxsize=7cm
\def\fpsangle{90}
\fpsbox[70 90 579 760]{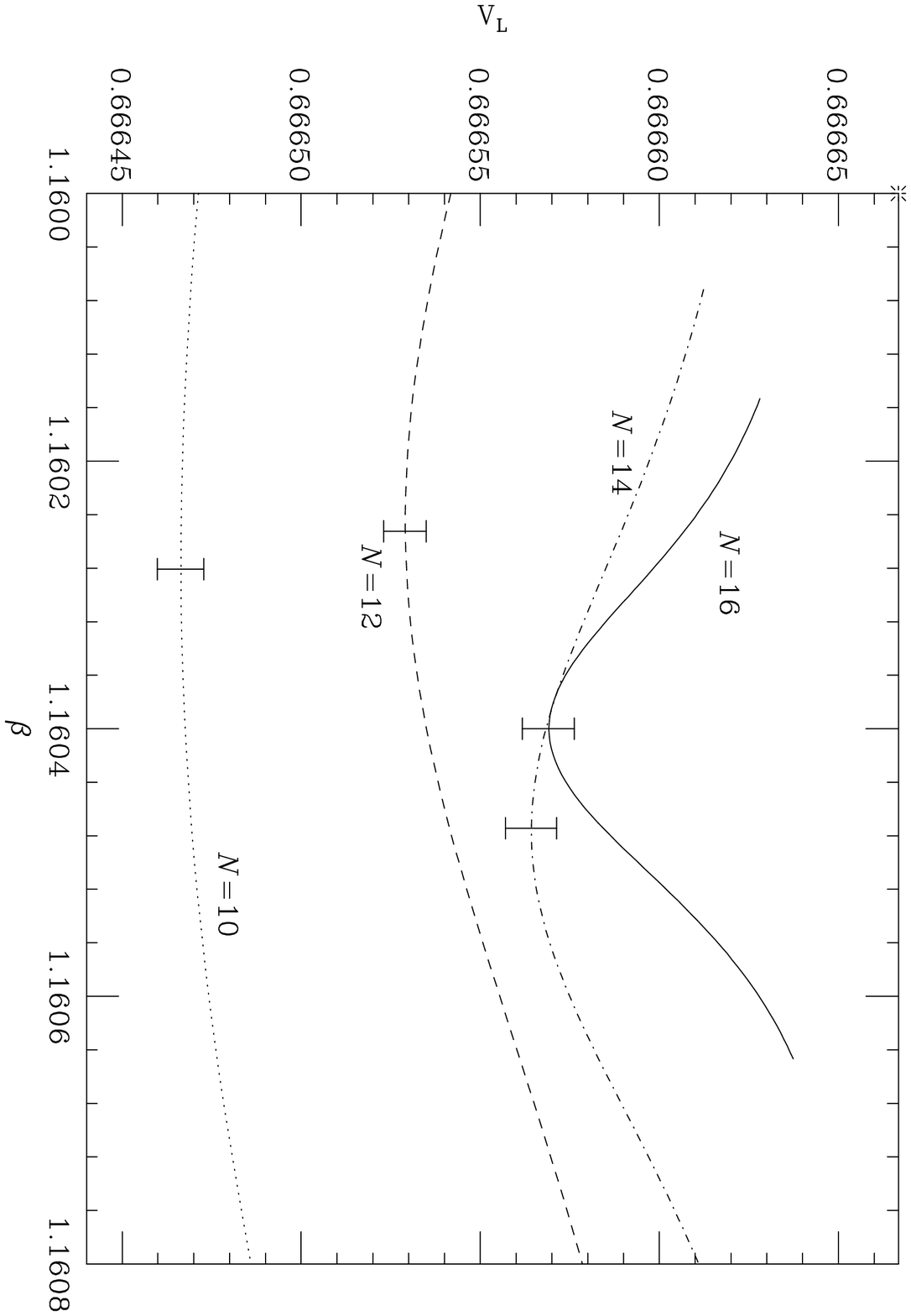} }
\caption{\small {Binder cumulant at $\gamma = -0.2$
on the $\HS$ topology in $N=~10,12,14,16$. The cross in the upper corner 
signals the second order value $2/3$.}}
\label{BINDER_02}
\end{figure}

The behavior of the maximum of the specific heat is shown in Figure 
\ref{PEAKS_02}. We observe the same trend than in the previous values
of $\gamma$ in the sphere, i.e., an increasingly fast divergence
of the specific heat with the lattice size, showing an effective
$\alpha/\nu \sim 4$ already when the histogram width becomes constant 
with increasing lattice size. 

At this value of $\gamma$ we have a worse estimation for the latent
heat to be, as the two peaks have not split enough to allow an accurate
measurement.

Again the behavior of the Binder cumulant is very significant 
(see Figure \ref{BINDER_02}. The
value $V_L^{\rm {min}}$ shows a very fast trend towards 2/3 in the small
lattices. When increasing the lattice size the rate gets slower, and finally
the value in $N=16$ is compatible with the value in $N=14$ preempting
the extrapolation to $2/3$. Unfortunately lattices larger than $N=16$
are unaccessible to our computers nowadays, and we cannot observe
a decreasing $V_L^{\rm {min}}$ as we did for the other $\gamma$ values.
However from the behavior between $N=14$ and $N=16$ an increasing
$V_L^{\rm {min}}$ for larger spheres seems to be very unlikely.

Concerning the effective critical exponent $\nu$ we have measured
the position of the first Fisher zero (see Table \ref{TABLA_ESFERA})
and compute $\nu_{\rm {eff}}$ (see Figure \ref{NUESFERA}). Again the
first order value $1/d$ is reached in the largest lattices.

\subsection{Toroidal versus Spherical topology}

\begin{figure}[!b]
\centerline{
\fpsxsize=7cm
\def\fpsangle{90}
\fpsbox[70 90 579 760]{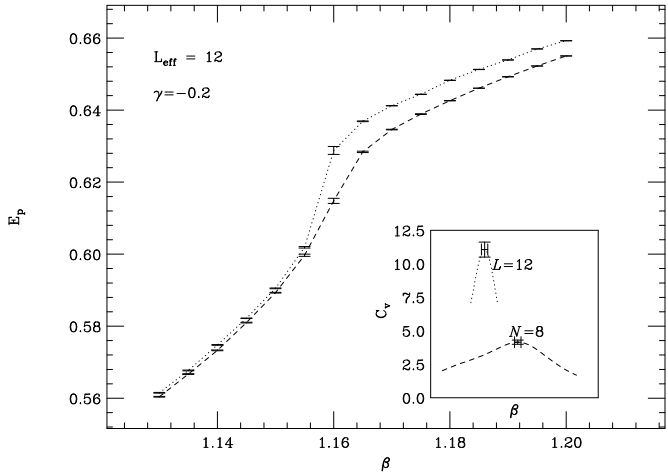} }
\caption{\small {$E_{\rm p}$ at $\gamma=-0.2$ in 
$L_{\rm {eff}} \sim 12$. The 
dotted line corresponds to the toroidal topology, the dashed one to the 
sphere $N=8$. The small window shows the difference in the 
specific heat between both cases.}}
\label{SALTO}
\end{figure}

\begin{figure}[!b]
\centerline{
\fpsxsize=7cm
\def\fpsangle{90}
\fpsbox[70 90 579 760]{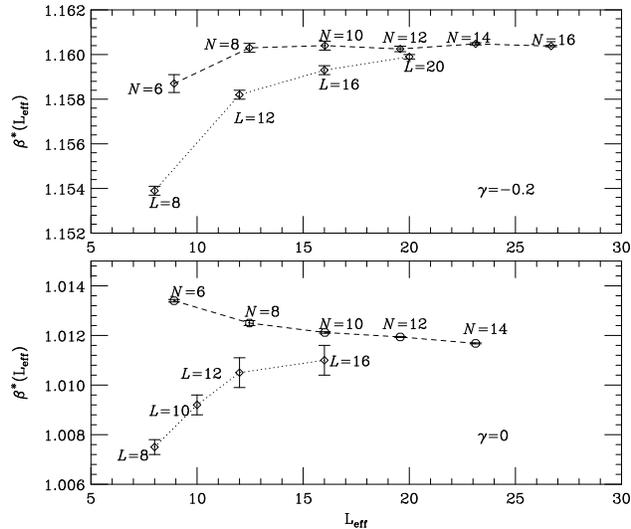} }
\caption{\small
{$\beta^{\ast}(L)$ for the different lattice sizes on the torus (dotted lines) 
and on the sphere (dashed lines) at $\gamma=0$ 
(lower window) and at $\gamma = -0.2$ 
(upper window). The couplings for $\gamma = 0$ on the torus have been taken
from \protect\cite{ALF}.}}
\label{BETACRIT}
\end{figure}

\begin{figure}[!t]
\centerline{
\fpsxsize=7cm
\def\fpsangle{90}
\fpsbox[70 90 579 760]{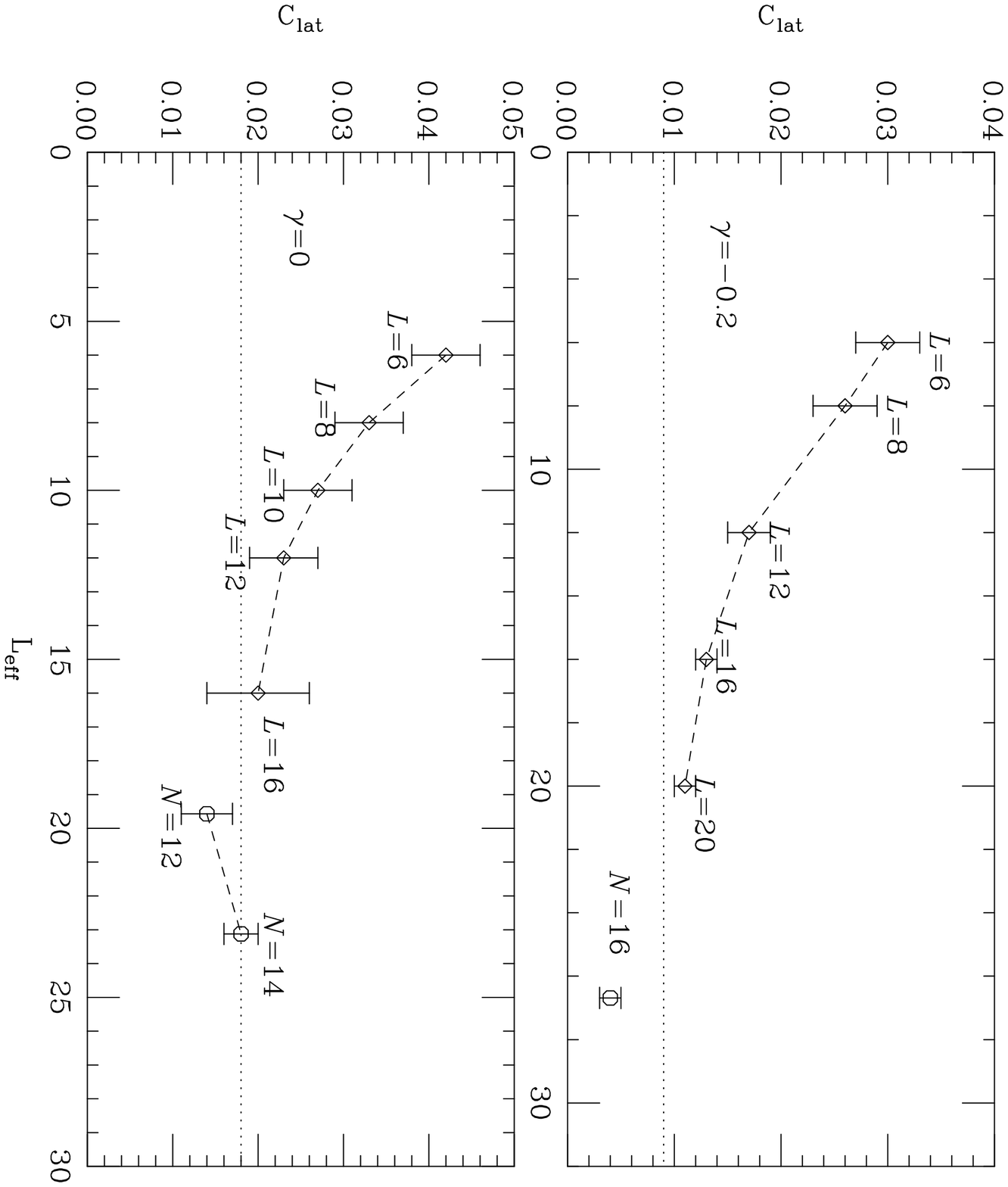} }
\caption{\small 
{Latent heat for the different lattice sizes on the torus and on the
sphere at $\gamma=0$ (lower window) and at $\gamma = -0.2$ (upper window).
The dotted line corresponds to $C_{\rm {lat}}(\infty)$ 
obtained by extrapolating
the values on the torus through a linear fit as function of $1/V$.
The values for $\gamma = 0$ on the torus have been taken
from \protect\cite{ALF}.}}
\label{LATCOMPA}
\end{figure}

Due to the fact that on the sphere there are a number of points
with less than the maximum connectivity, violations to standard
FSS in the form of uncontrolled finite size effects are expected
to appear on this topology. 

We have observed all along this work that working on the spherical
topology retards the onset of two-state signals. 
In terms of $L_{\rm {eff}}$, 
some qualitative prediction arising from our results would be that
at fixed $\gamma$, the minimum lattice
size required to observe a double peak structure, $L_{\rm {min}}$, is
around three times larger on the sphere than on the torus.
This one being the showier difference, is not certainly the only one.

When comparing $C_{\rm v}^{\rm {max}}(L_{\rm {eff}})$
on both topologies one finds always smaller values on the spherical
topology (see small window in Figure \ref{SALTO}).
We did run on $L=12$ on the torus and $N=8$ on the sphere, which
has an $L_{\rm {eff}} \sim 12$, sweeping an interval
of $\beta$ values including the phase transition. In Figure \ref{SALTO}
we have plotted $E_{\rm p}$ in both topologies.
In the region of low $\beta$ the system is disordered,
the entropy is higher and the system is
not so sensitive to inhomogeneities of the lattice. However, in the
region of high $\beta$ the energy is in general smaller than for the
homogeneous lattice since the system tends to be ordered, and the influence
of the sites with less than maximum connectivity is more evident.
This difference gets smaller when the lattice size is increased on the
sphere, and hence the energy jump is larger. This feature explains
the observed splitting of the two states on the spherical
lattices when increasing lattice size.

Altogether, it is not recommended to work on spherical lattices to check
the existence or not of two-state signals. However, there are a number
of facts that make this topology not so dissapointing in spite of those
uncontrolled finite size effects.
The first one concerns the behavior of $\beta^{\ast}(L)$ and the second
one the measure of the latent heat.

In Figure \ref{BETACRIT} we compare $\beta^{\ast}(L)$ on the torus
and on the sphere. A first observation is that despite they not
having the same values in finite lattices, both curves get 
closer when increasing the lattice size. This supports the idea
of a common thermodynamic limit for both topologies.
On the other hand, the shift in the apparent critical coupling with $L$ is much
less dramatic on the sphere than on the torus. It seems that finite size
corrections to $\beta_{\rm c}$ are smaller in the case of the spherical
topology. This behavior has also been observed in the Ising model
on spherical lattices \cite{JESUS}.

The behavior of the latent heat
is very significant. In Figure \ref{LATCOMPA} we compare the 
latent heat in finite volumes for both topologies. On the Wilson line,
at $\gamma = 0$ the asymptotic value of the latent heat is obtained
in the spherical topology. The inhomogeneity of the sphere has,
paradoxically, helped us to run lattices with $L_{\rm {eff}} \sim 24$
without having to worry about the tunneling rate. From this graph we
quote $C_{\rm {lat}} (\infty) = 0.018(1)$.
In the upper window the same is plotted for $\gamma = -0.2$. Unfortunately
we have just a single lattice size on the sphere to measure the latent
heat, however, from the behavior exhibited by this topology, 
an increase of the latent heat on spheres larger than $N=16$
seems rather likely.
Altogether, at $\gamma = -0.2$ a $C_{\rm {lat}} (\infty) \sim 0.009$
is plausible.

In what concerns the latent heat,
the spherical topology seems to afford an useful complement to the results
obtained for the the toroidal lattice. In fact, on the Wilson line
lattices larger than $L=16$ cannot be studied due to the technical problem
associated to the scarce tunneling. The spherical lattice
alleviates this technical problem, and the results
are supporting the value of the extrapolated latent heat as a 
function of the inverse of the volume, from the data on toroidal 
lattices up to $L=16$. One could consider the possibility of
simulating on spherical
lattices at negative $\gamma$  to solve the tunneling problem, but
in our opinion, the price to pay is too high.

\section{Conclusions}

The first order character of the deconfinement transition in pure
U(1) has been proved, up to the limits of a rather reasonable
numerical evidence, in the interval $\gamma \in [+0.2,-0.4]$.

In $\gamma = +0.2$ we have been able to stabilize the latent heat.
We are aware of no simulation on the torus showing a stable latent
heat due to the scarce tunneling in lattices larger than $L=6$.
The spherical topology has helped to solve this problem. However,
probably any lattice with inhomogeneities would produce the same
catalysing effects.

In $\gamma=0$ we have obeserved an increasing energy gap on the
spherical lattice, and also, we have been able to measure the 
asymptotic value of the latent heat. 
We have proved the suggestions of several authors about a quasi
stabilization on lattices larger than $L \sim 12$. The data on the
spherical topology have been crucial to discard the possibility
of a slowly vanishing latent heat. 
It follows that the discretization of pure compact U(1) LGT on
the lattice using the Wilson action exhibits a first order phase
transition with a latent heat in the thermodynamical limit
$C_{\rm {lat}} \sim 0.018$.

On the toroidal topology things happen qualitatively in the same
way than in $\gamma=0$ up to $\gamma = -0.4$. We have run on spherical
lattices in $\gamma = -0.2$ looking for an argument similar to the one
found in the Wilson case concerning the stabilization of the latent heat.
The simulation in $N=16$ gives an estimation for the latent heat though
rather imprecise because the splitting of the peaks is not good
enough for the measurement to be accurate. However, in view of the
behavior exhibited by the spherical topology concerning
the trend of the two
peaks on spherical lattices to separate, we are prone to consider the
value measured in $N=16$ as a lower bound for the latent heat
in $\gamma=-0.2$. On the other hand,
we would at present say that the latent heat extrapolated
from the data on the torus up to $L=20$ is rather accurate, in view
of the behavior for the Wilson case.
The possibility of running larger spheres surpasses our computer
resources. A highly parallelized version of the code should
be used to alleviate thermalization, which is a possibility
we do not discard completely at medium term.

As for the evolution of the latent heat along the transition
line as a function of $\gamma$. the scenario that follows from
our results is plotted in Figure \ref{LATGAM}. 

\begin{figure}[!t]
\centerline{
\fpsxsize=7cm
\def\fpsangle{90}
\fpsbox[70 90 579 760]{latenteg.ps} }
\caption{\small {Extrapolated value for the latent heat as a function
of $\gamma$. The values quoted for $\gamma < 0$ come from data
on toroidal lattices, the ones for $\gamma \geq 0$ from data
on spherical lattices.}}
\label{LATGAM}
\end{figure}

Either proving or discarding the possibility of a TCP at some
finite negative $| \gamma_{\rm {TCP}} | > 0.4$ will be a very
difficult task from the numerical point of view. An analytical
argument would be welcome.
A very tiny two-state signal, comparable with the one observed in $L=8$ at
$\gamma = -0.4$, is observed on the torus in $L=16$ at $\gamma = -0.8$,
which is the most negative value we have run on the torus.
Taking into account the factor 3 in $L_{\rm {eff}}$ one would
need a $N \sim 30$ sphere to observe a tiny double peak structure.
Simulation with spherical lattices in this range of $\gamma$ have no sense,
and nothing can be concluded from the absence of two state signals
from such negative $\gamma$ values.

In view of these difficulties to stabilize the latent heat for very
negative $\gamma$ values, the only chance to discern the order
of the phase transition is the study of effective critical exponents.
Cumulants of the energy, such as the Binder cumulant, have been
shown to behave like expected in the first order case
only when the stable two-peak structure is almost setting in, and there,
we do not need further evidences any longer. 
We could not expect additional information since by definition
the Binder cumulant
relies on the existence of stable latent heat in order to extrapolate
to $V_{\rm {L = \infty}}^{\rm {min}} < 2/3$. 

The advantage of studying the effective exponents (which is nothing
but studying the evolution of the histograms width) is that we do not
need a direct observation of latent heat to conclude that a transition
is first order \cite{SU2,ONAF}.
We have observed a $\nu_{\rm {eff}}$  which evolves
monotonically until it reaches the first order value, 
namely 0.25, in all cases.
The statistics needed to observe a monotonous behavior, 
are order $1000 \tau$ at the coupling $\beta^{\ast}(L)$, which has to be
located with high precision (four digits in our experience)
in order to accurately measure effective
exponents. 
However, one has to make every effort to observe the trend of the
effective exponents, since it is the only chance to discern the
order of such tricky transitions.

In pure U(1), for the Wilson case, an exponent $\nu \sim 1/3$ was widely
predicted \cite{LAU,BHA,BHA2} at the beginning of the eighties
when the lattice sizes where too small to reveal two peaks.
That $\nu$ was shown to become $\nu \sim 0.29$ when simulating
$L=14$ \cite{VIC2}. We have measured $\nu = 0.25$ on spherical lattices
and stated its first order character.
Within the approximation of effective potentials
it can be shown that along the transient region 
of a weak first order phase transition, everything goes
like in a second order one with a thermal exponent
$\alpha = 0.5$ \cite{LAF}. 
Together with Josephson law ($\alpha = 2 - \nu d$)
it implies $\nu \sim 0.37$. This statement has been checked in
2D Potts \cite{LAF}, 3D and 4D $O(N)$ models \cite{ISAF,ONAF,O4AF,O3AF}
and in the 4D SU(2)-Higgs at T=0 with fixed Higgs modulus \cite{SU2}.
Our results prove that pure compact U(1) theory behaves
in the same way.

\newpage

\chapter{Modelos bidimensionales de flujo de part\'{\i}culas \label{capRW}}
\thispagestyle{empty}
\markboth{\protect\small CAP\'ITULO \protect\ref{capRW}}
{\protect\small {\sl Modelos bidimensionales de flujo de part\'{\i}culas}}

\cleardoublepage        

\section{Introducci\'on}

En los \'ultimos a\~nos est\'a recibiendo una especial atenci\'on
la formulaci\'on de modelos para describir por ejemplo el paso de mensajes
a trav\'es de redes de ordenadores \cite{IEEETC4}, 
el tr\'afico en grandes ciudades \cite{EPSM4},
la adsorci\'on de mol\'eculas en un cristal \cite{EXSOL4}, conductores
de iones r\'apidos \cite{IONES4}.

Todos estos problemas tienen en com\'un que pueden ser estudiados desde
el punto de vista de la teor\'{\i}a de Random Walks. Sin embargo, el
elevado n\'umero de grados de libertad hace imposble extraer
informaci\'on relevante mediante c\'alculos anal\'{\i}ticos. La
alternativa es el uso de simulaciones num\'ericas.

Desde este punto de vista uno podr\'{\i}a pensar en formular modelos
tan pr\'oximos a la realidad como sea posible, incluyendo todos los grados
de libertad que se sea capaz de manejar. Sin embargo los modelos que
resultan aplicando esta filosof\'{\i}a son demasiado complicados de
estudiar, no s\'olo en lo que concierne a la descripci\'on de fen\'omenos
relevantes, sino tambi\'en porque carecen de poder predictivo en
general.

Una aproximaci\'on alternativa es formular modelos m\'as simples,
menos realistas pero manteniendo los rasgos fundamentales del sistema
f\'{\i}sico. Tales modelos est\'an descritos por unos pocos par\'ametros,
siendo por ello m\'as faciles de tratar y haciendo posible un estudio
global del espacio de par\'ametros sin perder intuici\'on f\'{\i}sica.

Mediante la formulaci\'on de estos modelos se pretende obtener resultados
relevantes sobre sistemas complejos, estudiando aproximaciones
relativamente simples, como se suele hacer en Mecanica Estad\'{\i}stica (ME)
donde sistemas tan complicados como redes ferromagn\'eticas comparten
muchas propiedades con modelos tan simples como el de Ising. En ME
la relaci\'on entre ambos sistemas se entiende a trav\'es del 
Grupo de Renormalizaci\'on. Desgraciadamente estamos lejos de poder
probar una relaci\'on de este tipo para modelos de flujo de part\'{\i}culas,
debido fundamentalmente a que no tienen un l\'{\i}mite termodin\'amico
bien definido. Sin embargo la modelizaci\'on de estos sistemas es necesaria.

Los problemas que se encuentran cuando se estudian sistemas que
envuelven flujo de part\'{\i}culas est\'an relacionados con procesos
de congesti\'on. Los sistemas realizan una transici\'on de una situaci\'on
de tr\'afico fluido a otra caracterizada por el atasco.

Podr\'{\i}amos utilizar las herramientas de la ME para estudiar estos
problemas si asimilamos este cambio a una transici\'on de fase.
Un primer paso es buscar las variables que gobiernan la transici\'on.

El flujo en la red est\'a condicionado por el n\'umero m\'aximo
de part\'{\i}culas que un nodo es capaz de contener, por la interacci\'on
entre las part\'{\i}culas y por la geometr\'{\i}a de la red (por ejemplo
el numero de coordinaci\'on).

En nuestro modelo hay una inyecci\'on continua de part\'{\i}culas
en la red, para cada una de ellas se elije al azar un nodo
que ser\'a su destino final. El movimiento de las part\'{\i}culas
se implementa permitiendoles moverse en la red de acuerdo con determinadas
reglas que definiremos m\'as tarde. En el tr\'afico real las
part\'{\i}culas encuentran a menudo obstaculos en su camino. Estos
obst\'aculos se simulan limitando el n\'umero de part\'{\i}culas
que un nodo de la red puede almacenar simultaneamente.

Volviendo a la discusi\'on desde el punto de vista de la ME, vamos
a enfocar la situaci\'on en la que la densidad de part\'{\i}culas
es alta. Intuitivamente es claro que una part\'{\i}cula podr\'{\i}a
evitar zonas de atasco si le permitimos rodear el obst\'aculo. El
sistema estar\'a globalmente m\'as descongestionado y esperamos
una mejora en el n\'umero de part\'{\i}culas que alcanzan finalmente
su destino. En t\'erminos de ME, la discusi\'on previa significa que
calentando el sistema ($T \neq 0$) se obtendran mejores resultados \cite{BETA4}

En lo que concierne a la interacci\'on entre las part\'{\i}culas, 
hemos introducido un t\'ermino de contacto, 
la existencia de obst\'aculos
en forma de nodos completamente ocupados. Las consecuencias de esta
interacci\'on se pueden ver como si existiera un potencial infinito
que actua en los nodos saturados. 

Un t\'ermino de interacci\'on apropiado se obtiene asignando {\sl carga
el\'ectrica} a cada part\'{\i}cula. La repulsi\'on alejar\'a a las
part\'{\i}culas de las regiones densamente cargadas. Consideraremos
este tipo de repulsi\'on actuando s\'olo sobre primeros vecinos
en la red. 

Este tipo de fuerzas
repulsivas se consideran a menudo en problemas de pol\'{\i}meros
\cite{REPULSION4}.

Hemos introducido un modelo en el cual hay un t\'ermino que controla
la magnitud de las fluctuaciones t\'ermicas y otro que simula una
fuerza repulsiva. Los par\'ametros que controlan estos t\'erminos
son respectivamente la temperatura y la {\sl carga}. Veremos como
ajustando ambos par\'ametros es posible mejorar el flujo de
part\'{\i}culas.

\newpage

\begin{center}

{\sl Abstract} 

\end{center}

We study a particle flow model which may be used to get insight into 
various real traffic problems.
The model is implemented using a discrete lattice, in which particles move
towards their destination, fluctuating about the minimal
distance path.  A repulsive interaction between particles is
introduced so as to avoid the appearance of a traffic jam.  We have
studied the parameter space finding regions of fluid traffic, and
saturated ones, both regions being separated by abrupt changes. The 
improvement of the system performance is also explored by 
introducing a non-constant potential acting on the particles.  
Finally, we deal with the behaviour of the system when temporary failures 
in transmission occur.

\section{The Model}

We consider a two-dimensional lattice with coordination number 4, and
periodic boundary conditions. The particles live in the lattice sites,
labeled by $n\equiv(n_0,n_1)$ and can move from its site to one of its four
nearest neighbors at every time step.  The maximum number of particles
that a site can contain will be denoted by $B$, and will be kept fixed
for all the simulation at a value $B=5$.  We denote the occupation number of
the site $n$ by $\sigma(n)$. In this notation, the particles are
prevented from moving to sites with $\sigma(n)=B$.  In addition we
consider an input queue at each site for the particles waiting to be
injected on the lattice. 

By analogy with statistical mechanics systems, we work with the inverse of the
temperature, $\beta$. We denote the charge of the particles by
$\kappa$ and the probability of particle injection to the lattice by
$p$.

The dynamic of the system is as follows:
\begin{enumerate}
\item{} A lattice site is chosen at random.
\item{} A particle is added, with fixed probability $p$, to the queue
of the site, waiting to be introduced to the lattice \footnote{The
size of the queue is as big as necessary, so that it contains all
particles waiting to be fed at that lattice site}
\item{}If the queue of the site is not empty, and $\sigma(n)<B$, a new
particle is introduced in the lattice, and an endpoint assigned to
it at random.
\item{} All particles at the considered site, try to move towards one
of its four neighbors.  For a given particle located at the
position $n$ we must
assign a probability of it jumping to each of its neighbouring locations. 
This probability is given by:
\begin{equation}
P(\pm \mu) = N \exp(\pm \beta {\rm sign}({n^f}_{\mu} - n_{\mu}) - 
\kappa \sigma(n_{\mu}))\ ,
\label{prob}
\end{equation}
where $n_{\mu}$ here signifies the $\mu$ coordinate of site $n$, $n^f$
is the endpoint of the particle being considered, and $N$ is the
normalization constant.  To choose between possible destinations we
use a $Heat$ $Bath$ \cite{SOKAL4} algorithm.

The factor multiplying $\beta$, is a potential term. It implies a
constant force acting on the particle driving it to its
endpoint. The $\kappa$ term produces a repulsion between particles
sitting in nearer neighbors sites.  Obviously there is a wide range of
potentials that could be considered, in order to produce more
effective forces, and partial improvements will be expected.

\item{} Movement is allowed if the chosen site has $\sigma(n) < B$,
otherwise the particle remains at its original site until a new 
movement is attempted in the next iteration.
\item{} A particle is removed from the lattice when it reaches its endpoint.
\end{enumerate}

The smaller $p$ is, the weaker is the effect of the interaction between
the particles. For fixed $(\beta,\kappa)$ the fluid flow of particles 
on the lattice only takes place for those values of $p$ being less 
than a certain threshold which is
$(\beta,\kappa)$ dependent.  Above this threshold, the density becomes
too high and the transmission process is prevented. We say the system
is saturated. The saturation mechanism begins with the appearance of
saturated domains, which grows in size making the movement of the particles 
more and more difficult.

Our purpose is to quantify this threshold density, as well as to
describe the flow properties along the parameter space
$(\beta,\kappa,p)$.

From the point of view of Telecommunication Networks, the lattice
sites can be thought of as nodes of such networks, and $particles$
here can be interpreted as $messages/packets$.
Within this approach, $B$ can be identified with the available
buffer space at the considered node.
In this way, our model can be considered as a first step towards 
an {\em abstract} modeling of packet/message Telecommunication Networks.
 In such networks, packets are routed by each node according to routing
tables dynamically maintained by the network so as to minimize some
{\em cost function} along the trajectory of the packets
\cite{DATANET4}. In the present model, this cost function is simply the
traveled distance, and, as the network state does not change during
simulations, routing tables need not be dynamically calculated. A
natural extension of this model would be to allow variable link
lengths between the nodes and to compute routing tables according to
the path lengths. The role of the temperature parameter is to quantify
how firmly the network will try to stick to its policy of minimizing
the cost along a trajectory: as will be shown below, implementing the
routing according to the cost minimization procedure
in a rigid way ({\em zero temperature}), although this is the best 
{\em naive} choice from an individual
user point of view, can be detrimental to the global behavior of the
network and thus to its collective utility (the total throughput
offered by the network).
 
Congestion is an unavoidable phenomenon in uncontrolled packet
networks, since packets introduced at one node have no guarantee of
finding the necessary resources (available buffer space) in the
transit nodes.
If uncontrolled, congestion result in packet losses which, in the
case of data transmission, are detected by upper layer protocols 
(the Transport Control Protocol, $TCP$, for instance in the case of the 
Internet) and are corrected
by retransmision request. Hence these packet losses will trigger
more packets to be introduced into the network, thus further
amplifying the congestion
Two approaches are possible to avoid this {\em vicious
circle}: reducing the source emission rate when a congestion is
detected (this is the $TCP$ approach) or trying to control
the congestion inside the network by preventing packets from accessing
overloaded areas (this approach does not rely on any source behavior:
in the extreme congestion case the source is simply denied any access
to the network). This last option, congestion control inside the
network, is presently a very active area of research in the networking
community \cite{CONGCONT4}. To implement such a control, the overload
information should be transmitted by a node to its neighbors at
least. This is precisely what is modeled by the repulsive charge term
in the present model: access to a loaded node is discouraged, and
access to an overloaded node is simply forbidden. It should be noted
that in the model as it stands, a node is instantaneously aware
of the load state of its neighbors. A more complex model should take
into account the latency introduced by the node-to-node transit time,
leading to a description of the congestion control by some kind of
{\em retarded potential}. It will be particularly important to take
this latency aspect into account for modeling high speed
networks \cite{LATENCY4}, but this is left for future work.

The system we are considering here is also related to a lattice
gas with fixed number of particles in a constant electric field
${\vec E}$ \cite{ISING4}.
The electric current is related to the traffic 
speed. Also in this model a transition appears, between a
disordered phase and another one were particles move collectively
in the ${\vec E}$ direction.
In our case the electric field is no longer constant, it depends
dynamically on the particle environment: the surrounding electrical
charge and the end point associated with the particle.
As is usual for traffic models, the number of particles is not constant; 
as a consequence, the transition mechanism is different.

\section{Numerical Simulation}

We have performed Monte Carlo (MC) simulations in order to study the
parameter space. Here we present results obtained using a $L\times L$
lattice with $L=32$ and periodic boundary conditions. 
The computations have been carried out on workstations.

The starting configuration is obtained by generating a
particle with probability $p$ at each site of the lattice, therefore, 
the total number of sites occupied initially is
about $p\times L^2$.  The random number generator is based on one
described in \cite{RANDOM4}.  The time step is identified with a MC
iteration, that is the update of $L^2$ lattice sites in the
way described in section 2.

The temporal evolution of the system exhibits a transient regime,
characterized by the instability in the observables (see next
section).  After this, the system falls into an extremely long-living
metastable state, where the flow properties do not change
significantly with the temporal evolution. We say the system has
reached the asymptotic regime.

The time the system spends in the transient regime depends on the
parameter space point $(\beta,\kappa,p)$.
In this regime, in the non-saturated region,
the particle density is initially low and grows with the particle
injection.

This transit time also grows near the parameter space points where an abrupt
change in the properties is observed (e.g. near the threshold
density), reaching in this case up to $2\times 10^4$.

For each value of the parameters, we have performed typically $8 \times
10^4$ MC iterations. For $\beta$ and $\kappa$ $\in(0,0.4)$ the
transient regime takes around 5000 iterations, while above $0.5$ for
both parameters, this time falls to $400-600$ iterations.  We have
also performed the simulations starting from different configurations,
allowing the system to evolve for up to $2 \times 10^5$ iterations. We
have computed the errors by calculating the dispersion between the
results obtained from starting from these different configurations.

\section{Observables}

\begin{figure}[!t]
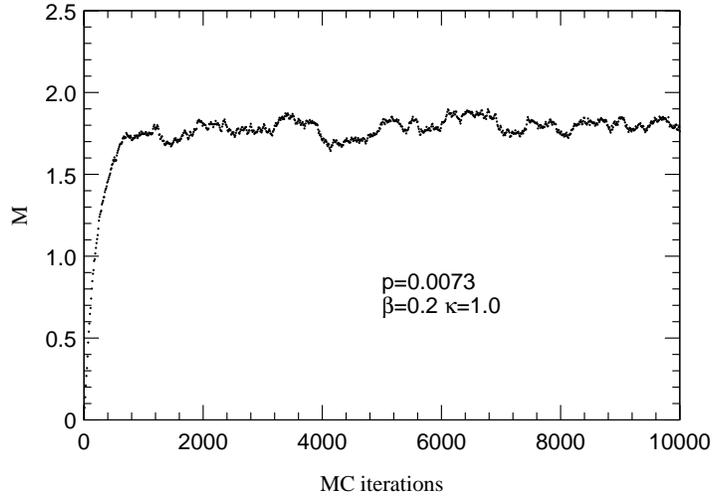

\centerline{
\fpsxsize=7cm
\def\fpsangle{270}
\fpsbox[70 90 579 760]{figure_1.eps} }
\caption{Temporal evolution of $M$ bellow the critical injection 
$p_c$.}
\label{Figure_1}
\end{figure}

A correct description of the system is obtained from the measurement
of the relevant observables.  From their temporal evolution we are able
to evaluate when the asymptotic regime is reached, or even whether 
this regime will or will not be saturated.  
From the averaged value of these magnitudes
we obtain a quantitative description of the flow process.

For a given configuration, we define the occupation $M$ as:
\begin{equation}
M=(1/V)\sum_{i=1}^V \sigma(i)\ .
\label{mag}
\end{equation}
Where V is $L \times L$.

The statistical average over the configurations (labeled by $j$) is:
\begin{equation}
\langle M \rangle =\lim_{N\to\infty}(1/N)\sum_{j=1}^N M_j\ .
\label{magmc}
\end{equation}
$N$ being the number of averaged configurations.

During the time interval $t_j - t_i \equiv \bigtriangleup t$, $n_e$
particles will reach their endpoint. We define the Band Width $(B_W)$
as the number of particles that arrive at the endpoint per time unit.
\begin{equation}
B_W (\bigtriangleup t)= \frac{n_e}{\bigtriangleup t}\ .
\label{bw}
\end{equation}
The statistical average for $B_W$ is obtained from its mean value over
a number of time intervals $N_T$:
\begin{equation}
\langle B_W \rangle=\lim_{N_T\to\infty}(1/N_T)\sum_{j=1}^{N_T}
B_W(j)\ .
\label{bwmc}
\end{equation}

We can obtain a good measure of the system performance from the mean
time taken by the particles to reache their endpoint, $T_M$.
We compute $T_M$ in the time interval $\bigtriangleup t$ by adding
the individual time spent by each particle (delay time), divided by
$n_f$:
\begin{equation}
T_M(\bigtriangleup t)=\frac{\sum_{i=1}^{n_f} T_i}{n_f}
\label{tmi}
\end{equation}
In the same way we define the statistical average as over $N_T$ 
intervals as:
\begin{equation}
\langle T_M \rangle=\lim_{N_T\to\infty}(1/N_T)\sum_{j=1}^{N_T} T_{M}(j)
\label{tmmc}
\end{equation}

The occupation frequency of a certain occupation number, $\sigma(n)$,
is defined as the number of times that the occupation $\sigma(n)$
appears at any lattice site.

\begin{figure}[!t]
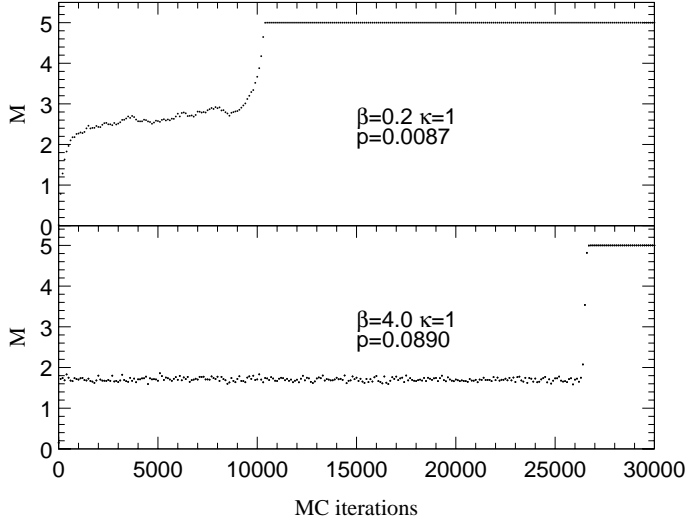

\centerline{
\fpsxsize=7cm
\def\fpsangle{270}
\fpsbox[70 90 579 760]{figure_2.eps} }
\caption{Temporal evolution of $M$ above $p_c$.}
\label{Figure_2}
\end{figure}

\begin{figure}[!t]
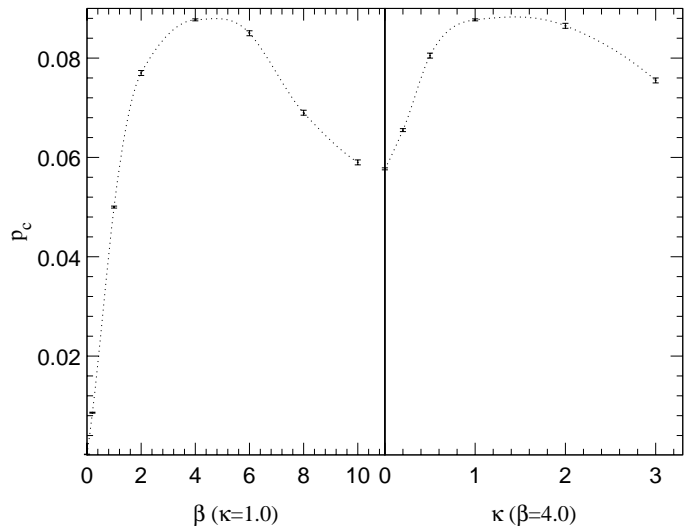

\centerline{
\fpsxsize=7cm
\def\fpsangle{270}
\fpsbox[70 90 579 760]{figure_3.eps} }
\caption{Sections through the phase space cube. On the left side is 
plotted
 $p_c$ versus $\beta$, on the right one, $p_c$ versus $\kappa$.}
\label{Figure_3}
\end{figure}

\section {Phase Diagram}

We examine the parameter space $(\beta,\kappa,p)$ searching for
regions where sharp changes in the temporal evolution arise.  At
each $(\beta,\kappa)$ value there is a $p$ value denoted by $p_c$
such that for $p<p_c$ the asymptotic regime presents a stationary
flow, and for $p>p_c$ the asymptotic regime is saturated and the flow
is no longer possible.  We plot the temporal evolution of $M$ 
below $p_c$ (Figure \ref{Figure_1} ) 
and above $p_c$ (Figure \ref{Figure_2}) for some values of the
parameters.

This change is similar to a phase transition. The temporal evolution
leads the system to one or another phase depending on the parameter
space point.  Once in the asymptotic regime, the non-saturated phase
exhibits a dynamic equilibrium: the number of injected particles, equals
the number of those arriving at their end position. 
This is reflected by $\langle B_W
\rangle =p \times L^2$.  In this phase, $\langle T_M \rangle$ is
constant, as well as $\langle M \rangle$ which is always less than
$B$.

The parameter space is divided into two regions by the surface defined
by $(\beta,\kappa,p_c)$.  Figure \ref{Figure_3} 
shows two sections, for fixed
$\beta$ and $\kappa$ respectively.

Above the surface, after the transient regime $\langle M \rangle=B$ and
$\langle B_W \rangle=0$, $\langle T_M \rangle$ diverges.

It is possible to give a simple interpretation of this congestion
phenomenon; when the load increases so does the probability that two
particles residing in two neighboring saturated sites will want to {\em
exchange} their sites, but of course they cannot do so since one of the
particles would have to move first and then have nowhere to go to
as no resource is will be available at the desired site. 
This mutual blocking then tends to
propagate to other neighboring sites and once above a given load a
complete congestion will develop at the scale of the network. 

Such mutual blocking is well-known in most {\em no loss} routing
schemes such as {\em wormhole} routing and is called {\em deadlock}
\cite{WORM4}. Deadlock avoidance is a very active research area,
specially for massively parallel system interconnections
\cite{DEADAVOID4}.

\subsubsection*{$\beta$ dependence}
 
The thermal fluctuations move particles away from their minimal
distance path. The higher $\beta$ is, the less the importance of these
fluctuations.

\begin{figure}[!t]
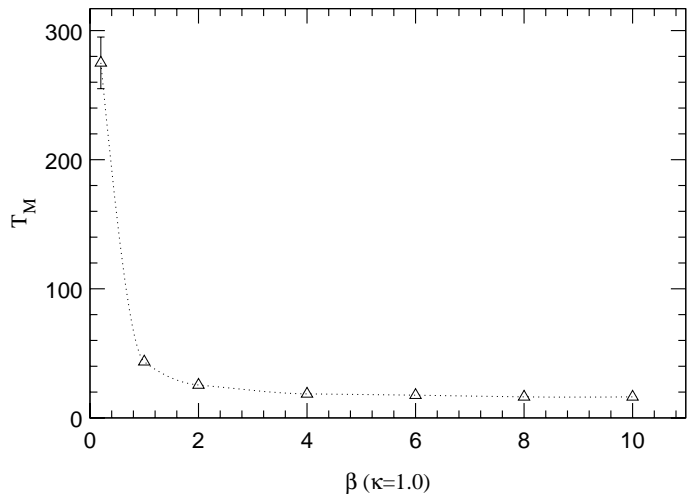

\centerline{
\fpsxsize=7cm
\def\fpsangle{270}
\fpsbox[70 90 579 760]{figure_4.eps} }
\caption{$\langle T_M \rangle$ dependence on $\beta$ for 
$p \approx p_c$.}
\label{Figure_4}
\end{figure}

As shown in Figure \ref{Figure_3} (left side), there is an interval of $\beta$
values where the reduction of fluctuations have a positive influence on
the throughput: $p_c$ rises, as well as $\langle B_W \rangle$, while
$\langle T_M \rangle$ decreases, as shown in Figure \ref{Figure_4}.

When going to higher $\beta$ values, the situation does not
persist. The absence of thermal fluctuations damages the throughput
because particles are not able of going round obstacles. As a
consequence $p_c$ and $\langle B_W \rangle$ decrease. $\langle T_M
\rangle$ does not appreciably change from its minimal value,
corresponding to the infinite directionality one.

Figure \ref{Figure_2} shows how the thermal fluctuations 
influence the saturation
mechanism. If they are important, $M$ grows slowly until it reaches
$B$. If they are not significant a sharp jump appears in the temporal
evolution of $M$, between its value in the transient regime and $B$.
We conclude from this that the saturated domains grow faster in the
absence of fluctuations, as would be expected from the earlier
discussions.

In particular, it is intuitive that a mutual deadlock will last 
longer if thermal fluctuations are disallowed (particles in
mutual deadlock will repeatedly attempt to exchange their sites) hence
creating a larger local congestion area which can eventually evolve
towards a global congestion. This was clearly observed in \cite{BETA4}
where only one buffer was available per site and this behavior is also
exhibited in this model, hence illustrating the well-known fact 
\cite{TANENBAUM4} that deadlock is a consequence
of our {\em no loss} routing scheme \footnote{movement is granted 
if and only if the available resource is available at the low end} 
and not a consequence of insufficient resources.

\begin{figure}[!]
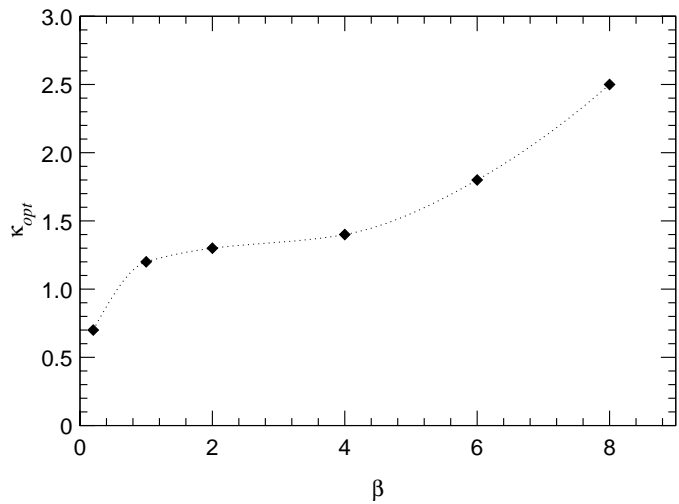

\centerline{
\fpsxsize=7cm
\def\fpsangle{270}
\fpsbox[70 90 579 760]{figure_5.eps} }
\caption{$(\beta,\kappa_{opt})$ line. The errors are of the
size of the $\kappa$ step measured, ($\bigtriangleup \kappa =0.5$.)
}
\label{Figure_5}
\end{figure}

\begin{figure}[!b]
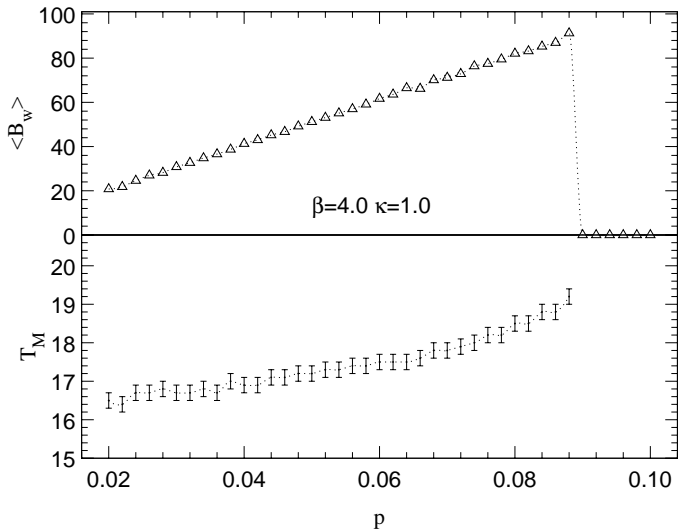

\centerline{
\fpsxsize=7cm
\def\fpsangle{270}
\fpsbox[70 90 579 760]{figure_6.eps} }
\caption{$\langle B_W \rangle$ (upper part) and $\langle T_M \rangle$
(lower part) versus $p$.}
\label{Figure_6}
\end{figure}

\begin{figure}[!t]
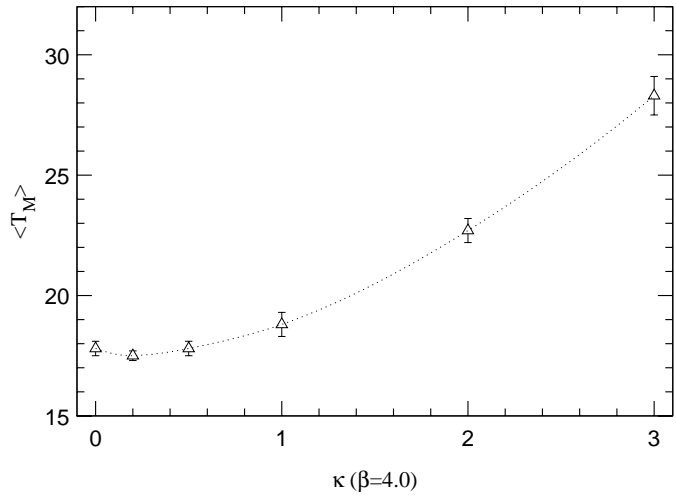

\centerline{
\fpsxsize=7cm
\def\fpsangle{270}
\fpsbox[70 90 579 760]{figure_7.eps} }
\caption{$\langle T_M \rangle$ dependence on $\kappa$.}
\label{Figure_7}
\end{figure}

\begin{figure}[!t]
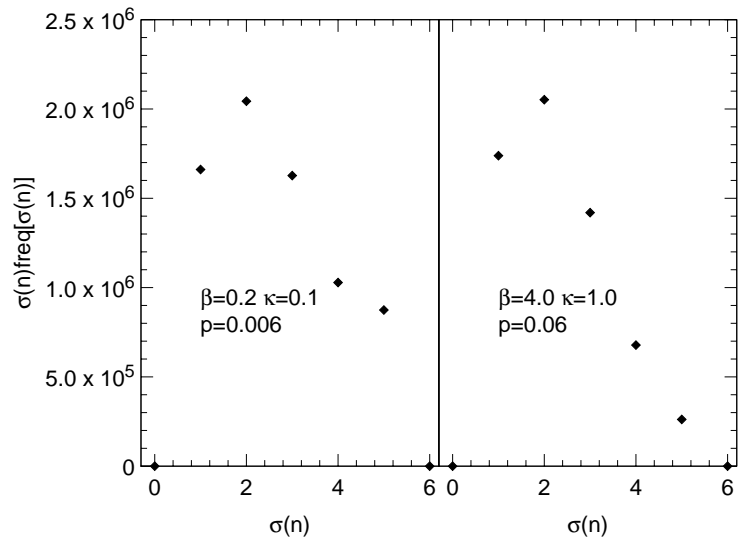

\centerline{
\fpsxsize=7cm
\def\fpsangle{270}
\fpsbox[70 90 579 760]{figure_8.eps} }
\caption{Distribution of the occupation number $\sigma(n)$ times the
frequency of this state versus $\sigma(n)$ after $5\times 10^5$ MC 
iterations. Transient regime contributions has been discarded.
}
\label{Figure_8}
\end{figure}

\begin{figure}[!t]
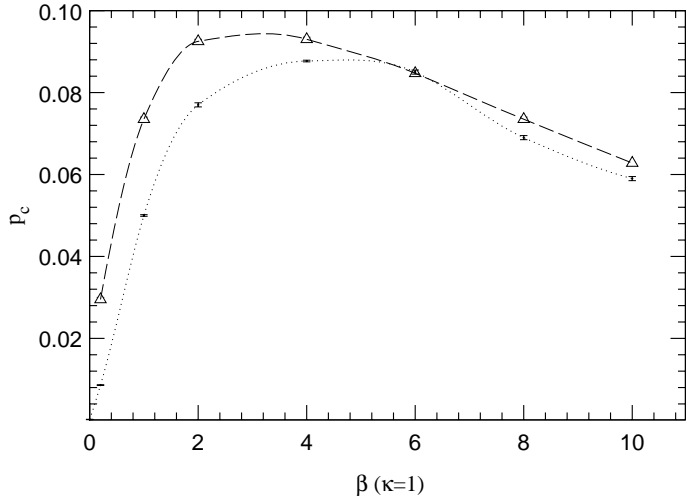

\centerline{
\fpsxsize=7cm
\def\fpsangle{270}
\fpsbox[70 90 579 760]{figure_9.eps} }
\caption{$\beta$ dependence of $p_c$ for the distance dependent 
force (dashed line). Dotted line represents the values obtained for 
the constant force.}
\label{Figure_9}
\end{figure} 

\begin{figure}[!]
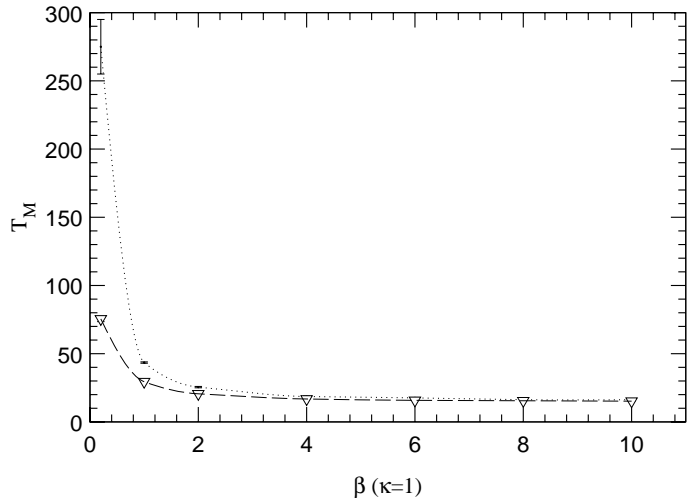

\centerline{
\fpsxsize=7cm
\def\fpsangle{270}
\fpsbox[70 90 579 760]{figure_10.eps} }
\caption{ $<T_M>$ versus $\beta$ for fixed
$\kappa=1$ in $p \approx p_c$ for the distance dependent force (dashed
line) and constant force (dotted line).}
\label{Figure_10}
\end{figure}

\subsubsection*{$\kappa$ dependence} 

The dependence of $p_c$ on $\kappa$ presents two different regions
(see Figure \ref{Figure_3}).

In the first one, $p_c$ rises with increasing $\kappa$. For these
$\kappa$ values, the inclusion of the repulsion term helps the system
to avoid congested regions.  
We denote by $\kappa_{opt}$ the value at which $p_c$ reaches its
maximal value.

In the second region, $\kappa > \kappa_{opt}$, $p_c$ decreases with
$\kappa$.  Above $\kappa_{opt}$ on, the repulsion is too strong and 
the particles move far away from their minimal paths. As a result,
the collapse appears for smaller injection densities.

In Figure \ref{Figure_5} we give the values of $\kappa_{opt}$ for some $\beta$
values. The more a particle is restricted to its minimal path, the
stronger the repulsion needed to avoid obstacles in the
lattice.

Therefore, there is a maximal value for the particle injection
supported by each $\beta$ value, which is reached for $\kappa$ in a
neighborhood of $\kappa_{opt}$.

\section{General Behavior and Optimization}

Let us focus on the behaviour of the other significant observables for
the traffic flow, when $\beta$ and $\kappa$ have been tuned 
to obtain a maximal $p_c$.

\subsubsection*{Behavior of $B_W$}

As we have already pointed out, $B_W$ in the asymptotic regime is
proportional to $p$ by a factor $L^2$, as it may be checked in figure
6. Hence,at each $(\beta,\kappa)$ value the maximum for $\langle B_W
\rangle$ is reached when $p=p_c$.
 
Above $p_c$, $\langle B_W \rangle$ remains constant during the
transient regime, and sharply falls to zero after it.  $B_W$ is a good
measure to determine when the system has reached the
asymptotic regime.

\subsubsection*{Behavior of $T_M$}

The greater $p$ the stronger the effect of the interaction over
the particles: they move further and further away from their minimal path with
increasing $\langle T_M \rangle$ (Figure \ref{Figure_6}).  
In Figure \ref{Figure_7}, we observe
how $\langle T_M \rangle$ slightly increases with $\kappa$ until
$\kappa_{opt}$ is reached 
From $\kappa_{opt}$ on, $\langle T_M \rangle$
increases with a high slope.

We conclude that better performances are obtained for the 
maximal injection
supported and for the bandwidth on the line
$(\beta,\kappa_{opt})$.  The repulsion term damages $T_M$: the greater
$\kappa$, the larger the time the particles take to reach their
destination.  However, it is only for $\kappa>\kappa_{opt}$ that
$\langle T_M \rangle$ increases in a dramatic fashion.

Figure \ref{Figure_8} shows the particle distribution in the simulation for
$\beta=0.2$ and $\beta=4.0$. The plot of $\sigma(n) {\rm
freq}[\sigma(n)]$ exhibits a maximum around $\langle M \rangle$, and
also reveals a wider distribution of particles for small $\beta$
values.

\section{Improving the throughput: distance dependent force.}

\begin{figure}[!]
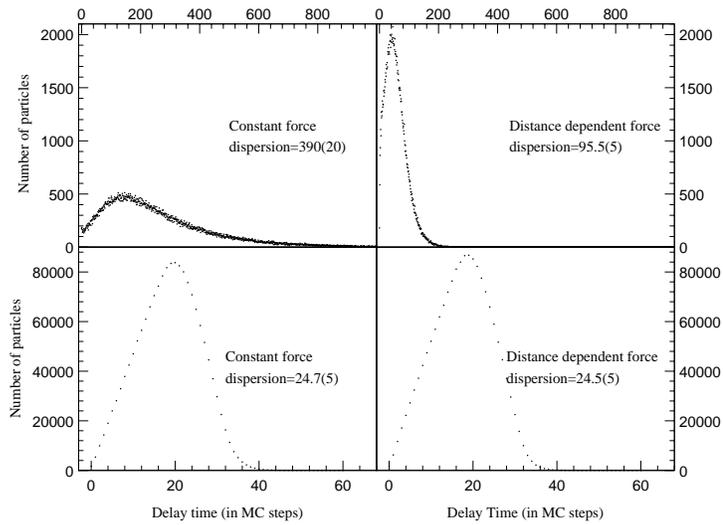

\centerline{
\fpsxsize=7cm
\def\fpsangle{270}
\fpsbox[70 90 579 760]{figure_11.eps} }
\caption{Delay distribution for the whole particles after $4 \times
10^4$ MC iterations. Graphics on the left (constant force) must be
compared with the right side ones (distance dependent force).  The
distributions in the top are calculated in
($\beta=0.2,\kappa=1,p=0.08$) in the bottom they correspond to
($\beta=4,\kappa=1,p=0.008$).
}
\label{Figure_11}
\end{figure}

Although our implementation of the physical system has been inspired
by the search for the simplest model accomplishing the desired
features, partial improvements are to be expected by taking into account
details not included up to this point.
As an example, we develop a possible improvement: 
quenching thermal fluctuations when they are no longer useful,
that is, at the neighborhood of the endpoints.

In the primary implementation, we have
used a constant force and therefore the particles support thermal
fluctuations of the same strength no matter what the remaining 
distance to the endpoint is.  We can expect a globally less
congested lattice
if particles which are only a few sites away from their destination are 
prevented from fluctuating.  We will see how the inclusion of this feature 
does not spoil the good properties of the throughput, and also that 
the general behaviour of the system is not altered.

A possible implementation to take this fact into account is obtained
by introducing a distance dependence in the probability
distribution, such that contributions of the fluctuations
decrease with decreasing distance to the endpoint.

A straightforward way of doing this is to include a dependence on the
relative distance to the endpoint:
the probability distribution now reads:
\begin{equation}
P(\pm \mu) = N \exp(\pm \beta {\rm sign}({n^f}_{\mu} - n_{\mu}) - 
\kappa \sigma(n_{\mu}))\left(\frac{r_n}{r_{n+\mu}}\right)\ .
\label{newprob}
\end{equation}

Where $r_n$ is defined by:
\begin{equation}
r_{(n_0,n_1)}=\sqrt{({n^f}_0 - n_0)^2 + ({n^f}_1 - n_1)^2}\ ,
\end{equation}

In Figure \ref{Figure_9} the evolution of $p_c$ with $\beta$ 
is shown.  A global
throughput improvement is reflected by higher $p_c$ values. This
effect is more remarkable when the size of the thermal fluctuations is
important (small $\beta$ values), corroborating our first intuition on
the effect of fluctuations in the steps preceding the endpoint.

In Figure \ref{Figure_10} we plot the $\beta$ dependence of $T_M$. 
We observe a global decrease in this time for all $\beta$ values.

The improvements concerning the delay time, are not restricted to a
smaller $\langle T_M \rangle$. Figure \ref{Figure_11} 
shows the delay distribution
for all particles, compared with the delay distribution obtained
with the constant potential.  We see at $\beta=0.2$ how the dispersion
of the distribution strongly decreases when using a distance
dependent force.  So, the particles arrive in more similar times,
increasing the uniformity and the reliability of the traffic flow.

\begin{table}[!b]
{
\begin{center}
{
\begin{tabular}{|c|c|c|c|}\hline
Shift  &$n=10$  &$n=100$  &$n=200$  \\ \hline
\multicolumn{4}{|c|}{$\beta=4.0$ $\kappa=1.0$} \\ \hline
$\bigtriangleup_r p_c$                    &0.09    &0.40  &0.58 \\ 
\hline
$\bigtriangleup_r \langle T_M \rangle(\approx p_c)$
  &-0.05   &-0.38 &-0.89  \\ \hline
$\bigtriangleup_r \langle M \rangle (\approx p_c)$   
  &0.00    &0.37  &0.55  \\ \hline
\multicolumn{4}{|c|}{$\beta=0.2$ $\kappa=1.0$} \\ \hline
$\bigtriangleup_r p_c$                      &0.04   &0.24  &0.41 \\ 
\hline
$\bigtriangleup_r \langle T_M \rangle (\approx p_c)$  
  &-0.00  &-0.08 &-0.37  \\ \hline
$\bigtriangleup_r \langle M \rangle (\approx p_c)$   
  &0.00   &0.43  &0.50  \\ \hline
\end{tabular}
}
\end{center}
}
\caption[a]{ Shifted values of relevant observables for 
some values of the number of failures $n$.}
\protect\label{table_shift}

\end{table}

\section{Fault tolerance}

\begin{figure}[!t]
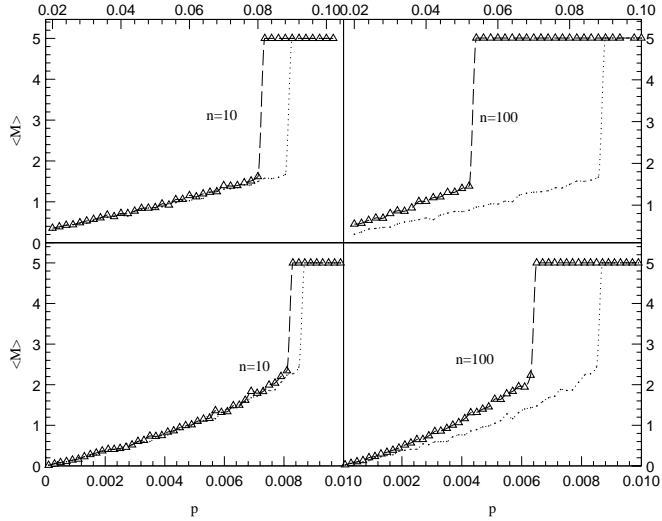

\centerline{
\fpsxsize=7cm
\def\fpsangle{270}
\fpsbox[70 90 579 760]{figure_12.eps} }
\caption{$\langle M \rangle$ versus p for values of the fault
percentage. Dotted lines represent $\langle M \rangle$ values
obtained for an ideal lattice. On the top 
diagram the parameter space point
is ($\beta=4,\kappa=1$), and on the bottom frames the plotted point is
($\beta=0.2,\kappa=1$).}
\label{Figure_12}
\end{figure}

As we have pointed out, the lattice sites may simulate the control nodes
of an information flow system. Needless to say the sites here are 
idealizations of nodes in real system because the possibility of
communication failures is not allowed.  Real nodes are subject to
external factors, such as technical constraints or outside influences,
that often damage the communication ability between some nodes.  These
nodes are then temporarily out of order, and cannot communicate
or receive information from other nodes.  It is therefore of
interest to have an estimate of how robust a system is when temporary
problems in the transmission occur.

To implement the ocurrence of communication problems, 
information
exchange is prevented at $n$ randomly chosen sites during an interval
of $\bigtriangleup t \equiv 10$ $MC$ iterations in the simulation.  After
this time interval these $n$ sites are again allowed to communicate
and another $n$ sites are broken at random \footnote{In this section 
the probability distribution (\ref{prob}) is used}.

In general, the measured values for observables will be shifted by
an amount depending on $n$ and in the parameter region.  For each $p$
value, we define the shift in the observable $O$ as its value relative
to the one obtained in the ideal system:
\begin{equation}
\bigtriangleup_r O(p) = \frac{O(p,n=0)-O(p,n)}{O(p,n=0)}\ .
\label{shift}
\end{equation} 

We have studied the influence on the relevant parameters of
communication failures for $n=10$, $100$ and $200$ with $\beta=0.2$ and
$\beta=4.0$.

Figure \ref{Figure_12} shows the comparative evolution of $\langle M \rangle$
for $n=10$ and $n=100$.

The first consequence of a bad transmission is that the system must
support globally higher occupation numbers. We measure smaller
$\bigtriangleup_r p_c$ values for lower $\beta$, because, as has
been already demonstrated, the lattice supports higher occupation numbers
when the fluctuations are important.

$\bigtriangleup_r \langle M \rangle$ is almost zero for both $\beta$
values with $n=10$ (around 1\% of nodes out of order), while with
$n=100$ and $200$ $\bigtriangleup_r \langle M \rangle$ is always large.

In Table \ref{table_shift} we give the results obtained.  As would be expected
$\langle T_M \rangle$ increases for all $n$ values even though it is
almost zero for $n=10$.

\section{Summary and Outlook}
  
We have studied a model useful for describing the relevant processes
occurring in traffic flow systems with immediate applications to
network message passing and traffic problems in general.

The introduction of the parameters $\beta$ and $\kappa$ as controllers
of the system behavior allows us to go a step further from purely
descriptive models, because we are able to give prescriptions to
improve the performance of the flow process.

Deeper studies are also possible.
Concretely an accurate study of the scaling with $L$ of the relevant
magnitudes as well as a detailed description of how the saturation
time behaves, could lead us to the definition of quantities analogous
to critical exponents.  Also non-equilibrium states could be studied
in order to monitor the parameters controlling the saturation
process.

The introduction of the $\kappa$ parameter has implied that the
particles are able to avoid congested regions. The study has been limited
to short-ranged interactions, because the particles at a site 
only see the occupation
of its nearest neighbors. By informing the particles about the
occupation of wider surrounding regions improvements in the
throughput are expected.

\newpage

\chapter*{Conclusiones}
\thispagestyle{empty}

Las conclusiones de esta memoria se han expuesto individualmente
a lo largo de cada cap\'{\i}tulo. A continuaci\'on pasamos
a resumir y comentar los resultados m\'as importantes que
se han obtenido.

\begin{itemize}

\item{{\bf El Modelo O(4) Anti-ferromagn\'etico}}

La introducci\'on de un acoplo negativo a segundos vecinos en el
modelo $\sigma$ no lineal hace que aparezca frustraci\'on
en los vac\'{\i}os encontrados en el diagrama de fases. La consecuencia
de la frustraci\'on en este modelo es la aparici\'on de antiferromagnetismo
en distintas dimensionalidades. 
La forma que hemos encontrado m\'as simple para introducir antiferromagnetismo
en $4d$ es trabajar en una red $F_4$ con acoplos \-negativos. La transici\'on
del vac\'{\i}o desordenado paramagn\'etico al vac\'{\i}o AF a planos
es la m\'as interesante puesto que los datos num\'ericos son
compatibles con segundo orden
y los exponentes cr\'{\i}ticos no son los de campo medio.

Se han expuesto las explicaciones alternativas, razonables dentro
de la evidencia num\'erica, que pueden dar cuenta de los resultados
num\'ericos.
Probablemente la explicaci\'on menos plausible es la posibilidad
de tener trivialidad logar\'{\i}tmica puesto que la dimensi\'on
an\'omala del campo que hemos encontrado es demasiado distinta
de cero, y este tipo de correcciones dan cuenta de peque\~nas desviaciones
a las predicciones de campo medio.

Como se ha se\~nalado, a la vista de los resultados obtenidos
en O(2) y O(3) es muy dif\'{\i}cil mantener esperanzas de que
la transici\'on en O(4) pueda ser de segundo orden. El escenario
de primer orden d\'ebil parece ser el m\'as plausible.

\item{{\bf La transici\'on de fase en el modelo SU(2)-Higgs}}

Esta transici\'on de fase de nuevo es un ejemplo 
del problema que se encuentra en $d=4$ a la hora de discernir entre
transiciones de primer orden muy d\'ebiles y transiciones
continuas.

Hemos visto que en SU(2)-Higgs la prueba m\'as directa que
demuestra el caracter de primer orden de la transici\'on,
a saber la medici\'on del calor latente, 
no es accesible a un tiempo de c\'alculo razonable.
Se ha estudiado el problema en un espacio de par\'ametros extendido,
lo cual ha ayudado a tener una visi\'on global del mecanismo de 
debilitamiento de la transici\'on de fase. 
La observaci\'on de la tendencia de los
exponentes cr\'{\i}ticos efectivos hacia los valores de primer
orden constituye una evidencia del caracter de primer
orden de la transici\'on.

\item{{\bf U(1) compacto con topolog\'{\i}as toroidal y esf\'erica}}

Este es un problema podr\'{\i}amos decir que cl\'asico en el estudio
de TCC en la red. Las cuestiones abiertas son dos: el orden de la transici\'on
de fase y los mecanismos que la producen.
En concreto se ha venido conjeturando sobre una posible influencia
de las condiciones de contorno en el orden de la transici\'on de 
fase. 

En esta memoria se ha expuesto el estudio de la transici\'on de fase
desconfinante en U(1) puro gauge en redes con topolog\'{\i}a toroidal
y esf\'erica (homot\'opica a $S^4$). Se han estudiado los efectos
de tama\~no finito asociados a ambas topolog\'{\i}as encontrando que en la
red esf\'erica hay m\'as efectos de tama\~no finito incontrolados que
en la toroidal debido a las inhomogeneidades. 

En cuanto al orden de la transici\'on de fase, se ha chequeado la 
existencia de gap de energ\'{\i}a para redes suficientemente grandes
en ambas topolog\'{\i}as, lo cual caracteriza a la transici\'on
como de primer orden. As\'{\i} mismo se ha dado una estimaci\'on
del calor latente en el l\'{\i}mite de volumen infinito.

\item{{\bf Modelos bidimensionales de flujo de part\'{\i}culas}}

Se ha formulado un modelo para describir, en general, sistemas 
que envuelven un flujo de informaci\'on. 

Se ha encontrado que se
pueden obtener mejoras en el flujo permitiendo a las part\'{\i}culas
fluctuar alrededor de la trayectoria de m\'{\i}nima distancia.
La introducci\'on del par\'ametro $\kappa$ permite al sistema
evitar regiones altamente congestionadas. 

La combinaci\'on apropiada de flexibilidad en la
direccionalidad del movimiento de las part\'{\i}culas hacia su
destino, es decir fluctuaciones t\'ermicas, y posibilidad de evitar
zonas congestionadas, es decir el par\a'metro $\kappa$, es lo que
permite un flujo \'optimo de part\'{\i}culas en la red.

\end{itemize}

\newpage

\chapter*{Relaci\'on de publicaciones}
\thispagestyle{empty}                             

\begin{enumerate}

\item{  {\large {\it ``The Confining Higgs phase transition 
	in U(1)-Higgs \-Lattice Gauge Theory ''}} \\
	Colaboraci\'on RTN (J.L. Alonso et al.) \\
	{\sl Phys. Lett.} {\bf B296} (1992) p 154 }
\vspace{0.7cm}
\item{  {\large {\it ``The U(1)-Higgs model:critical behavior in the confining
	Higgs region ''}} \\
	Colaboraci\'on RTN (J.L. Alonso et al.) \\
	{\sl Nuc. Phys.} {\bf B405} (1993) p 574 }
\vspace{0.7cm}
\item{  {\large {\it ``The U(1)-Higgs model: study of the confining Higgs 
	transition ''}} \\
	Colaboraci\'on RTN (J.L. Alonso et al.) \\
	{\sl Nuc. Phys.} {\bf B30} (Proc. Suppl.) (1993) p 701}
\vspace{0.7cm}
\item{  {\large {\it ``Instanton like contribution to the dynamics of the
	Yang-Mills fields on the twisted torus ''}} \\
	Colaboraci\'on RTN (J.L. Alonso et al.) \\
	{\sl Phys. Lett.} {\bf B305} (1993) p 366 }
\vspace{0.7cm}
\item{  {\large {\it ``Critical behavior of Random Walks''}} \\
	I. Campos y A. Taranc\'on, \\
	{\sl Phys. Rev.} {\bf E50} (1994) p 91 }
\vspace{0.7cm}
\item{  {\large {\it ``Thermal and repulsive traffic flow''}} \\
	I. Campos, F. Clerot, L.A. Fern\'andez y A. Taranc\'on, \\
	{\sl Phys. Rev.} {\bf E52} (1995) p 5946}
\vspace{0.7cm}
\item{  {\large {\it ``Anti-ferromagnetic 4D O(4) model''}} \\
	I. Campos, L.A. Fern\'andez y A. Taranc\'on \\
	{\sl Phys. Rev.} {\bf D55} (1997) p 2965 }
\vspace{0.7cm}
\item{  {\large {\it ``Anti-ferromagnetism in four dimensions: the search for
	non triviality''}} \\
	J.L. Alonso et al. \\
	{\sl Nuc. Phys.} {\bf B53} (Proc. Suppl.) (1997) p 680}
\vspace{0.7cm}
\item{  {\large {\it ``On the SU(2)-Higgs phase transition''}} \\
	I. Campos \\
	{\sl Nuc. Phys.} {\bf B514} (1998) p 336}
\vspace{0.7cm}
\item{  {\large {\it ``The order of the SU(2)-Higgs phase transition''}} \\
	I. Campos \\
	{\sl Nuc. Phys.} {\bf B63} (Proc. Suppl.) (1998) p 676}
\vspace{0.7cm}
\item{  {\large {\it ``First order signatures in the 4D pure compact U(1) gauge
	\-theory with toroidal and spherical topologies''}} \\
	I. Campos, A. Cruz y A. Taranc\'on \\
	{\sl Phys. Lett.} {\bf B424} (1998) p 328 }
\vspace{0.7cm}
\item{  {\large{\it ``A study of the phase transition in 4D pure compact U(1)
	LGT on toroidal and spherical lattices''}} \\
	I. Campos, A. Cruz y A. Taranc\'on \\
	{\sl Aceptado en Nuc. Phys. B}	}

\end{enumerate}

\newpage

\chapter*{${\mathcal EPILOGO}$}

\vspace{1.5cm}

{\sl {\Large 

En efecto, rematado ya su juicio, vino a dar en el m\'as
extra\~no pensamiento que jam\'as di\'o loco en el mundo,
y fue que le pareci\'o convenible y necesario, as\'{\i}
para el aumento de su honra como para el servicio de su rep\'ublica,
hacerse caballero andante e irse por todo el mundo con sus armas y
su caballo a buscar aventuras y a ejercitarse en todo el aquello
que \'el hab\'{\i}a le\'{\i}do que los caballeros andantes se
ejercitaban, desfaciendo todo g\'enero de agravio, y poni\'endose
en ocasiones y peligros, donde, acab\'andolos, cobrase eterno
nombre y fama.

}
}

\vspace{1cm}
\begin{flushright}
{\sl Cap\'{\i}tulo primero: Que trata de la condici\'on y ejercicio
del famoso hidalgo Don Quijote de la Mancha} \\
{\bf ``El Ingenioso hidalgo Don Quijote de la Mancha''} \\
{\sl (Miguel de Cervantes Saavedra)}
\end{flushright}

\end{document}